\documentclass[%
 reprint,
nofootinbib,
 amsmath,amssymb,
 aps,
 prd,
]{revtex4-2}

\usepackage[usenames,dvipsnames]{xcolor}
\clearpage{}%

\newcommand\bigO{\mathcal{O}}

\usepackage{graphicx}%
\graphicspath{{./}{figures/}}
\usepackage[subrefformat=parens,labelformat=parens,caption=false]{subfig}
\usepackage{dcolumn}%
\usepackage{bm}%
\usepackage{hyperref}%
\hypersetup{
	colorlinks=true,
	citecolor=NavyBlue,
	linkcolor=NavyBlue,
	urlcolor=NavyBlue,
}
\usepackage[acronym]{glossaries}
\newacronym{GSN}{GSN}{Generalized Sasaki-Nakamura}
\newacronym{SN}{SN}{Sasaki-Nakamura}
\newacronym{MST}{MST}{Mano-Suzuki-Takasugi}
\newacronym{BH}{BH}{black hole}
\newacronym{GW}{GW}{gravitational wave}
\newacronym{ODE}{ODE}{ordinary differential equation}
\newacronym{QNM}{QNM}{quasinormal mode}
\newacronym{EMRI}{EMRI}{extreme mass-ratio inspiral}
\newacronym{LIGO}{LIGO}{Laser Interferometer Gravitational-Wave Observatory}
\newacronym{LISA}{LISA}{Laser Interferometer Space Antenna}
\newacronym{DECIGO}{DECIGO}{Deci-hertz Interferometer Gravitational wave Observatory}
\newacronym{AD}{AD}{automatic differentiation}

\makeatletter
\newcommand*{\glsplainhyperlink}[2]{%
  \colorlet{currenttext}{.}%
  \colorlet{currentlink}{\@linkcolor}%
  \hypersetup{linkcolor=currenttext}%
  \hyperlink{#1}{#2}%
  \hypersetup{linkcolor=currentlink}%
}
\let\@glslink\glsplainhyperlink
\makeatother

\begin{document}

\title{Recipes for computing radiation from a Kerr black hole using a generalized Sasaki-Nakamura formalism: Homogeneous solutions}

\author{Rico K.~L. Lo}
\email{kalok.lo@nbi.ku.dk}
\affiliation{%
Niels Bohr International Academy, Niels Bohr Institute, Blegdamsvej 17, DK-2100 Copenhagen, Denmark
}%
\affiliation{%
LIGO, California Institute of Technology, Pasadena, California 91125, USA
}%

\date{\today}%

\begin{abstract}
Central to black hole perturbation theory calculations is the Teukolsky equation that governs the propagation and the generation of radiation emitted by Kerr black holes. However, it is plagued by a long-ranged potential associated with the perturbation equation and hence a direct numerical integration of the equation is challenging. Sasaki and Nakamura devised a formulation that transforms the equation into a new equation that is free from the issue for the case of outgoing gravitational radiation. The formulation was later generalized by Hughes to work for any type of radiation. In this work, we revamp the Generalized Sasaki-Nakamura (GSN) formalism and explicitly show the transformations that convert solutions between the Teukolsky and the GSN formalism for both in- and outgoing radiation of scalar, electromagnetic, and gravitational type. We derive all necessary ingredients for the GSN formalism to be used in numerical computations. In particular, we present a new numerical implementation of the formalism, \texttt{GeneralizedSasakiNakamura.jl}, that computes homogeneous solutions to both perturbation equations in the Teukolsky and the GSN formalism. The code works well at low frequencies and is even better at high frequencies by leveraging the fact that black holes are highly permeable to waves at high frequencies. This work lays the foundation for an efficient scheme to compute gravitational radiation from Kerr black holes and an alternative way to compute quasinormal modes of Kerr black holes.

\end{abstract}

\maketitle

\section{Introduction \label{sec:intro}}
The first detection of a binary black hole merger by the two detectors of the \gls{LIGO} in 2015 \cite{LIGOScientific:2016aoc} marked the beginning of a new era in physics where scientists can directly observe gravitational radiation emitted from collisions of compact objects such as \glspl{BH}, allowing the strong field regime of gravity to be probed. Subsequent observing runs of the Advanced \gls{LIGO} \cite{LIGOScientific:2014pky}, Advanced Virgo \cite{VIRGO:2014yos}, and KAGRA \cite{Somiya:2011np, Aso:2013eba, KAGRA:2020tym} detectors have unveiled about a hundred more such \glspl{GW} coming from the collisions of compact objects \cite{LIGOScientific:2016dsl, LIGOScientific:2018mvr, LIGOScientific:2020ibl, LIGOScientific:2021djp}. With planned updates to the current detectors \cite{KAGRA:2013rdx} and constructions of new detectors \cite{Reitze:2019iox, Punturo:2010zza}, some targeting different frequency ranges such as the \acrlong{LISA} \cite{LISA:2017pwj} and the \acrlong{DECIGO} \cite{Kawamura:2020pcg}, we will be observing \glspl{GW} coming from various kinds of sources on a regular basis.

In order to identify \gls{GW} signals from noisy data and characterize properties of their sources, it is imperative to have theoretical understanding of what those waveforms look like so that we can compare them with observations.
Gravitational waveforms can be computed using a number of approaches, such as numerically solving the full nonlinear Einstein field equation or solving a linearized field equation as an approximation. \gls{BH} perturbation theory is one such approximation scheme where the dynamical spacetime is decomposed into a stationary background spacetime and a small radiative perturbation on top of it. The metric of the background spacetime is known exactly, and we only need to solve, usually numerically, for the metric perturbation. See, for example Refs.~\cite{Mino:1997bx, Sasaki:2003xr, Pound2020, GWPT} for a comprehensive review on \gls{BH} perturbation theory.

At the core of \gls{BH} perturbation theory is the Teukolsky formalism \cite{PhysRevLett.29.1114, Teukolsky:1973ha, Press:1973zz, Teukolsky:1974yv} where a rotating (and uncharged) \gls{BH} of mass $M$ and angular momentum per unit mass $a$ is used as the background spacetime. The metric for such a spacetime is known as the Kerr metric \cite{Kerr:1963ud}, and in the Boyer-Lindquist coordinates $(t,r,\theta, \phi)$ the exact line element $ds$ is given by \cite{Boyer:1966qh, Misner:1973prb}
\begin{multline}
	ds^2 = 	-\left(1 - \dfrac{2Mr}{\Sigma} \right) dt^2 - \dfrac{4Mar\sin^2 \theta}{\Sigma} dt d\phi + \dfrac{\Sigma}{\Delta} dr^2 \\
	+ \Sigma d\theta^2 + \sin^2 \theta \left( r^2 + a^2 + \dfrac{2Ma^2 r\sin^2 \theta}{\Sigma} \right) d\phi^2,
\end{multline}
where $\Sigma \equiv r^2 + a^2 \cos^2 \theta$ and $\Delta \equiv r^2 - 2Mr + a^2 = (r - r_{+})(r - r_{-})$ with $r_{+} = M + \sqrt{M^2 - a^2}$ as the outer event horizon and $r_{-} = M - \sqrt{M^2 - a^2}$ as the inner Cauchy horizon.
In the Teukolsky formalism, instead of solving directly the perturbed radiative field (e.g., the metric for gravitational radiation and the electromagnetic field tensor for electromagnetic radiation), we solve for its (gauge-invariant) scalar projections onto a tetrad. For instance, the (Weyl) scalars $\psi_0$ and $\psi_4$ contain information about the in- and the outgoing gravitational radiation, respectively \cite{PhysRevLett.29.1114}. Teukolsky showed that these scalar quantities all follow the same form of the master equation (aptly named the Teukolsky equation), and it is given by \cite{PhysRevLett.29.1114}
\begin{multline}
\label{eq:fullTeukolskyeqn}
\left[ \dfrac{\left( r^2 + a^2 \right)^2}{\Delta} - a^2 \sin^2 \theta \right] \dfrac{\partial^2 \psi}{\partial t^2} + \dfrac{4Mar}{\Delta} \dfrac{\partial^2 \psi}{\partial t \partial \phi} \\
+ \left[ \dfrac{a^2}{\Delta} - \dfrac{1}{\sin^2 \theta} \right] \dfrac{\partial^2 \psi}{\partial \phi^2} - \Delta^{-s} \dfrac{\partial}{\partial r} \left( \Delta^{s+1} \dfrac{\partial \psi}{\partial r} \right) \\
- \dfrac{1}{\sin \theta} \dfrac{\partial}{\partial \theta} \left( \sin \theta \dfrac{\partial \psi}{\partial \theta} \right) - 2s \left[ \dfrac{a\left(r-M\right)}{\Delta}  + \dfrac{i \cos \theta}{\sin^2 \theta} \right] \dfrac{\partial \psi}{\partial \phi} \\
- 2s \left[ \dfrac{M \left( r^2 - a^2 \right)}{\Delta} - r - ia\cos \theta \right] \dfrac{\partial \psi}{\partial t} \\
+ \left( s^2 \cot^2 \theta - s \right) \psi = 4\pi \Sigma T,
\end{multline}
where $T$ is a source term for the Teukolsky equation, and $\psi$ can correspond to different scalar projections with different spin weights $s$. In particular, $s=0$ for scalar radiation, $s=\pm 1$ for in- and outgoing electromagnetic radiation, respectively, and $s=\pm 2$ for in- and outgoing gravitational radiation, respectively. For example, $\psi_0$ satisfies Eq.~\eqref{eq:fullTeukolskyeqn} by setting $\psi \equiv \psi_0$ and $s = 2$, whereas $\psi_4$ satisfies the equation by setting $\psi \equiv (r - ia\cos \theta)^4 \psi_{4}$ and $s=-2$.

Despite its fearsome look, Eq.~\eqref{eq:fullTeukolskyeqn} is actually separable by writing $\psi(t,r,\theta, \phi) = R(r) S(\theta, \phi)e^{-i\omega t}$. The separation of variables gives one \gls{ODE} for the angular part in $\theta$ (since the $\phi$ dependence must be $\psi \sim e^{im\phi}$ with $m$ being an integer due to the azimuthal symmetry of a Kerr \gls{BH}) and another \gls{ODE} for the radial part in $r$. We discuss the angular part of the Teukolsky equation and the recipes for solving the equation numerically more in depth in Appendix~\ref{app:swsh}.
Limiting ourselves to consider the source-free ($T=0$) case for now,\footnote{We consider the $T \neq 0$ case in a subsequent paper (see Sec.~\ref{subsec:inhomogeneous_usingGSN}).} the \gls{ODE} for the radial part is given by \cite{PhysRevLett.29.1114} 
\begin{equation}
\label{eq:radialTeukolskyeqn}
	\Delta^{-s} \dfrac{d}{dr}\left( \Delta^{s+1} \dfrac{dR}{dr} \right) - V_{\mathrm{T}}(r) R = 0,
\end{equation}
with
\begin{equation}
\label{eq:VT}
	V_{\mathrm{T}}(r) = \lambda - 4is\omega r - \dfrac{K^2 - 2is(r-M)K}{\Delta},
\end{equation}
where $K \equiv (r^2 + a^2)\omega - ma$, and $\lambda$ is a separation constant related to the angular Teukolsky equation [see Appendix~\ref{app:swsh}, and in particular, Eq.~\eqref{eq:lambda_const}]. The general solution of $\psi(t,r,\theta,\phi)$ can then be written as
\begin{equation}
	\psi(t, r, \theta, \phi) = \sum_{\ell m \omega} {}_{s}R_{\ell m \omega}(r) {}_{s}S_{\ell m \omega}(\theta, \phi)e^{-i\omega t},
\end{equation}
where $\ell$ labels an eigenfunction of the angular Teukolsky equation (cf. Appendix~\ref{app:swsh}).

While the radial Teukolsky equation in Eq.~\eqref{eq:radialTeukolskyeqn} looks benign, it is challenging to solve it numerically in that form because the potential associated with the \gls{ODE} is long ranged. To see this, we can recast Eq.~\eqref{eq:radialTeukolskyeqn} into the Schrödinger equation form that is schematically given by 
\begin{equation}
\label{eq:Schrodingereqn}
	\dfrac{d^2 Y}{dr_{*}^2} + \left( \omega^2 - V_{Y} \right) Y = 0,
\end{equation}
with $r_{*}$ being the tortoise coordinate for Kerr \glspl{BH} defined by
\begin{equation}
\label{eq:drstardr}
    \dfrac{dr_*}{dr} = \frac{r^2 + a^2}{\Delta},
\end{equation}
where $Y$ is some function transformed from the Teukolsky function $R$, and $V_{Y}$ is the potential associated with the \gls{ODE} \cite{Teukolsky:1973ha}.
For the radial Teukolsky equation, the potential $V_{Y}$ is long ranged\footnote{A prime example of a long-ranged potential is the Coulomb potential in electrostatics.} in the sense that $V_{Y} \sim -2is\omega/r$ as $r \to \infty$, as opposed to a short-ranged potential that falls at $1/r^{n}$ with $n \geq 2$ (for an illustration, see Fig.~\ref{fig:canonical_potentials}). The long rangedness of the potential $V_{\mathrm{T}}$ implies that the two wavelike ``leftgoing'' and ``rightgoing'' solutions of Eq.~\eqref{eq:radialTeukolskyeqn} will have different power-law dependences of $r$ in their wave amplitudes as $r \to \infty$ \cite{Teukolsky:1973ha, Hughes:2000pf}. A direct numerical integration of Eq.~\eqref{eq:radialTeukolskyeqn} will suffer from the problem where the solution with a higher power of $r$ in its asymptotic amplitude will overwhelm the other solution and eventually take over the entire numerical solution due to finite precision in computation when $r$ becomes large \cite{Teukolsky:1973ha, Hughes:2000pf}. In fact, the same problem arises when $r \to r_{+}$ (equivalently when $\Delta \to 0$) where the left- and the right-going waves have again different power-law dependences of $\Delta$ in their wave amplitudes and the solution with a smaller power of $\Delta$ in its asymptotic amplitude will overwhelm the other one numerically as $\Delta \to 0$ \cite{Teukolsky:1973ha, Hughes:2000pf}.\footnote{Refer to Sec.~\ref{subsec:asymptotic_behaviors} for more details and the explicit dependence in $r$ and $\Delta$ for the asymptotic wave amplitudes of $R$ approaching infinity and the horizon, respectively.} Therefore, a direct numerical integration, at least with the Boyer-Lindquist coordinates, is not suitable for solving the radial Teukolsky equation accurately.

Fortunately, there are other techniques that can get around this issue and allow us to solve for $R(r)$ accurately. One such technique is the \gls{MST} method \cite{Mano:1996vt}, originally as a low-frequency expansion and later extended by Fujita and Tagoshi \cite{10.1143/PTP.112.415, 10.1143/PTP.113.1165} as a numerical method for solving the homogeneous radial Teukolsky equation at arbitrary frequency. The \gls{SN} formalism \cite{SASAKI198185, SASAKI198268, 10.1143/PTP.67.1788}, which is the main topic of this paper (and subsequent papers), also enables accurate and efficient numerical computations of homogeneous solutions to the radial Teukolsky equation. In short, Sasaki and Nakamura devised a class of transformations, originally only for $s=-2$, that convert the radial Teukolsky equation with the long-ranged potential $V_{\mathrm{T}}$ into another \gls{ODE} with a short-ranged potential. One can then solve the numerically better-behaved \gls{ODE} instead. The transformations were later generalized by Hughes \cite{Hughes:2000pf} to work for arbitrary integer spin weight $s$.

Comparing to the \gls{MST} method, the \gls{GSN} formalism is conceptually simpler and thus easier to implement. Practically speaking, the \gls{MST} method expresses a homogeneous solution to the radial Teukolsky solution $R(r)$ in terms of special functions, which makes it ideal for analytical work. However, for numerical work there are no closed-form expressions for these special functions and oftentimes the evaluations of these special functions involve solving some \glspl{ODE} numerically \cite{10.5555/1403886}. Thus, efficiencywise, the \gls{GSN} formalism is not inferior, at the very least, to the \gls{MST} method even at low frequencies. On the other hand, while the extension of the \gls{MST} method by Fujita and Tagoshi \cite{10.1143/PTP.112.415, 10.1143/PTP.113.1165} allows the method to, in principle, compute homogeneous solutions at arbitrary frequency, practically the authors of Refs.~\cite{10.1143/PTP.112.415, 10.1143/PTP.113.1165} reported that it was numerically challenging to find solutions when wave frequencies become somewhat large. The \gls{GSN} formalism, as we will show later, becomes even \emph{more} efficient in those cases at high frequencies.

Another appealing capability of the \gls{SN} formalism has to do with computing solutions to the inhomogeneous radial Teukolsky equation. The solutions encode the physical information about the radiation emitted by a perturbed \gls{BH}, say, for example, the \gls{GW} emitted when a test particle plunges toward a \gls{BH}. Based on the \gls{SN} transformation (for the source-free case), the \gls{SN} formalism has a prescription to convert a Teukolsky source term that could be divergent, near infinity or the horizon (or both), into a well-behaved source term.\footnote{For more discussions on solving the inhomogeneous radial Teukolsky equation using the \gls{SN} formalism, see Sec.~\ref{subsec:inhomogeneous_usingGSN}.}

In this paper, we revamp the \gls{GSN} formalism for the source-free case to take full advantages of the formalism for computing radiation from a Kerr \gls{BH}. We explicitly show the \gls{GSN} transformations for physically relevant radiation fields ($s=0,\pm 1, \pm 2$) that transform the radial Teukolsky equation with a long-ranged potential into a new \gls{ODE}, referred to as the \gls{GSN} equation, which has a short-ranged potential instead. To aid numerical computations using the \gls{GSN} formalism, we derive expressions for the higher-order corrections to the asymptotic solutions of the \gls{GSN} equation, improving the accuracy of numerical solutions. We also derive expressions for the frequency-dependent conversion factors that convert asymptotic amplitudes of \gls{GSN} solutions to that of their corresponding Teukolsky solutions, which are needed in wave scattering problems and computations of inhomogeneous solutions.

Furthermore, we describe an open-source implementation of the aforementioned \gls{GSN} formalism that is written in \texttt{julia} \cite{Julia-2017}, a modern programming language designed with numerical analysis and scientific computing in mind. The numerical implementation leverages the reformulation of the \gls{GSN} equation, which is a second-order linear \gls{ODE}, into a form of first-order nonlinear \gls{ODE} known as a Riccati equation to gain additional performance. Our new code is validated by comparing results with an established code \texttt{Teukolsky} \cite{BHPToolkit} that implements the \gls{MST} method.

The paper is structured as follows: In Sec.~\ref{sec:formalism}, we first review the \gls{GSN} formalism for the source-free case. We then derive the asymptotic behaviors and the appropriate boundary conditions for solving the \gls{GSN} equation. In Sec.~\ref{sec:numerical_implementation}, we describe our numerical implementation of the \gls{GSN} formalism and compare it with the \gls{MST} method. Finally, in Sec.~\ref{sec:conclusion_and_future_work} we summarize our results and briefly discuss two applications of the \gls{GSN} formalism developed in this paper, namely laying the foundation for an efficient procedure to compute gravitational radiation from \glspl{BH} near both infinity and the horizon and as an alternative method for determining \glspl{QNM}. For busy readers, in Appendix~\ref{app:explicitGSN} we give ``ready-to-use'' expressions for both the \gls{GSN} transformations and the asymptotic solutions to the corresponding \gls{GSN} equation, as well as the conversion factors to convert between the Teukolsky and the \gls{GSN} formalism.

Throughout this paper, we use geometric units $c=G=M=1$, and a prime to denote differentiation with respect to $r$.

\section{Generalized Sasaki-Nakamura formalism \label{sec:formalism}}
In this section, we first review, following Ref.~\cite{Hughes:2000pf} closely, the core idea behind the \gls{GSN} formalism, i.e. performing a transformation, which is different for each spin weight $s$, from the Teukolsky function $R(r)$ into a new function $X(r_{*})$. This new function $X(r_{*})$ is referred to as the \gls{GSN} function, expressed in the tortoise coordinate $r_{*}$ (for Kerr \glspl{BH}) instead of the Boyer-Lindquist $r$ coordinate. A defining feature of the $r_{*}$ coordinate is that it maps the horizon to $r_{*} \to -\infty$ and infinity to $r_{*} \to \infty$.
The \gls{GSN} transformations were chosen such that the new \gls{ODE} that $X(r_{*})$ satisfies, which is referred to as the \gls{GSN} equation, is more suitable for numerical computations than the original radial Teukolsky equation in Eq. \eqref{eq:radialTeukolskyeqn}. We then study the leading asymptotic behaviors, approaching the horizon $r \to r_{+} \, (r_{*} \to -\infty)$ and approaching infinity $r \to \infty \, (r_{*} \to \infty)$, of both the \gls{GSN} equation and the \gls{GSN} transformations to establish the boundary conditions to be imposed, as well as the conversion factors for converting the complex amplitude of a \gls{GSN} function to that of the corresponding Teukolsky function at the two boundaries. To aid numerical computations when using numerically finite inner and outer boundaries (in place of negative and positive infinity, respectively, in the $r_{*}$ coordinate), we also derive the higher-order corrections to the asymptotic boundary conditions.

\subsection{Generalized Sasaki-Nakamura transformation \label{subsec:transformation}}
The \gls{GSN} transformation can be broken down into two parts. The first part transforms the Teukolsky function $R(r)$ and its derivative $R'(r)$ into a new set of functions $\left(\chi(r), \chi'(r)\right)$ as an intermediate step. In general, we write such a transformation as
\begin{equation}
\label{eq:GDTeqn}
    \chi(r) = \tilde{\alpha}(r) R(r) + \tilde{\beta}(r) R'(r),
\end{equation}
where $\tilde{\alpha}(r)$ and $\beta(r)$ are weighting functions that generate the transformation. This kind of transformation is also known as a generalized Darboux transformation \cite{Glampedakis:2017rar}, but differs from a ``conventional'' Darboux transformation in that the weighting function $\tilde{\beta}(r)$ for a conventional Darboux transformation is a constant instead of a function of $r$. For later convenience, we rescale $\tilde{\beta}$ by $\Delta^{s+1}$ and write $\alpha(r) = \tilde{\alpha}(r)$ and $\beta(r) = \tilde{\beta}(r) \Delta^{-(s+1)}$. Differentiating Eq.~\eqref{eq:GDTeqn} with respect to $r$ and packaging them into a matrix equation, we have \cite{Hughes:2000pf}
\begin{equation}
\label{eq:forwardtransformation}
    \begin{pmatrix}
        \chi \\ \chi'
    \end{pmatrix} = \begin{pmatrix}
        \alpha & \beta \Delta^{s+1} \\
        \alpha' + \beta V_{\mathrm{T}}\Delta^s & \alpha + \beta' \Delta^{s+1} \\
    \end{pmatrix} \begin{pmatrix}
        R \\ R'
    \end{pmatrix},
\end{equation}
where we have used Eq.~\eqref{eq:radialTeukolskyeqn} to write $R^{\prime\prime}$ in terms of $R,R'$ as
\begin{equation}
\label{eq:Rpp_intermsof_RRp}
	R''(r) = \dfrac{V_{\mathrm{T}}}{\Delta}R(r) - \dfrac{2(s+1)(r-1)}{\Delta}R'(r).
\end{equation}

The inverse transformation going from $\left(\chi(r), \chi'(r)\right)$ to $\left(R(r),R'(r)\right)$ is obtained by inverting Eq.~\eqref{eq:forwardtransformation} and is given by \cite{Hughes:2000pf}
\begin{equation}
\label{eq:inversetransformation}
	\begin{pmatrix}
		R \\ R'
	\end{pmatrix} = \dfrac{1}{\eta} \begin{pmatrix}
		\alpha + \beta' \Delta^{s+1} & -\beta \Delta^{s+1} \\
		-(\alpha' + \beta V_{\mathrm{T}} \Delta^{s}) & \alpha \\
	\end{pmatrix} \begin{pmatrix}
		\chi \\ \chi'
	\end{pmatrix},
\end{equation}
where $\eta(r)$ is the determinant of the above matrix, which is given by \cite{Hughes:2000pf}
\begin{equation}
\label{eq:eta}
    \eta = \alpha \left( \alpha + \beta' \Delta^{s+1} \right) - \beta \Delta^{s+1} \left( \alpha' + \beta V_{\mathrm{T}}\Delta^s \right).
\end{equation}
In the second step of the \gls{GSN} transformation, we further rescale $\chi(r)$ to $X(r_*)$ (the motivation of doing so can be found in Ref.~\cite{Hughes:2000pf}) by
\begin{equation}
\label{eq:Xfromchi}
    X(r_*(r)) = \chi(r) \sqrt{(r^2 + a^2) \Delta^s},
\end{equation}
where an analytical expression of $r_{*}(r)$ can be obtained by integrating Eq.~\eqref{eq:drstardr} (with a particular choice of the integration constant) such that the transformation from $r$ to $r_*$ is given by
\begin{equation}
\label{eq:rstar_from_r}
    r_{*}(r) = r + \frac{2r_{+}}{r_{+}-r_{-}} \ln \left( \frac{r - r_+}{2} \right) - \frac{2r_{-}}{r_{+}-r_{-}} \ln \left( \frac{r - r_-}{2} \right).
\end{equation}
It should be noted that there is no simple analytical expression for the inverse transformation $r = r(r_*)$ and one has to invert $r_*$ numerically, typically using root-finding algorithms (for example see Appendix~\ref{app:rstar_inversion}).

In short, the \gls{GSN} transformation amounts to acting a linear differential operator ${}_{s}\Lambda$ on the Teukolsky radial function $R(r)$ that transforms it into the \gls{GSN} function $X(r_{*})$.\footnote{This is a generalization of the $\Lambda$ operator introduced in Ref.~\cite{GWPT} for $s=-2$ to any integer $s$.} Schematically this means
\begin{equation}
\label{eq:X_from_R}
	X(r_{*}(r)) = {}_{s}\Lambda\left[ R(r) \right].
\end{equation}
Using Eqs.~\eqref{eq:forwardtransformation} and~\eqref{eq:Xfromchi} we see that the ${}_{s}\Lambda$ operator is given by
\begin{equation}
\label{eq:X_Lambda_R}
	{}_{s}\Lambda \left[ R(r) \right] = \sqrt{\left( r^2 + a^2 \right) \Delta^{s}} \left[ \left( \alpha + \beta \Delta^{s+1} \dfrac{d}{dr} \right) R(r) \right],
\end{equation}
while the inverse \gls{GSN} transformation amounts to acting the inverse operator ${}_{s}\Lambda^{-1}$ on the \gls{GSN} function that gives back the Teukolsky function. Again, schematically this can be written as
\begin{equation}
\label{eq:R_InvLambda_X}
	R(r(r_{*})) = {}_{s}\Lambda^{-1} \left[ X(r_{*}) \right].
\end{equation}
Using Eqs.~\eqref{eq:inversetransformation} and~\eqref{eq:Xfromchi} we see that ${}_{s}\Lambda^{-1}$ is given by
\begin{multline}
	{}_{s}\Lambda^{-1} \left[ X(r_{*}) \right] = \\ \dfrac{1}{\eta} \left\{ \left[ \left( \alpha + \beta' \Delta^{s+1} \right) - \beta \Delta^{s+1} \dfrac{d}{dr} \right] \dfrac{X(r_{*})}{\sqrt{\left( r^2 + a^2 \right) \Delta^{s}}} \right\}.
\end{multline}

Equipped with the transformation, one can show that by substituting $R(r), R'(r)$ given by Eq.~\eqref{eq:inversetransformation} into Eq.~\eqref{eq:radialTeukolskyeqn}, the intermediate function $\chi(r)$ satisfies the following \gls{ODE}, which is given by \cite{Hughes:2000pf}
\begin{equation}
\label{eq:chieqn}
	\Delta^{-s} \left( \Delta^{s+1} \chi' \right)' - \Delta F_{1} \chi' - U_{1} \chi = 0,
\end{equation}
with
\begin{align}
	F_{1}(r) & = \dfrac{\eta'}{\eta}, \\
	U_{1}(r) & = V_{\mathrm{T}} + \dfrac{1}{\beta \Delta^{s}} \left[ \left( 2\alpha + \beta' \Delta^{s+1} \right)' - F_{1}\left(\alpha + \beta' \Delta^{s+1} \right) \right].
\end{align}
Further rewriting Eq.~\eqref{eq:chieqn} in terms of $X$ and its first and second derivatives with respect to $r_{*}$ using Eqs.~\eqref{eq:Xfromchi} and~\eqref{eq:drstardr}, one can show that $X(r_{*})$ satisfies the \gls{GSN} equation, which is given by \cite{Hughes:2000pf}
\begin{equation}
\label{eq:GSNeqn}
    \dfrac{d^2 X}{d r_{*}^2} - \mathcal{F}(r)\dfrac{dX}{dr_{*}} - \mathcal{U}(r)X = 0,
\end{equation}
with the \gls{GSN} potentials $\mathcal{F}(r)$ and $\mathcal{U}(r)$ given by \cite{Hughes:2000pf}
\begin{align}
	\mathcal{F}(r) = & \dfrac{\Delta F_{1}}{r^2 + a^2}, \\
	\mathcal{U}(r) = & \dfrac{\Delta U_{1}}{\left( r^2 + a^2 \right)^2} + G^2 + \dfrac{\Delta G'}{r^2 + a^2} - \dfrac{\Delta G F_{1}}{r^2 + a^2},
\end{align}
where
\[
	G = \dfrac{r\Delta}{\left( r^2 + a^2 \right)^2} + \dfrac{s(r-1)}{r^2 + a^2}.
\]
While the GSN equation given by Eq.~\eqref{eq:GSNeqn} looks significantly more complicated than the original radial Teukolsky equation given by Eq.~\eqref{eq:radialTeukolskyeqn}, Eq.~\eqref{eq:GSNeqn} actually represents a collection of \glspl{ODE} equivalent to Eq.~\eqref{eq:radialTeukolskyeqn} that we can engineer so that the resulting \gls{ODE} has a short-ranged potential and thus can be solved more easily and efficiently with numerical algorithms.

Up to this point, the weighting functions $\alpha(r)$ and $\beta(r)$ are arbitrary, apart from being continuous and differentiable [so that Eqs.~\eqref{eq:forwardtransformation} and~\eqref{eq:inversetransformation} make sense]. However, in order to generate useful transformations, these functions have to satisfy certain criteria. For example, they can be constrained by requiring that, when $a \to 0$, the function $X(r_{*})$ satisfies the Regge-Wheeler equation \cite{SASAKI198185, SASAKI198268, 10.1143/PTP.67.1788, Hughes:2000pf}. Transformations for fields with different spin weight $s$ that satisfy such a constraint were first given in Ref.~\cite{Hughes:2000pf} and can be written in the form of
\begin{widetext}
\begin{equation}
\label{eq:Rtochigeneric_s}
\chi =
\begin{cases}
	\left(\sqrt{(r^2 + a^2)\Delta}\right)^{|s|} g_0(r) J_- \left[ g_1(r) J_- \left[ g_2(r) \dots J_- \left[ g_{|s|}(r) \left( \dfrac{1}{\sqrt{r^2 + a^2}}\right)^{|s|} R \right] \right] \right], & s < 0 \\
	g_0(r) R, & s = 0 \\
	\left( \sqrt{\dfrac{r^2 + a^2}{\Delta}} \right)^{s} g_0(r) J_+ \left[ g_1(r) J_+ \left[ g_2(r) \dots J_+ \left[ g_s(r) \left( \dfrac{\Delta}{\sqrt{r^2 + a^2}} \right)^{s} R \right] \right] \right], & s > 0 \\
\end{cases},
\end{equation}
\end{widetext}
where $J_{\pm}$ are two linear differential operators defined by
\begin{equation}
\label{eq:Jpm_operator}
    J_{\pm} = \dfrac{d}{dr} \pm i \frac{K}{\Delta}.
\end{equation}
Inspecting Eq.~\eqref{eq:Rtochigeneric_s}, we see that for a spin-$|s|$ field, the operator $J_{\pm}$ will act on $R(r)$ $|s|$-many times, leading to an expression relating $\chi(r)$ \textit{linearly} to $R(r), R'(r), \dots, R^{(|s|)}(r)$. Higher-order derivatives $R^{(n)}(r)$ can be evaluated in terms of $R(r), R'(r)$ by using Eq.~\eqref{eq:Rpp_intermsof_RRp} successively for $n \geq 2$. Therefore, by comparing Eqs.~\eqref{eq:forwardtransformation} and~\eqref{eq:Rtochigeneric_s}, one can extract the appropriate $\alpha(r)$ and $\beta(r)$ for different $s$ modulo some functions $g_{i}(r)$ that remain unspecified.

These functions $g_{i}(r)$ should reduce to nonvanishing constants when $a \to 0$ such that Eq.~\eqref{eq:GSNeqn} is exactly the Regge-Wheeler equation for Schwarzschild \glspl{BH}. In practice, it was found that choosing $g_{i}(r)$ as simple rational functions of $r$ leads to desirable short-ranged \gls{GSN} potentials. With some particular choices of $g_{i}(r)$, which we explicitly show in Appendix~\ref{app:explicitGSN} for fields with spin weight $s=0,\pm 1, \pm 2$, the expressions for $\alpha(r)$ and $\beta(r)$ can be quite concise, and we can write $\eta(r)$ in a compact form as 
\begin{equation}
\label{eq:eta_1overr_polynomial}
	\eta(r) = c_0 + c_1/r + c_2/r^2 + c_3/r^3 + c_4/r^4.
\end{equation}
It should be noted that if one chooses instead $g_{i}(r) = 1$, while the associated \gls{GSN} potentials are still short ranged, the corresponding expression for $\eta(r)$ \emph{cannot} be written in the form of Eq.~\eqref{eq:eta_1overr_polynomial} and the weighting functions $\alpha(r)$ and $\beta(r)$ are long (except for $s=0$).

\subsection{Asymptotic behaviors and boundary conditions of the generalized Sasaki-Nakamura equation \label{subsec:asymptotic_behaviors}}
Before studying the asymptotic behaviors of the \gls{GSN} equation, it is educational to first revisit the asymptotic behaviors of the radial Teukolsky equation so that we can compare the behaviors of the two equations and understand the reasons why it is preferred to use the \gls{GSN} equation instead of the Teukolsky equation when performing numerical computations.

Unless otherwise specified, we focus only on nonstatic ($\omega \neq 0$) solutions hereinafter, as the behaviors of nonstatic solutions are different from those of static solutions. Discussions on exact and analytical static mode solutions of the Teukolsky equation can be found in Appendix~\ref{app:static_modes_for_Teukolsky}.

\subsubsection{Teukolsky equation}
It can be shown that (for example, see Refs.~\cite{Teukolsky:1973ha, GWPT}) when $r \to \infty \left( r_{*} \to \infty \right)$ the radial Teukolsky equation admits two (linearly independent) asymptotic solutions that go like $R \sim r^{-1} e^{-i\omega r_{*}}$ or $R \sim r^{-\left(2s+1\right)} e^{i\omega r_{*}}$. Similarly, when $r \to r_{+} \left( r_{*} \to -\infty \right)$ the equation admits two (linearly independent) asymptotic solutions $R \sim \Delta^{-s} e^{-ipr_{*}}$ or $R \sim e^{ipr_{*}}$, where we define a new wave frequency
\begin{equation}
\label{eq:wavefreqp}
	p \equiv \omega - m \Omega_{\mathrm{H}},
\end{equation}
with $\Omega_{\mathrm{H}} \equiv a/(2r_{+})$ being the angular velocity of the horizon (therefore, intuitively speaking $p$ is the ``effective'' wave frequency near the horizon).

\begin{figure}[t]
\centering
\subfloat[IN solution\label{fig:potential_barrier_IN}]{\includegraphics[width=\columnwidth]{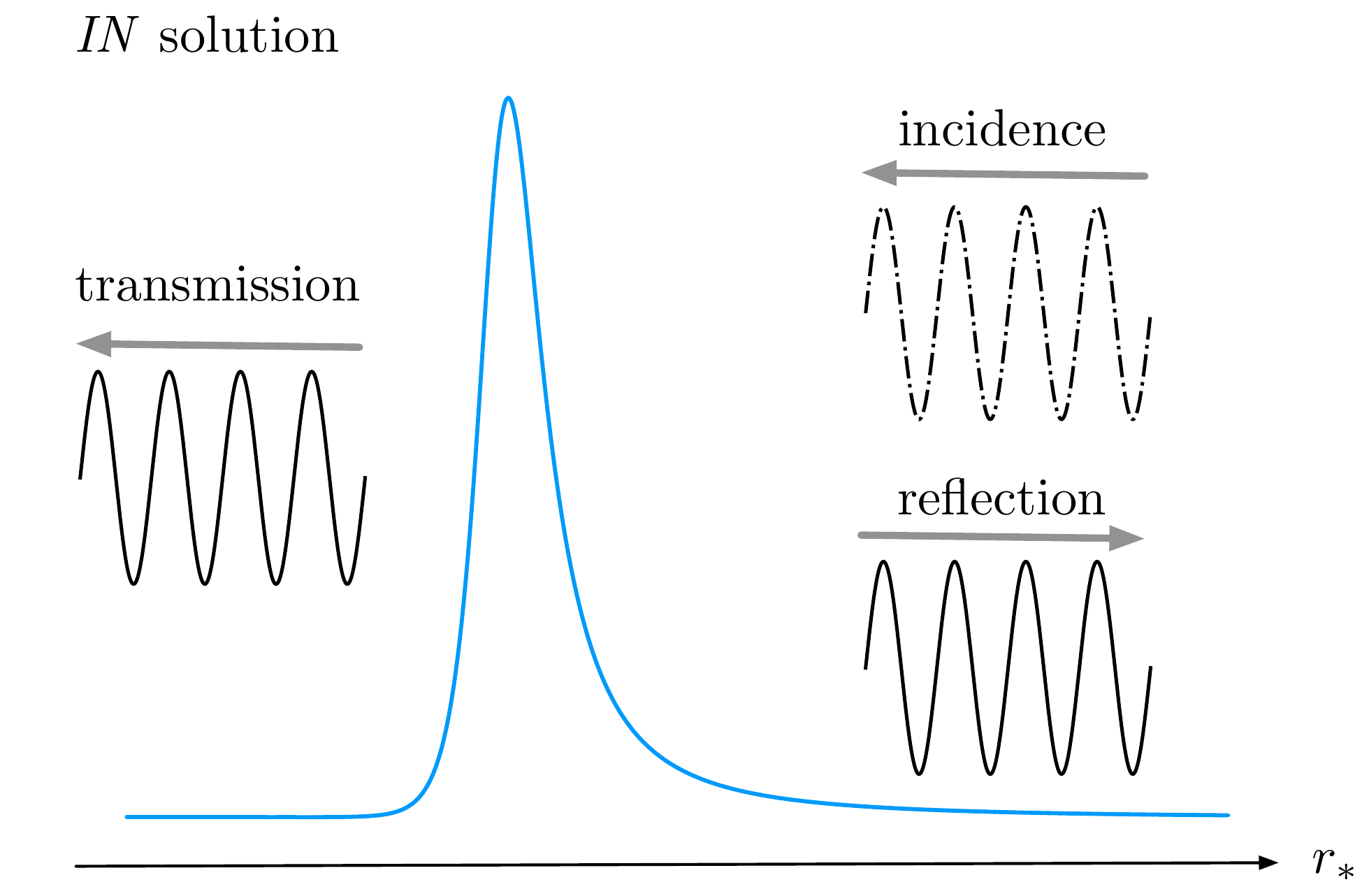}}\\
\subfloat[UP solution\label{fig:potential_barrier_UP}]{\includegraphics[width=\columnwidth]{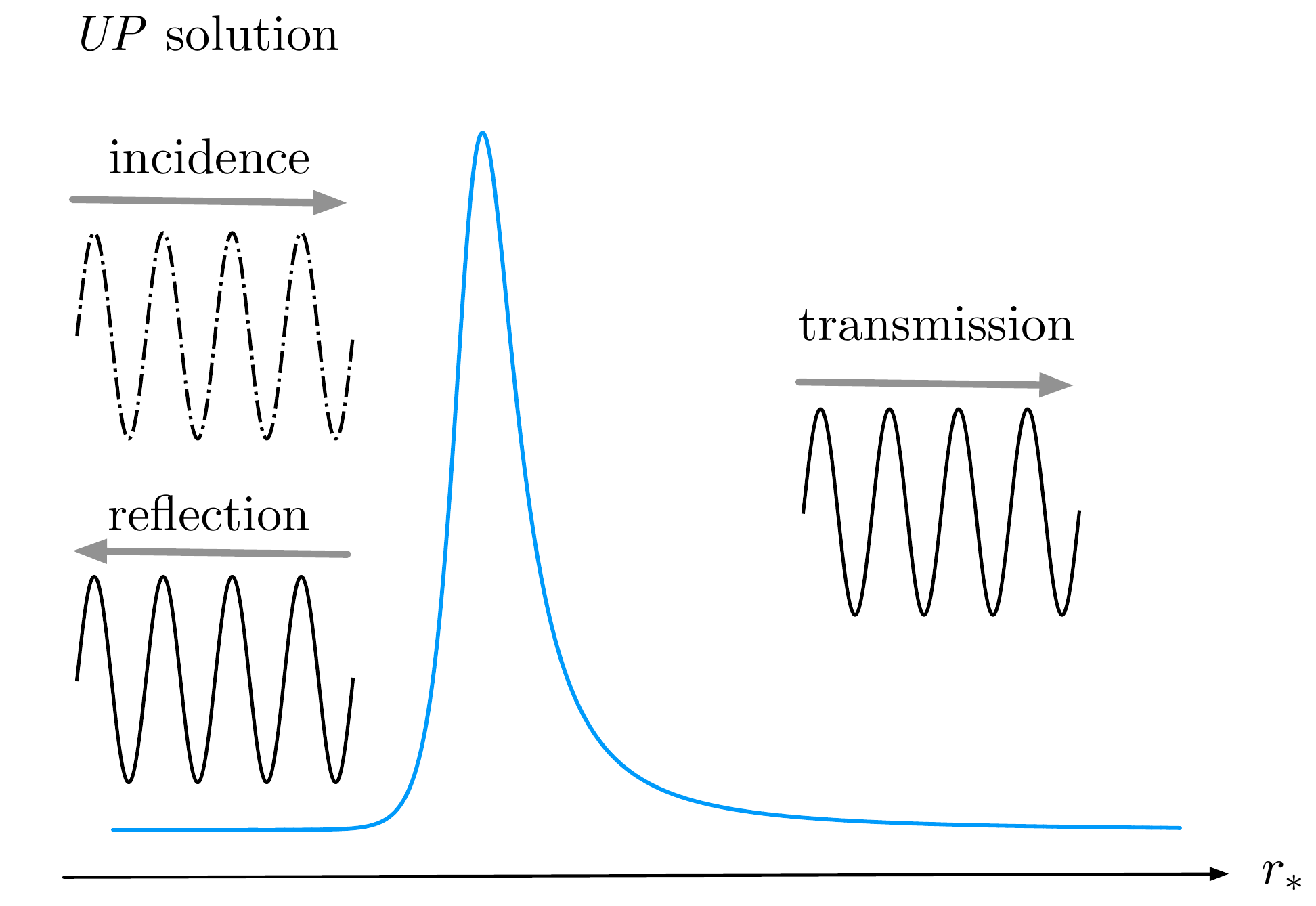}}
\caption{\label{fig:potential_barrier}Physical interpretations of the amplitudes in front of each of the asymptotic solutions, for the IN solution (a) and for the UP solution (b).}
\end{figure}

Using these asymptotic solutions at the two boundaries, we can construct pairs of linearly independent solutions. A pair that is commonly used in literature (and is physically motivated) is $\left\{ R^{\mathrm{in}}, R^{\mathrm{up}}\right\}$\footnote{In some literature, for example Ref.~\cite{Hughes:2000pf}, $R^{\mathrm{in}}$ is also denoted by $R^{\mathrm{H}}$ and $R^{\mathrm{up}}$ is also denoted by $R^{\infty}$.} with $R^{\mathrm{in}}$ satisfying a purely ingoing boundary condition at the horizon and $R^{\mathrm{up}}$ satisfying a purely outgoing boundary condition at infinity.\footnote{See Appendix~\ref{app:out-down} for a discussion of an alternative pair of linearly independent solutions $\left\{R^{\mathrm{out}}, R^{\mathrm{down}}\right\}$.} Mathematically,
\begin{align}
\label{eq:Rin}
	R^{\mathrm{in}}(r) & = \begin{cases}
		B^{\mathrm{trans}}_{\mathrm{T}}  \Delta^{-s} e^{-ipr_*}, & r \to r_+ \\
		B^{\mathrm{inc}}_{\mathrm{T}} \dfrac{e^{-i\omega r_*}}{r} + B^{\mathrm{ref}}_{\mathrm{T}} \dfrac{e^{i\omega r_*}}{r^{2s+1}}, & r \to \infty \\
	\end{cases}, \\
\label{eq:Rup}
	R^{\mathrm{up}}(r) & = \begin{cases}
		C^{\mathrm{ref}}_{\mathrm{T}} \Delta^{-s} e^{-i p r_*} + C^{\mathrm{inc}}_{\mathrm{T}} e^{i p r_*}, & r \to r_+ \\
		C^{\mathrm{trans}}_{\mathrm{T}} \dfrac{e^{i\omega r_*}}{r^{2s+1}}, & r \to \infty
	\end{cases}.
\end{align}
Here we follow mostly Ref.~\cite{Sasaki:2003xr} in naming the coefficients/amplitudes in front of each of the asymptotic solutions (except renaming $C^{\mathrm{up}}$ in Ref.~\cite{Sasaki:2003xr} to $C^{\mathrm{inc}}$ for a more symmetric form and adding a subscript $\mathrm{T}$ for Teukolsky formalism). These amplitudes carry physical interpretations. Conceptually for the $R^{\mathrm{in}}$ ($R^{\mathrm{up}}$) solution, imagine sending a ``leftgoing'' wave from infinity toward the horizon (a ``rightgoing'' wave from the horizon toward infinity)\footnote{As we have assumed a harmonic time dependence of $\exp\left(-i\omega t\right)$, radial functions of the form $\exp\left(i\omega r_{*}\right)$ are said to be traveling to the right since the waves would depend on the combination $t - r_{*}$. Similarly, for radial functions of the form $\exp \left( -i\omega r_{*} \right)$, they are said to be traveling to the left since the waves would depend on the combination $t + r_{*}$.} with an amplitude $B^{\mathrm{inc}}_{\mathrm{T}}$ ($C^{\mathrm{inc}}_{\mathrm{T}}$). As the wave propagates through the potential barrier (see Fig.~\ref{fig:potential_barrier}), part of the incident wave is transmitted through the barrier and continues to travel with an amplitude $B^{\mathrm{trans}}_{\mathrm{T}}$ ($C^{\mathrm{trans}}_{\mathrm{T}}$), while part of the incident wave is \emph{ref}lected by the barrier and travels in the opposite direction with an amplitude $B^{\mathrm{ref}}_{\mathrm{T}}$ ($C^{\mathrm{ref}}_{\mathrm{T}}$). This setup is reminiscent of a potential well problem in quantum mechanics.\footnote{However, unlike a potential well problem in quantum mechanics, the square of the reflection amplitude and the square of the transmission amplitude (each normalized by the incidence amplitude) does not have to add up to unity. This is known as superradiance where energy is being extracted from the black hole.}

In numerical computations, however, instead of starting with an incident wave, it is easier to start with a transmitted wave and then integrate outward (inward) for $R^{\mathrm{in}}$ ($R^{\mathrm{up}}$) to extract the corresponding incidence and reflection amplitude at infinity (at the horizon). Inspecting Eqs.~\eqref{eq:Rin} and \eqref{eq:Rup}, we can see why it is challenging to accurately read off those amplitudes if one solves the Teukolsky equation numerically using Eq.~\eqref{eq:radialTeukolskyeqn} directly as the amplitude of the incident and the reflected wave are of different orders of magnitude. For the $R^{\mathrm{in}}$ solution as $r \to \infty$, the ratio of the amplitude of the rightgoing wave to that of the leftgoing wave is $\sim 1/r^{2s}$ (which becomes infinitely large for $s < 0$ and infinitely small for $s > 0$). While for the $R^{\mathrm{up}}$ solution as $r \to r_{+}$, that ratio is $\sim \Delta^s$ (which again becomes infinitely large for $s < 0$ and infinitely small for $s > 0$ as $\Delta \to 0$ when $r \to r_{+}$).
This implies that, when solving Eq.~\eqref{eq:radialTeukolskyeqn} numerically with a finite precision, the numerical solution will be completely dominated by the rightgoing wave and thus impossible to extract the amplitude for the leftgoing wave.

To see that $R^{\mathrm{in}}$ and $R^{\mathrm{up}}$ are indeed linearly independent, we can calculate the scaled Wronskian $\mathcal{W}_{R}$ of the two solutions, which is given by
\begin{equation}
\label{eq:WR_def}
	\mathcal{W}_{R} = \Delta^{s+1} \left( R^{\mathrm{in}} {R^{\mathrm{up }}}^{\prime} - R^{\mathrm{up }}{R^{\mathrm{in}}}^{\prime}  \right).
\end{equation}
Substituting the asymptotic forms of the two solutions $R^{\mathrm{in, up}}$ when $r \to \infty$ in Eqs.~\eqref{eq:Rin} and \eqref{eq:Rup}, respectively,  gives the relation
\begin{equation}
\label{eq:WR_at_inf}
	\mathcal{W}_R = 2 i \omega C^{\mathrm{trans}}_{\mathrm{T}} B^{\mathrm{inc}}_{\mathrm{T}},
\end{equation}
which is a nonzero constant\footnote{Scaled Wronskians are by construction constants and are not functions of the independent variable. For more details, see Appendix~\ref{app:WR_WX_identity}.} (when $\omega \neq 0$) and thus they are indeed linearly independent.
If instead we substitute the asymptotic forms of $R^{\mathrm{in, up}}$ when $r \to r_+ $ into Eq.~\eqref{eq:WR_def}, we obtain another relation for $\mathcal{W}_{R}$, which is
\begin{equation}
\label{eq:WR_at_hor}
	\mathcal{W}_R = \left[ 2ip(r_{+}^2 + a^2) + 2s(r_{+} - 1)\right] B^{\mathrm{trans}}_{\mathrm{T}} C^{\mathrm{inc}}_{\mathrm{T}}.
\end{equation}
By equating Eqs.~\eqref{eq:WR_at_inf} and Eq.~\eqref{eq:WR_at_hor}, we get an identity relating $(B^{\mathrm{inc}}_{\mathrm{T}}/B^{\mathrm{trans}}_{\mathrm{T}})$ with $(C^{\mathrm{inc}}_{\mathrm{T}}/C^{\mathrm{trans}}_{\mathrm{T}})$. From a numerical standpoint, we can use this identity as a sanity check of numerical solutions. More explicitly, the identity is given by
\begin{equation}
\label{eq:WR_identity}
	\dfrac{B^{\mathrm{inc}}_{\mathrm{T}}}{B^{\mathrm{trans}}_{\mathrm{T}}} = \dfrac{p(r_{+}^2 + a^2) - is(r_{+} - 1)}{\omega}\dfrac{C^{\mathrm{inc}}_{\mathrm{T}}}{C^{\mathrm{trans}}_{\mathrm{T}}}.
\end{equation}
It also means that we technically only need to read off $\left\{ B^{\mathrm{ref}}_{\mathrm{T}}, B^{\mathrm{inc}}_{\mathrm{T}}, C^{\mathrm{ref}}_{\mathrm{T}}\right\}$ or $\left\{ B^{\mathrm{ref}}_{\mathrm{T}}, C^{\mathrm{ref}}_{\mathrm{T}}, C^{\mathrm{inc}}_{\mathrm{T}} \right\}$ from numerical solutions since the rest of the amplitudes are either fixed by the normalization convention (which will be covered shortly below) or by the constant scaled Wronskian, which can be computed at an arbitrary location within the domain of the numerical solutions.

\subsubsection{Generalized Sasaki-Nakamura equation}
\begin{figure}[h]
\centering
\includegraphics[width=\columnwidth]{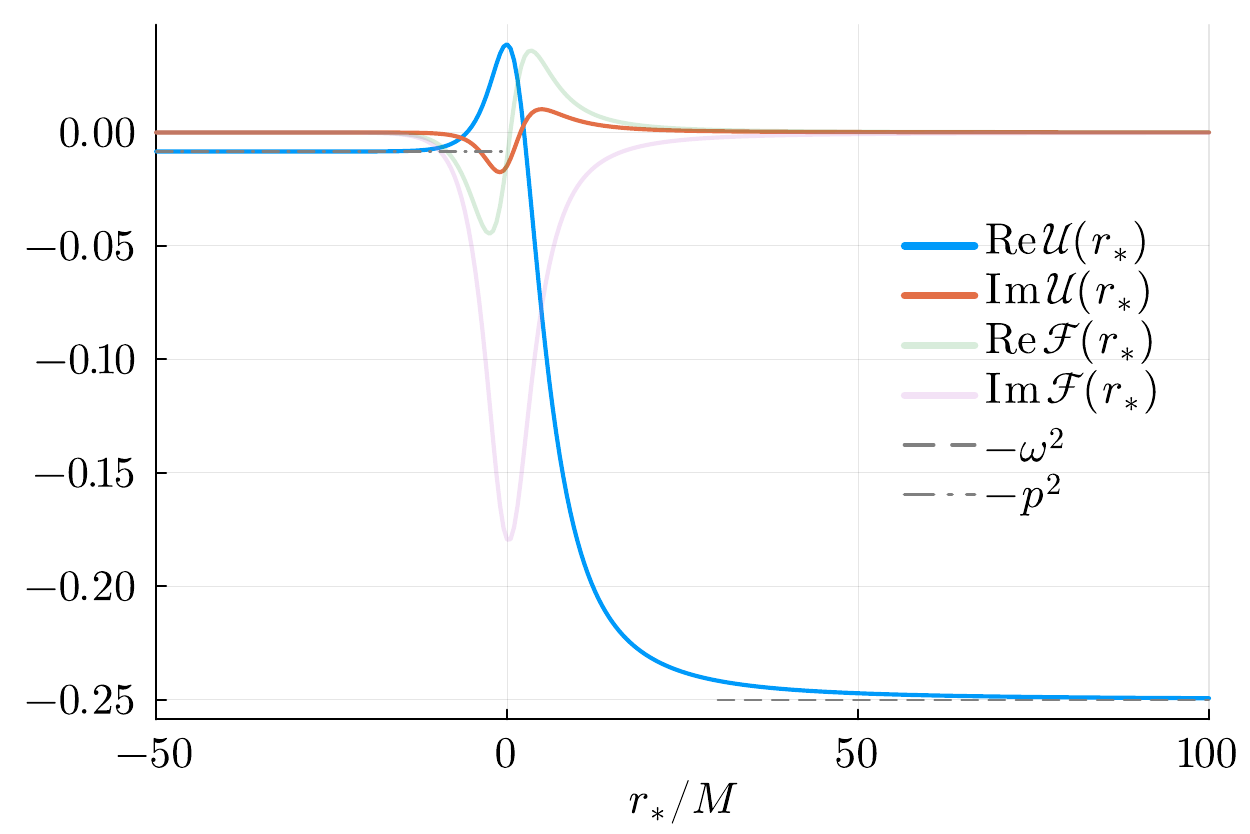}
\caption{Asymptotic behaviors of the \gls{GSN} potentials $\mathcal{F}(r)$ and $\mathcal{U}(r)$. Both potentials quickly approach to their corresponding asymptotic values. In particular, $\mathcal{U}(r)$ approaches to $-p^2$ near the horizon and $-\omega^2$ near infinity, respectively. \label{fig:GSN_potentials}}
\end{figure}

\begin{figure}[h]
\centering
\includegraphics[width=\columnwidth]{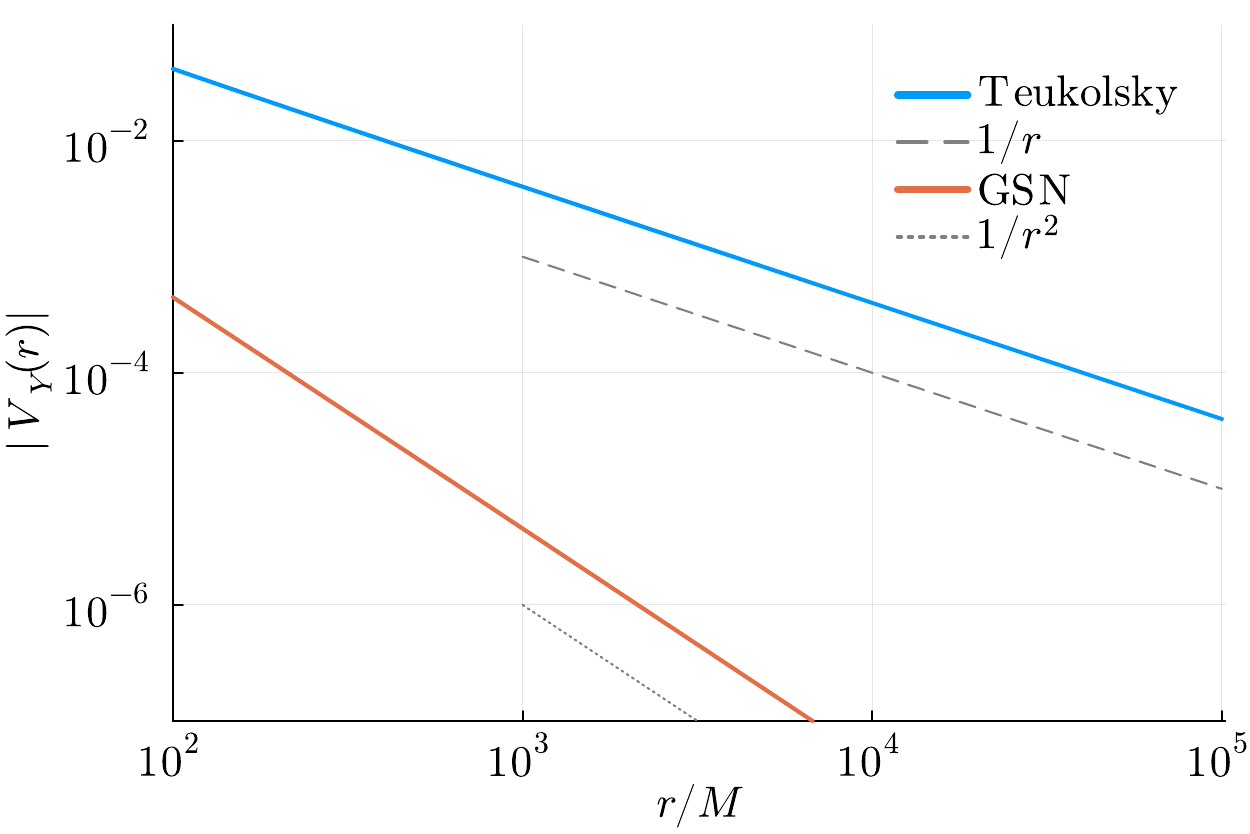}
\caption{Potential $V_{Y}(r)$ associated with the Teukolsky equation and the \gls{GSN} equation. As $r \to \infty$, the potential for the Teukolsky equation decays at $1/r$ and thus it is long ranged, while the potential for the GSN equation decays at a steeper $1/r^2$ and hence it is short ranged.\label{fig:canonical_potentials}}
\end{figure}

Now we turn to the \gls{GSN} equation. Suppose the \gls{GSN} transformation is of the form of Eq.~\eqref{eq:Rtochigeneric_s} and satisfies Eq.~\eqref{eq:eta_1overr_polynomial}, the \gls{GSN} potentials $\mathcal{F}(r)$ and $\mathcal{U}(r)$ then have the following asymptotic behaviors (see Fig.~\ref{fig:GSN_potentials} for a visualization)
\begin{align}
	\mathcal{F}(r) \sim \begin{cases}
 	0 + \bigO(r-r_+) & r \to r_+ \\
 	\dfrac{-c_{1}/c_{0}}{r^2} + \bigO(r^{-3}) & r \to \infty \\
 \end{cases},\\
 	\mathcal{U}(r) \sim \begin{cases}
 	-p^2 + \bigO(r-r_+) & r \to r_+ \\
 	-\omega^2 + \bigO(r^{-2}) & r \to \infty \\
 \end{cases}.
\end{align}
To see more clearly that the \gls{GSN} potentials are indeed short ranged, we recast the \gls{GSN} equation into the same form as Eq.~\eqref{eq:Schrodingereqn} by writing $Y \equiv X/\sqrt{\eta}$. Figure~\ref{fig:canonical_potentials} shows the magnitude of the potential $V_{Y}(r)$ associated with the Teukolsky equation (blue) and the \gls{GSN} equation (orange), respectively. Specifically we are showing the potentials of the $s=-2,\ell=2,m=2$ mode with $a=0.7$ and $\omega = 1$ as examples. We can see that the potential for the Teukolsky equation decays only at $1/r$ when $r \to \infty$ (and hence long ranged) while the potential for the \gls{GSN} equation decays at $1/r^2$ when $r \to \infty$ (and hence short ranged).

The asymptotic behaviors of the \gls{GSN} potentials imply that as $r \to r_{+}$, the \gls{GSN} equation behaves like a simple wave equation
$d^2 X/d r_{*}^2 + p^2 X = 0$, admitting simple plane-wave solutions $e^{\pm ipr_*}$. Similarly, when $r \to \infty$, the \gls{GSN} equation behaves like $d^2 X/d r_{*}^2 + \omega^2 X = 0$, again admitting plane-wave solutions $e^{\pm i\omega r_*}$.\footnote{Exact analytical solutions to the \gls{GSN} equation when $\omega = 0$ can be obtained by transforming the exact solutions to the radial Teukolsky equation, covered in Appendix~\ref{app:static_modes_for_Teukolsky}, using the \gls{GSN} transformation. These exact solutions to the \gls{GSN} equation can also be expressed using Gauss hypergeometric functions since the ${}_{s}\Lambda$ operator in Eq.~\eqref{eq:X_Lambda_R} is a linear differential operator and derivatives of these functions are simply rescaled Gauss hypergeometric functions of different parameters.} Therefore, we can similarly construct the pair of linearly independent solutions $\left\{ X^{\mathrm{in}}, X^{\mathrm{up}} \right\}$ that satisfies the purely ingoing boundary condition at the horizon and the purely outgoing boundary condition at infinity, respectively, using these asymptotic solutions.\footnote{See Appendix~\ref{app:out-down} for a discussion of an alternative pair of linearly independent solutions $\left\{X^{\mathrm{out}}, X^{\mathrm{down}}\right\}$.} Mathematically, 
\begin{align}
\label{eq:Xin}
	X^{\mathrm{in}}(r_*) & = \begin{cases}
		B^{\mathrm{trans}}_{\mathrm{SN}}  e^{-ipr_*} & r_* \to -\infty \\
		B^{\mathrm{inc}}_{\mathrm{SN}} e^{-i\omega r_*} + B^{\mathrm{ref}}_{\mathrm{SN}} e^{i\omega r_*} & r_* \to \infty \\
	\end{cases}, \\
\label{eq:Xup}
	X^{\mathrm{up}}(r_*) & = \begin{cases}
		C^{\mathrm{ref}}_{\mathrm{SN}} e^{-i p r_*} + C^{\mathrm{inc}}_{\mathrm{SN}} e^{i p r_*} & r_* \to -\infty \\
		C^{\mathrm{trans}}_{\mathrm{SN}} e^{i\omega r_*} & r_* \to \infty
	\end{cases}.
\end{align}
Here the amplitudes in front of each of the asymptotic solutions have the same physical interpretations as in Eqs.~\eqref{eq:Rin} and \eqref{eq:Rup} (cf. Fig.~\ref{fig:potential_barrier}).\footnote{Note, however, when $p = 0$ or, equivalently, $\omega = m\Omega_{\text{H}}$, the incidence amplitude $C^{\mathrm{inc}}_{\mathrm{SN}}$ and the reflection amplitude $C^{\mathrm{ref}}_{\mathrm{SN}}$ for the $X^{\mathrm{up}}$ solution are not well defined and lose their meanings as $\exp(ipr_{*}) = \exp(-ipr_{*}) = 1$. That said, the $X^{\mathrm{in,up}}$ solutions constructed in Eqs.~\eqref{eq:Xin} and \eqref{eq:Xup} are still linearly independent to each other, evident from their different asymptotic behaviors near infinity as $r_{*} \to \infty$.} Again by inspecting Eqs.~\eqref{eq:Xin} and \eqref{eq:Xup}, we can see that it is easy to accurately read off those amplitudes as the ratio of the asymptotic amplitude of the incident wave to that of the reflected wave at both boundaries is $\sim \bigO(1)$, instead of being infinitely large or infinitely small in the Teukolsky formalism.

Similar to the case of Teukolsky functions, we can also define a scaled Wronskian $\mathcal{W}_{X}$ for the \gls{GSN} functions, namely,
\begin{equation}
\label{eq:WX_def}
	\mathcal{W}_{X} = \dfrac{1}{\eta} \left[ X^{\mathrm{in}} (dX^{\mathrm{up}}/dr_{*}) - (dX^{\mathrm{in}}/dr_{*}) X^{\mathrm{up}} \right],
\end{equation}
which is also a constant.
Substituting the asymptotic forms of $X^{\mathrm{in, up}}$ in Eqs.~\eqref{eq:Xin} and \eqref{eq:Xup}, respectively, as $r_{*} \to \infty$, and the fact that $\eta(r) \to c_{0}$ as $r \to \infty$, it can be shown that
\begin{equation}
\label{eq:WX_at_inf}
	\mathcal{W}_{X} = \dfrac{2i\omega C^{\mathrm{trans}}_{\mathrm{SN}} B^{\mathrm{inc}}_{\mathrm{SN}}}{c_0}.
\end{equation}
Equivalently, we can also use the asymptotic forms of $X^{\mathrm{in, up}}$ as $r_{*} \to -\infty$ and the fact that $\eta(r \to r_+) \sim \bigO(1)$ to show that
\begin{equation}
\label{eq:WX_at_hor}
	\mathcal{W}_{X} = \dfrac{2ip B^{\mathrm{trans}}_{\mathrm{SN}} C^{\mathrm{inc}}_{\mathrm{SN}}}{\eta(r_+)}.
\end{equation}
We can again equate Eqs.~\eqref{eq:WX_at_inf} and \eqref{eq:WX_at_hor} to get an identity relating $B^{\mathrm{inc}}_{\mathrm{SN}}/B^{\mathrm{trans}}_{\mathrm{SN}}$ with $C^{\mathrm{inc}}_{\mathrm{SN}}/C^{\mathrm{trans}}_{\mathrm{SN}}$ to check the sanity of numerical solutions. Explicitly, the identity is given by
\begin{equation}
\label{eq:WX_identity}
	\dfrac{B^{\mathrm{inc}}_{\mathrm{SN}}}{B^{\mathrm{trans}}_{\mathrm{SN}}} = \dfrac{pc_0}{\omega\eta(r_{+})}\dfrac{C^{\mathrm{inc}}_{\mathrm{SN}}}{C^{\mathrm{trans}}_{\mathrm{SN}}}.
\end{equation}
An interesting and useful relation between the scaled Wronskians for GSN functions $\mathcal{W}_{X}$ and that for Teukolsky functions $\mathcal{W}_{R}$ (with the same $s, \ell, m, a, \omega$) is that, despite having different definitions [see Eq.~\eqref{eq:WR_def} for $\mathcal{W}_{R}$ and Eq.~\eqref{eq:WX_def} for $\mathcal{W}_{X}$], they are actually identical, i.e.,
\begin{equation}
\label{eq:WR_WX_identity}
	\mathcal{W}_{X} = \mathcal{W}_{R},
\end{equation}
where we give a derivation in Appendix~\ref{app:WR_WX_identity}. This means that \gls{GSN} transformations (not limited only to our particular choices of $g_{i}$) are scaled-Wronskian preserving. This also means that one can compute the \gls{QNM} spectra of Kerr \glspl{BH} using either the Teukolsky formalism or the \gls{GSN} formalism (see Sec.~\ref{subsec:QNMusingGSN}).

Since one can freely rescale a homogeneous solution by a constant factor, we use this freedom to set $B^{\mathrm{trans}}_{\mathrm{SN}} = C^{\mathrm{trans}}_{\mathrm{SN}} = 1$, i.e., we normalize our solutions to the \gls{GSN} equation to have a unit SN transmission amplitude. However, the common normalization convention in literature is to normalize $R^{\mathrm{in}}(r)$ and $R^{\mathrm{up}}(r)$ to each have a unit transmission amplitude, i.e., $B^{\mathrm{trans}}_{\mathrm{T}} = C^{\mathrm{trans}}_{\mathrm{T}} = 1$. In fact, one can relate incidence/reflection/transmission amplitudes in the \gls{GSN} formalism to that in the Teukolsky formalism and vice versa by frequency-dependent conversion factors. To see why this is the case and to obtain the conversion factors, note that when going from a Teukolsky function to the corresponding \gls{GSN} function, we have the ${}_{s}\Lambda$ operator that satisfies
\begin{equation}
\label{eq:LambdaRelation}
	{}_{s}\Lambda \left[ f(r) e^{\pm ik r_{*}} \right] \propto e^{\pm ik r_{*}},
\end{equation}
and vice versa with the inverse operator ${}_{s}\Lambda^{-1}$ that satisfies
\begin{equation}
\label{eq:InvLambdaRelation}
	{}_{s}\Lambda^{-1} \left[ f(r) e^{\pm ik r_{*}} \right] \propto e^{\pm ik r_{*}},
\end{equation}
for any differentiable function $f(r)$ and $k$ is any nonzero constant, since both ${}_{s}\Lambda$ and ${}_{s}\Lambda^{-1}$ are linear differential operators. This means that we can simply match the asymptotic solution in one formalism with the corresponding asymptotic solution with the same exponential dependence in another formalism transformed by either ${}_{s}\Lambda$ or ${}_{s}\Lambda^{-1}$ at the appropriate boundary.

For example, to get the conversion factor $C^{\mathrm{trans}}_{\mathrm{T}}/C^{\mathrm{trans}}_{\mathrm{SN}}$, we match the asymptotic solution as $r \to \infty$ for the Teukolsky and the \gls{GSN} formalism like
\begin{multline}
\label{eq:relating_Ctrans_with_Lambda}
	C^{\mathrm{trans}}_{\mathrm{SN}} \left[ 1 + \bigO\left(\dfrac{1}{r} \right)\right] e^{i\omega r_{*}} = \\
	C^{\mathrm{trans}}_{\mathrm{T}} {}_{s}\Lambda \left\{ \dfrac{1}{r^{2s+1}} \left[ 1 + \bigO\left(\dfrac{1}{r}\right) \right] e^{i\omega r_*} \right\},
\end{multline}
where the expression on the rhs, to the leading order, should be $\sim \bigO\left(1\right) e^{i\omega r_{*}}$. We can then obtain the desired conversion factor by taking the limit as
\begin{equation}
\label{eq:Ctrans_convfactor_Lambda}
	\dfrac{C^{\mathrm{trans}}_{\mathrm{SN}}}{C^{\mathrm{trans}}_{\mathrm{T}}} = \lim_{r \to \infty} {}_{s}\Lambda \left\{ \dfrac{1}{r^{2s+1}} \left[ 1 + \bigO\left(\dfrac{1}{r}\right) \right] e^{i\omega r_*} \right\} e^{-i\omega r_{*}},
\end{equation}
and we know that the expression on the rhs does not depend on $e^{\pm i\omega r_*}$ using Eq.~\eqref{eq:LambdaRelation} so that the limit could be determinate. Equivalently, we can also match the asymptotic solution as $r \to \infty$ in the two formalisms like this instead
\begin{multline}
\label{eq:relating_Ctrans_with_InvLambda}
	C^{\mathrm{trans}}_{\mathrm{T}} \dfrac{1}{r^{2s+1}} \left[ 1 + \bigO \left( \dfrac{1}{r} \right)  \right] e^{i\omega r_*} = \\
	C^{\mathrm{trans}}_{\mathrm{SN}} {}_{s}\Lambda^{-1} \left\{ \left[ 1 + \bigO \left( \dfrac{1}{r} \right) \right] e^{i\omega r_{*}} \right\},
\end{multline}
where the expression on the rhs, to the leading order, should be $\sim \bigO\left(1\right) r^{-\left(2s+1\right)}e^{i\omega r_{*}}$.
Similarly, we can obtain
\begin{equation}
\label{eq:Ctrans_convfactor_InvLambda}
	\dfrac{C^{\mathrm{trans}}_{\mathrm{T}}}{C^{\mathrm{trans}}_{\mathrm{SN}}} = \lim_{r \to \infty} {}_{s}\Lambda^{-1} \left\{ \left[ 1 + \bigO \left( \dfrac{1}{r} \right) \right] e^{i\omega r_{*}} \right\} r^{2s+1}  e^{-i\omega r_{*}},
\end{equation}
and again we know that the rhs of the expression does not depend on $e^{\pm i\omega r_*}$ using Eq.~\eqref{eq:InvLambdaRelation} so that the limit could be determinate.

We find that sometimes it is more convenient to compute the limit in the form of Eq.~\eqref{eq:Ctrans_convfactor_Lambda} than to use the limit in the form of Eq.~\eqref{eq:Ctrans_convfactor_InvLambda} in order to find the same conversion factor, and in some cases the reverse is true even though formally both expressions should give the same answer. In fact, using the identity between the scaled Wronskian of the \gls{GSN} functions $\mathcal{W}_{X}$ and that of the Teukolsky functions $\mathcal{W}_{R}$, we can simplify expressions for these conversion factors by equating expressions of $\mathcal{W}_{X}$ in terms of the incidence and transmission amplitudes in the \gls{GSN} formalism with expressions of $\mathcal{W}_{R}$ in terms of those amplitudes in the Teukolsky formalism. In particular, we get identities relating these conversion factors as
\begin{align}
	 \dfrac{C^{\mathrm{trans}}_{\mathrm{T}}}{C^{\mathrm{trans}}_{\mathrm{SN}}} \dfrac{B^{\mathrm{inc}}_{\mathrm{T}}}{B^{\mathrm{inc}}_{\mathrm{SN}}} & = \dfrac{1}{c_0}, \label{eq:CtransBinc_identity}\\
	 \dfrac{B^{\mathrm{trans}}_{\mathrm{T}}}{B^{\mathrm{trans}}_{\mathrm{SN}}} \dfrac{C^{\mathrm{inc}}_{\mathrm{T}}}{C^{\mathrm{inc}}_{\mathrm{SN}}} & = \dfrac{2ip}{\eta(r_+)\left[ 2ip(r_{+}^2 + a^2) + 2s(r_{+} - 1)\right]}. \label{eq:BtransCinc_identity}
\end{align}
These identities imply that we only need to derive either $C^{\mathrm{trans}}_{\mathrm{T}}/C^{\mathrm{trans}}_{\mathrm{SN}}$ or $B^{\mathrm{inc}}_{\mathrm{T}}/B^{\mathrm{inc}}_{\mathrm{SN}}$ and either $B^{\mathrm{trans}}_{\mathrm{T}}/B^{\mathrm{trans}}_{\mathrm{SN}}$ or $C^{\mathrm{inc}}_{\mathrm{T}}/C^{\mathrm{inc}}_{\mathrm{SN}}$.

\subsubsection{Higher-order corrections to asymptotic behaviors}
In Eqs.~\eqref{eq:Xin} and \eqref{eq:Xup}, we use the asymptotic solutions of the \gls{GSN} equation only to their leading order [i.e., $\bigO(r^0)$]. However, in order to obtain accurate numerical solutions solved on a numerically finite interval (e.g., $\left[ r^{\mathrm{in}}_{*}, r^{\mathrm{out}}_{*} \right]$), it is more efficient to include higher-order corrections to the asymptotic solutions than to simply set $r^{\mathrm{in}}_{*}$ as a small number and $r^{\mathrm{out}}_{*}$ as a large number. To find such higher-order corrections, we use an ansatz of the form 
\begin{equation}
\label{eq:Xansatz}
	X(r_*) \sim \begin{cases}
		f^{\infty}_{\pm}(r) e^{\pm i\omega r_{*}}, & r_* \to \infty \\
		g^{\mathrm{H}}_{\pm}(r) e^{\pm i p r_{*}}, & r_* \to -\infty
		\end{cases},
\end{equation}
where the plus (minus) sign corresponds to the out/rightgoing (in/leftgoing) mode, and the superscript $\infty$ ($\mathrm{H}$) corresponds to the outer (inner) boundary at infinity (the horizon). Substituting Eq.~\eqref{eq:Xansatz} back to the \gls{GSN} equation in Eq.~\eqref{eq:GSNeqn}, we get four second-order \glspl{ODE} for each of the functions $f^{\infty}_{\pm}(r)$ and $g^{\mathrm{H}}_{\pm}(r)$ [cf. Eq.~\eqref{eq:ODE_for_fansatz}]. We look for their formal series expansions of the form
\begin{align}
\label{eq:finfpm}
	f^{\infty}_{\pm}(r) = & \sum_{j=0}^{\infty} \dfrac{\mathcal{C}^{\infty}_{\pm, j}}{\left( \omega r\right)^j}, \\
\label{eq:fhorpm}
	g^{\mathrm{H}}_{\pm}(r) = & \sum_{j=0}^{\infty} \mathcal{C}^{\mathrm{H}}_{\pm, j} \left[ \omega (r - r_+) \right]^j,
\end{align}
where $\mathcal{C}^{\infty/ \mathrm{H}}_{\pm, j}$ are the expansion coefficients. In Appendix~\ref{app:recurrence_relations}, we show how one can compute these coefficients using recurrence relations. Such recurrence relations for some of the spin weights ($s=0$ and $s=-2$) can also be found in literature (e.g., Refs.~\cite{Shah:2012gu, Gralla:2015rpa, Piovano:2020zin}).\footnote{Unfortunately the expansion coefficients given in Refs.~\cite{Hughes:2000pf} are incorrect except for the case with $s=0$ because the author made an incorrect assumption that the \gls{GSN} potentials are purely real, which is not true in general.} In Appendix~\ref{app:explicitGSN}, we show explicitly the expressions of the expansion coefficients $\mathcal{C}^{\infty}_{\pm,j}$ for $j=0, 1,2,3$.

With the explicit \gls{GSN} transformation and hence the \gls{GSN} potentials and the \gls{GSN} equation as discussed in Sec.~\ref{subsec:transformation}, as well as the asymptotic solutions to the \gls{GSN} equation and the conversion factors for converting asymptotic amplitudes between the Teukolsky and the \gls{GSN} formalism as discussed in Sec.~\ref{subsec:asymptotic_behaviors}, we now have all the necessary ingredients to use the \gls{GSN} formalism to perform numerical computations. In the next section, we describe the recipes to use those ingredients to get homogenous solutions to both the Teukolsky and the \gls{GSN} equations.

\section{Numerical implementation\label{sec:numerical_implementation}}
In principle, a frequency-domain Teukolsky/\gls{GSN} equation solver can be implemented in any programming language with the help of the ingredients in Sec.~\ref{sec:formalism} and Appendix~\ref{app:explicitGSN}. Here we describe an open-source implementation of the \gls{GSN} formalism that is written in \texttt{julia} \cite{Julia-2017}, namely, \texttt{GeneralizedSasakiNakamura.jl}.\footnote{\url{https://github.com/ricokaloklo/GeneralizedSasakiNakamura.jl}.} Instead of fixing a particular choice of an numerical integrator for solving Eq.~\eqref{eq:GSNeqn}, the code can be used in conjunction with other \texttt{julia} packages, such as \texttt{DifferentialEquations.jl} \cite{rackauckas2017differentialequations}, which implements a suite of \gls{ODE} solvers. The \gls{GSN} potentials $\mathcal{F}(r), \mathcal{U}(r)$ for $s=0, \pm 1, \pm 2$ are implemented as pure functions in \texttt{julia}, and can be evaluated to arbitrary precision. This also allows us to use \gls{AD} to compute corrections to the asymptotic boundary conditions at arbitrary order (see Appendix~\ref{app:recurrence_relations}).\footnote{In particular, we use two variants of \gls{AD}. The first type is referred to as the forward-mode \gls{AD} as implemented in \texttt{ForwardDiff.jl} \cite{RevelsLubinPapamarkou2016}. However, the computational cost of using the forward-mode \gls{AD} to compute higher-order derivatives scales exponentially with the order. Therefore, for computing corrections to the asymptotic boundary conditions we switch to the second type, which is based on Taylor expansion as implemented in \texttt{TaylorSeries.jl} \cite{Benet2019}, where the cost only scales linearly with the order of the derivatives.}

\subsection{Numerical solutions to the generalized Sasaki-Nakamura equation}
\subsubsection{Rewriting generalized Sasaki-Nakamura functions as complex phase functions \label{subsubsec:complexphasefunc}}
Instead of solving directly for the \gls{GSN} function $X(r_{*})$, we follow Ref.~\cite{Finn:2000sy} and introduce a complex phase function $\Phi(r_{*})$ such that
\begin{equation}
\label{eq:complexphase}
	X(r_{*}) \equiv \exp \left[ i \Phi \left( r_{*} \right) \right].
\end{equation}
Substituting Eq.~\eqref{eq:complexphase} into Eq.~\eqref{eq:GSNeqn}, we obtain a first-order nonlinear differential equation\footnote{Unlike what was claimed in Appendix~3 of Ref.~\cite{Finn:2000sy}, we find that the \gls{ODE} for both the real and the imaginary part of $\Phi$ can be integrated immediately to first-order (nonlinear) differential equations in $\left( d\Phi_{\mathrm{Re}}/dr_{*}, d\Phi_{\mathrm{Im}}/dr_{*} \right)$, which is expected since solutions to a homogeneous \gls{ODE} are determined only up to a multiplicative factor. Combining the differential equations for $d\Phi_{\mathrm{Re}}/dr_{*}$ and $d\Phi_{\mathrm{Im}}/dr_{*}$ such that $d\Phi/dr_{*} = \left( d\Phi_{\mathrm{Re}}/dr_{*} + i d\Phi_{\mathrm{Im}}/dr_{*} \right)$ will give Eq.~\eqref{eq:Riccatieqn}.} as
\begin{equation}
\label{eq:Riccatieqn}
	\dfrac{d}{dr_{*}} \left( \dfrac{d\Phi}{dr_{*}} \right) = -i\,\mathcal{U} + \mathcal{F} \left( \dfrac{d\Phi}{dr_{*}} \right) - i \left( \dfrac{d\Phi}{dr_{*}} \right)^2.
\end{equation}
Such a differential equation is also known as a Riccati equation. Furthermore, the conversion between $\left(X, dX/dr_{*}\right)$ and $\left( \Phi, d\Phi/dr_{*}\right)$ is given by
\begin{align}
	\Phi & = -i \log \left( X \right), \label{eq:PhiFromXXprime} \\
	\dfrac{d\Phi}{dr_{*}} & = -i \dfrac{dX/dr_{*}}{X}. \label{eq:PhiprimeFromXXprime}
\end{align}

While at first glance it may seem unwise to turn a linear problem into a non-linear problem, solving Eq.~\eqref{eq:Riccatieqn} numerically presents no additional challenge compared to solving directly Eq.~\eqref{eq:GSNeqn}. In fact, there are advantages in writing the \gls{GSN} function in the form of Eq.~\eqref{eq:complexphase}, especially when $|\omega|$ is large. Recall that asymptotically (both near infinity and near the horizon) \gls{GSN} functions behave like plane waves, i.e., $X$ oscillates like $\exp \left( \pm ikr_{*} \right)$ where $|k|$ is the oscillation frequency (assuming $k$ is real, and recall that $|k| \to |\omega|$ when $r_{*} \to \infty$ and $|k| \to |p|$ when $r_{*} \to -\infty$). Therefore, in order to properly resolve the oscillations, the step size $\delta r_{*}$ for the numerical integrator needs to be much less than the wavelength, i.e., $\delta r_{*} \ll 1/|k|$. This can get quite small for large $|k|$, which results in taking a longer time to integrate Eq.~\eqref{eq:GSNeqn} for a fixed accuracy.

Fortunately, this is not the case when solving for the complex phase function $\Phi(r_{*})$ since it is varying much slower (spatially) than the \gls{GSN} function $X(r_{*})$. Intuitively this is because the complex exponential in Eq.~\eqref{eq:complexphase} accounts for most of the oscillatory behaviors. This is especially true if we consider the asymptotic plane-wave-like solutions of the \gls{GSN} equation, where the real part of the phase function $\Phi_{\mathrm{Re}}(r_{*}) \sim kr_{*}$ is linear in $r_{*}$, and the imaginary part of the phase function $\Phi_{\mathrm{Im}}(r_{*})$ is constant in $r_{*}$.

However, this might not be the case when we consider general solutions to the \gls{GSN} equation where the left- and the rightgoing modes are superimposed, for example the $X^{\mathrm{in, up}}$ pair as shown in Eqs.~\eqref{eq:Xin} and \eqref{eq:Xup}. That being said, the variation of the complex phase function due to the beating or interference
between the left-/rightgoing modes depends on their relative amplitude (which is, in general, a complex number and hence introduces a phase shift). In particular, physically Kerr \glspl{BH} are much more permeable to waves at high frequencies (see Fig.~\ref{fig:reflectivity}). This means that at those high frequencies, the relative amplitudes of the left-/rightgoing modes are going to be extreme and hence the beating will be suppressed.

\subsubsection{Solving $X^{\mathrm{in, up}}$ as initial value problems}
Recall that there is a pair of linearly independent solutions to the \gls{GSN} equation that is of particular interest, namely, $\left\{ X^{\mathrm{in}}, X^{\mathrm{up}} \right\}$, where $X^{\mathrm{in}}$ satisfies the boundary condition that it is purely ingoing at the horizon as given by Eq.~\eqref{eq:Xin}, and $X^{\mathrm{up}}$ satisfies the boundary condition that it is purely outgoing at infinity as given by Eq.~\eqref{eq:Xup}, respectively.

Despite the use of the term ``boundary condition,'' what we are really enforcing is the asymptotic form of a solution at one of the two boundaries, $X^{\mathrm{in}}$ at the horizon and $X^{\mathrm{up}}$ at infinity, respectively. This can be formulated as an initial value problem. Explicitly for $\hat{X}^{\mathrm{in}}$, where a hat denotes a numerical solution hereafter, we integrate Eq.~\eqref{eq:Riccatieqn} outward from the (finite) inner boundary $r_{*}^{\mathrm{in}}$ to the (finite) outer boundary $r_{*}^{\mathrm{out}}$ with
\begin{align}
	\hat{X}(r^{\mathrm{in}}_{*}) & =  g^{\mathrm{H}}_{-} \left(r(r^{\mathrm{in}}_{*})\right)e^{-ip r^{\mathrm{in}}_{*}}, \\
	\dfrac{d\hat{X}(r^{\mathrm{in}}_{*})}{dr_{*}} & = -ip\hat{X}(r^{\mathrm{in}}_{*}) + \left. \dfrac{dr}{dr_{*}}\dfrac{dg^{\mathrm{H}}_{-}(r)}{dr}\right|_{r=r(r^{\mathrm{in}}_{*})} e^{-ip r^{\mathrm{in}}_{*}},
\end{align}
as the initial values at $r_{*} = r_{*}^{\mathrm{in}}$ after converting them to $\hat{\Phi}^{\mathrm{in}}$ and $d\hat{\Phi}^{\mathrm{in}}/dr_{*}$ using Eqs.~\eqref{eq:PhiFromXXprime} and~\eqref{eq:PhiprimeFromXXprime}, respectively.
Similarly for $\hat{X}^{\mathrm{up}}$, we integrate Eq.~\eqref{eq:Riccatieqn} inward from the outer boundary $r_{*}^{\mathrm{out}}$ to the inner boundary $r_{*}^{\mathrm{in}}$ with
\begin{align}
	\hat{X}(r^{\mathrm{out}}_{*}) & =  f^{\infty}_{+} \left(r(r^{\mathrm{out}}_{*})\right)e^{i\omega r^{\mathrm{out}}_{*}}, \\
	\dfrac{d\hat{X}(r^{\mathrm{out}}_{*})}{dr_{*}} & = i\omega\hat{X}(r^{\mathrm{out}}_{*}) + \left. \dfrac{dr}{dr_{*}}\dfrac{df^{\infty}_{+}(r)}{dr}\right|_{r=r(r^{\mathrm{out}}_{*})} e^{i\omega r^{\mathrm{out}}_{*}},
\end{align}
as the initial values at $r_{*} = r_{*}^{\mathrm{out}}$ after converting them to $\hat{\Phi}^{\mathrm{up}}$ and $d\hat{\Phi}^{\mathrm{up}}/dr_{*}$ using again Eqs.~\eqref{eq:PhiFromXXprime} and~\eqref{eq:PhiprimeFromXXprime}, respectively. Note that for both $\hat{X}^{\mathrm{in}}$ and $\hat{X}^{\mathrm{up}}$, we have chosen the normalization convention of a unit transmission amplitude, i.e., $B^{\mathrm{trans}}_{\mathrm{SN}} = C^{\mathrm{trans}}_{\mathrm{SN}} = 1$. After solving Eq.~\eqref{eq:Riccatieqn} numerically for a complex phase function $\hat{\Phi}(r_{*})$ and its derivative $d\hat{\Phi}/dr_{*}$ on a grid of $r_{*} \in [r_{*}^{\mathrm{in}}, r_{*}^{\mathrm{out}}]$, we first convert them back to $\hat{X}$ and $d\hat{X}/dr_{*}$ using Eqs.~\eqref{eq:complexphase} and~\eqref{eq:PhiprimeFromXXprime}, respectively.

\subsubsection{Transforming generalized Sasaki-Nakamura functions to Teukolsky functions \label{subsubsec:Transform_GSN_to_Teukolsky_functions}}
In principle, if we want to transform a \gls{GSN} function $\hat{X}$ back to a Teukolsky function, we simply need to apply the inverse operator ${}_{s}\Lambda^{-1}$ on the numerical \gls{GSN} function.
Since we have the numerical solutions to both $\hat{X}$ and $d\hat{X}/dr_{*}$, the inverse operator can actually be written as a matrix multiplication to the column vector $\left( \hat{X}, d\hat{X}/dr_{*} \right)^T$.

First, consider the conversion from $\left( \hat{X}, d\hat{X}/dr_{*} \right)^T$ to $\left( \hat{X}, \hat{X}' \right)^T$. This can be done by left multiplying the column vector with the matrix
\begin{equation}
	M_{1} = \begin{pmatrix}
		1 & 0 \\
		0 & \dfrac{r^2 + a^2}{\Delta}
	\end{pmatrix}.
\end{equation}
Next, consider the transformation from $\left( \hat{X}, \hat{X}' \right)^T$ to $\left( \hat{\chi}, \hat{\chi}' \right)^T$ using Eq.~\eqref{eq:Xfromchi}. Again this can be done by left multiplying the column vector $\left( \hat{X}, \hat{X}' \right)^T$ by the matrix
\begin{equation}
	M_{2} = \begin{pmatrix}
		\dfrac{1}{\sqrt{\left(r^2 + a^2\right)\Delta^s}} & 0 \\
		\left( \dfrac{1}{\sqrt{\left(r^2 + a^2\right)\Delta^s}} \right)' & \dfrac{1}{\sqrt{\left(r^2 + a^2\right)\Delta^s}}
	\end{pmatrix}.
\end{equation}
At last, the transformation from $\left( \hat{\chi}, \hat{\chi}' \right)^T$ to $\left( R, R' \right)^T$ is given by the matrix equation as shown in Eq.~\eqref{eq:inversetransformation}, where we now explicitly define the matrix as
\begin{equation}
M_3 = \dfrac{1}{\eta} \begin{pmatrix}
		\alpha + \beta' \Delta^{s+1} & -\beta \Delta^{s+1} \\
		-(\alpha' + \beta V_{\mathrm{T}} \Delta^{s}) & \alpha \\
	\end{pmatrix}.
\end{equation}

The overall transformation from $\hat{X}$ and $d\hat{X}/dr_{*}$ to $R$ and $R'$ is thus given by the matrix equation
\begin{equation}
	\begin{pmatrix}
		\hat{R} \\ \hat{R}'
	\end{pmatrix} = M_{3} M_{2} M_{1} \begin{pmatrix}
		\hat{X} \\ \dfrac{d\hat{X}}{dr_{*}}
	\end{pmatrix}.
\end{equation}
By multiplying $\left( \hat{X}, d\hat{X}/dr_{*} \right)^T$ with the overall transformation matrix $M_{3}M_{2}M_{1}$ that we explicitly simplified, we facilitate cancellations between terms. This allows us to accurately convert numerical \gls{GSN} functions to Teukolsky functions close to the horizon ($\Delta \to 0$) when some of the terms, such as $\alpha(r)$, diverge near the horizon.

\subsection{Extracting incidence and reflection amplitudes from numerical solutions}
Apart from evaluating a \gls{GSN} or a Teukolsky function numerically on a grid of $r$ or $r_{*}$ coordinates, it is also useful to be able to determine the incidence and the reflection amplitude at a particular frequency $\omega$ (see Sec.~\ref{subsec:asymptotic_behaviors} for a theoretical discussion) from a numerical solution accurately. This is essential for constructing inhomogeneous solutions using Green's function method (e.g., calculating gravitational waveforms observed at infinity) and for scattering problems (e.g., calculating the gray-body factor of a \gls{BH} as a function of the wave frequency $\omega$).

Since we only have numerical solutions on a finite grid of $r_{*} \in [r_{*}^{\mathrm{in}}, r_{*}^{\mathrm{out}}]$, in order to determine the reflection amplitude $\hat{B}^{\mathrm{ref}}_{\mathrm{SN}}$ and the incidence amplitude $\hat{B}^{\mathrm{inc}}_{\mathrm{SN}}$ of a $\hat{X}^{\mathrm{in}}$ solution in the \gls{GSN} formalism, we solve the system of linear equations at the outer boundary $r_{*}^{\mathrm{out}}$ that
\begin{equation}
\begin{aligned}
	& \left. \begin{pmatrix}
		f^{\infty}_{+}(r) e^{i\omega r_*} & f^{\infty}_{-}(r) e^{-i\omega r_*} \\[1em]
		(df^{\infty}_{+}/dr_{*} + i\omega f^{\infty}_{+})e^{i\omega r_*} & 
		(df^{\infty}_{-}/dr_{*} - i\omega f^{\infty}_{-})e^{-i\omega r_*}
	  \end{pmatrix} \right|_{r^{\mathrm{out}}_{*}}
	\\
	& \times 
	\begin{pmatrix}
		\hat{B}^{\mathrm{ref}}_{\mathrm{SN}} \\[1em]
		\hat{B}^{\mathrm{inc}}_{\mathrm{SN}}
	\end{pmatrix} = 
	\left. \begin{pmatrix}
		\hat{X}^{\mathrm{in}} \\
		\dfrac{d\hat{X}^{\mathrm{in}}}{dr_{*}}
	\end{pmatrix} \right|_{r^{\mathrm{out}}_{*}},
\end{aligned}
\end{equation}
where we impose continuity of the numerical solution $(\hat{X}^{\mathrm{in}}, d\hat{X}^{\mathrm{in}}/dr_{*})$ with the analytical asymptotic solution near infinity at $r_{*} = r_{*}^{\mathrm{out}}$.
Similarly, we use the same scheme to determine the reflection amplitude $\hat{C}^{\mathrm{ref}}_{\mathrm{SN}}$ and the incidence amplitude $\hat{C}^{\mathrm{inc}}_{\mathrm{SN}}$ of a $\hat{X}^{\mathrm{up}}$ solution in the \gls{GSN} formalism at the inner boundary $r_{*}^{\mathrm{in}}$ by solving
\begin{equation}
\begin{aligned}
	& \left. \begin{pmatrix}
		g^{\mathrm{H}}_{+}(r) e^{ip r_*} & g^{\mathrm{H}}_{-}(r) e^{-ip r_*} \\[1em]
		(dg^{\mathrm{H}}_{+}/dr_{*} + ip g^{\mathrm{H}}_{+})e^{ip r_*} & 
		(dg^{\mathrm{H}}_{-}/dr_{*} - ip g^{\mathrm{H}}_{-})e^{-ip r_*}
	  \end{pmatrix} \right|_{r^{\mathrm{in}}_{*}}
	\\
	& \times 
	\begin{pmatrix}
		\hat{C}^{\mathrm{inc}}_{\mathrm{SN}} \\[1em]
		\hat{C}^{\mathrm{ref}}_{\mathrm{SN}}
	\end{pmatrix} = 
	\left. \begin{pmatrix}
		\hat{X}^{\mathrm{up}} \\
		\dfrac{d\hat{X}^{\mathrm{up}}}{dr_{*}}
	\end{pmatrix} \right|_{r^{\mathrm{in}}_{*}},
\end{aligned}
\end{equation}
where again we impose continuity of the numerical solution $(\hat{X}^{\mathrm{up}}, d\hat{X}^{\mathrm{up}}/dr_{*})$ to the asymptotic solution near the horizon at $r_{*} = r_{*}^{\mathrm{in}}$.\footnote{This matching procedure at the two numerical boundaries actually allows us to obtain ``semianalytical'' \gls{GSN} functions (and by extension Teukolsky functions) that are accurate everywhere, even outside the grid $[r_{*}^{\mathrm{in}}, r_{*}^{\mathrm{out}}]$. Using $X^{\mathrm{in}}$ as an example, for $r_{*} < r_{*}^{\mathrm{in}}$ the analytical ansatz $g^{\mathrm{H}}_{-}(r(r_{*}))e^{-ipr_{*}}$ can be used. This is because the numerical solution $\hat{X}^{\mathrm{in}}$ was constructed by using that ansatz to compute the appropriate initial conditions. While for $r_{*} > r_{*}^{\rm out}$, the linear combination of the analytical ans\"{a}tze $\hat{B}^{\mathrm{ref}}_{\mathrm{SN}} f^{\infty}_{+}(r(r_{*})) e^{i\omega r_*} + \hat{B}^{\mathrm{inc}}_{\mathrm{SN}} f^{\infty}_{-}(r(r_{*})) e^{-i\omega r_*}$ can be used, where the reflection and the incidence coefficient were constructed to ensure continuity with the numerical solution.}

Indeed, the inclusion of the higher-order corrections $f^{\infty}_{\pm}$ at the outer boundary and $g^{\mathrm{H}}_{\pm}$ at the inner boundary, respectively, allow us to get very good agreements on the incidence and the reflection amplitudes over a range of frequencies with the \gls{MST} method, which we will show in the next subsection.

\subsection{Numerical results \label{subsec:numerical_results}}
Here we showcase some numerical results obtained using our \texttt{GeneralizedSasakiNakamura.jl} implementation. Unless otherwise specified, we use the \gls{ODE} solver \texttt{Vern9} \cite{2010NuAlg..53..383V} as implemented in \texttt{DifferentialEquations.jl} \cite{rackauckas2017differentialequations}, and we include corrections to the asymptotic solutions at infinity up to the third order [i.e., truncating the sum in Eq.~\eqref{eq:finfpm} at $j=3$] and that at the horizon only to the zeroth order [i.e., taking only the leading term $j=0$ in the sum in Eq.~\eqref{eq:fhorpm}]. We set the numerical inner boundary at $r_{*}^{\mathrm{in}} = -50M$\footnote{More concretely, this corresponds to $\left(r^{\mathrm{in}} - r_{+}\right)/M \approx 8 \times 10^{-10}$ when $a/M = 0.7$. This difference is a monotonically increasing function in $|a|/M$ (for a similar discussion but for $r_{*}/M = 0$, see Fig.~\ref{fig:h_upperbound}).} and the outer boundary at $r_{*}^{\mathrm{out}} = 1000M$. We use double-precision floating-point numbers throughout, and both the ``absolute tolerance'' \texttt{abstol} (roughly the error around the zero point) and the ``relative tolerance'' \texttt{reltol} (roughly the local error) passed to the numerical \gls{ODE} solver are set to $10^{-12}$.

\subsubsection{Numerical solutions \label{subsubsec:numerical_solutions}}
\begin{figure}[ht]
\centering
\subfloat[\label{fig:Xin_vs_Phiin_Momega_0p5}]{\includegraphics[width=\columnwidth]{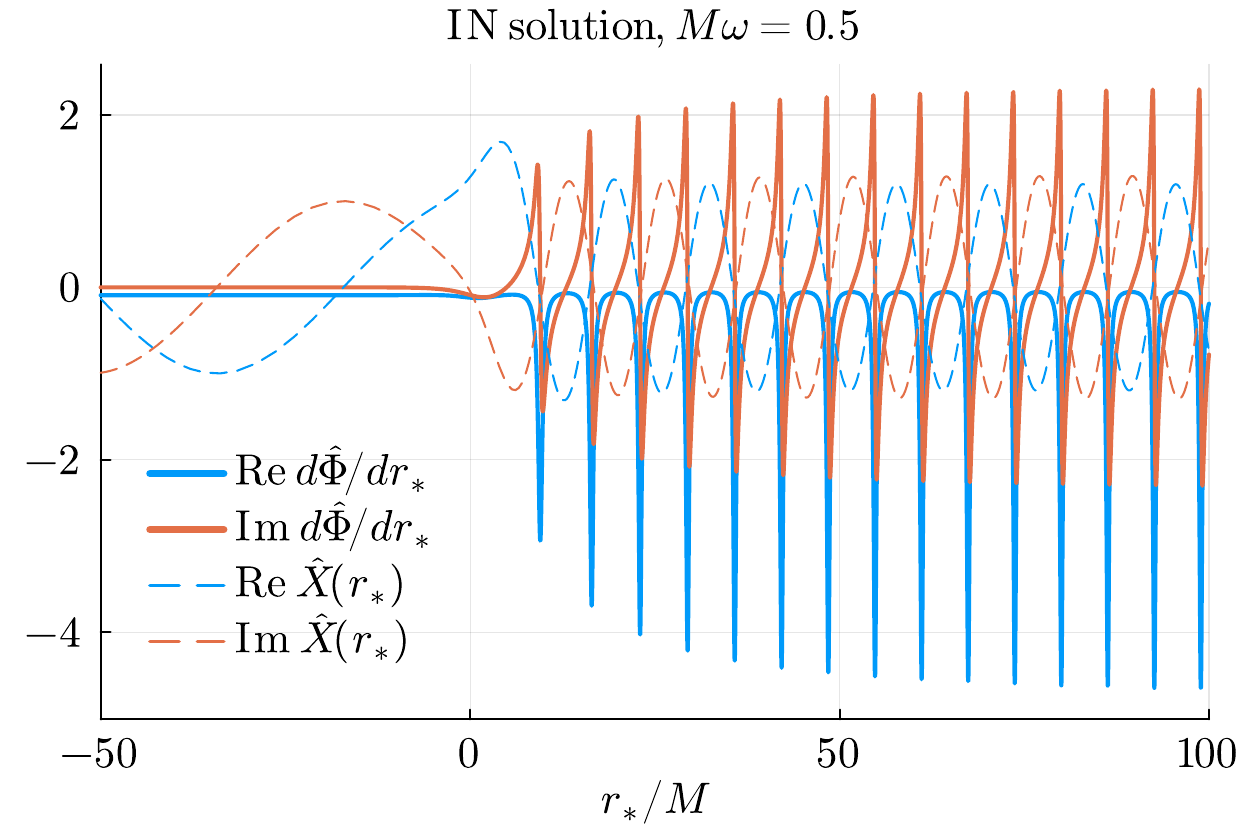}}\\
\subfloat[\label{fig:Xin_vs_Phiin_Momega_1}]{\includegraphics[width=\columnwidth]{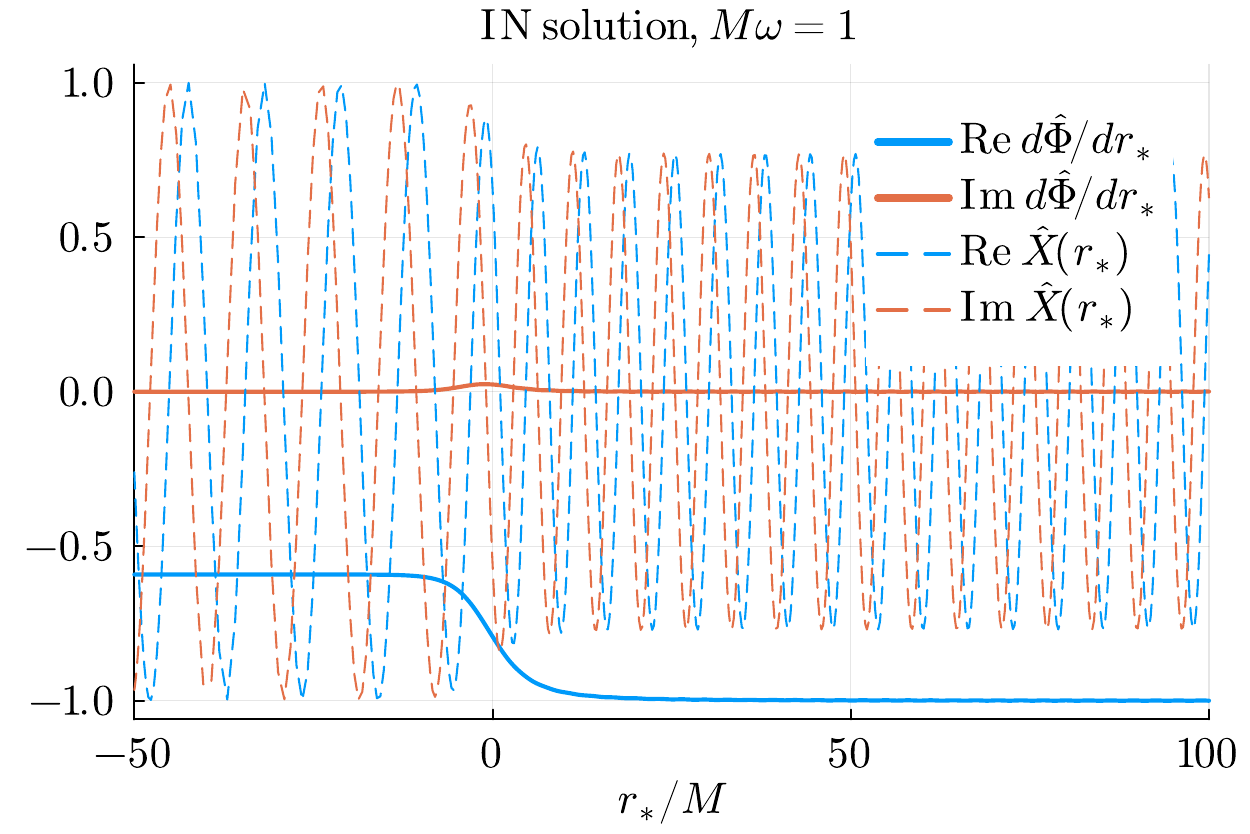}} 
\caption{\gls{GSN} IN solution of the $s=-2, \ell=2, m=2$ mode of a \gls{BH} with $a/M = 0.7$ and two different values of $\omega$ [(a): $M\omega = 0.5$, (b): $M\omega = 1$], in terms of the \gls{GSN} function $X$ and the complex frequency function $d\Phi/dr_{*}$. \label{fig:Xin_vs_Phiin}}
\end{figure}

Figure~\ref{fig:Xin_vs_Phiin} shows the IN solution in the \gls{GSN} formalism of the $s=-2, \ell=2,m=2$ mode for a \gls{BH} with $a/M = 0.7$ and two different values of $\omega$, in terms of the \gls{GSN} function $X$ and the complex frequency function $d\Phi/dr_{*}$. Recall that for an IN solution, it is purely ingoing at the horizon. We see from the figure that for both $M\omega=0.5$ (upper panel) and $M\omega=1$ (lower panel) near the horizon, $d\hat{\Phi}/dr_{*}$ is flat and approaches the imposed asymptotic value $-p^2$, while $\hat{X}$ is oscillating with the frequency $p$.
On the other hand, when $r_{*} \to \infty$, the IN solution is an admixture of the left- and the rightgoing modes where their relative amplitude $B^{\mathrm{ref}}_{\mathrm{SN}}/B^{\mathrm{inc}}_{\mathrm{SN}}$ is $\omega$ dependent. We see from Fig.~\ref{fig:Xin_vs_Phiin_Momega_0p5} that both $\hat{X}$ and $d\hat{\Phi}/dr_{*}$ exhibit oscillatory behaviors, and that the oscillation frequency for $d\hat{\Phi}/dr_{*}$ from beating is twice that for $\hat{X}$. Whereas we see from Fig.~\ref{fig:Xin_vs_Phiin_Momega_1} that $\hat{X}$ is oscillatory but $d\hat{\Phi}/dr_{*}$ is flat, as the ratio of the left- and rightgoing mode is extreme and hence beating is heavily suppressed. 

\begin{figure}[h]
\centering
\subfloat[\label{fig:Phiin_Momega_0p5_secondD}]{\includegraphics[width=\columnwidth]{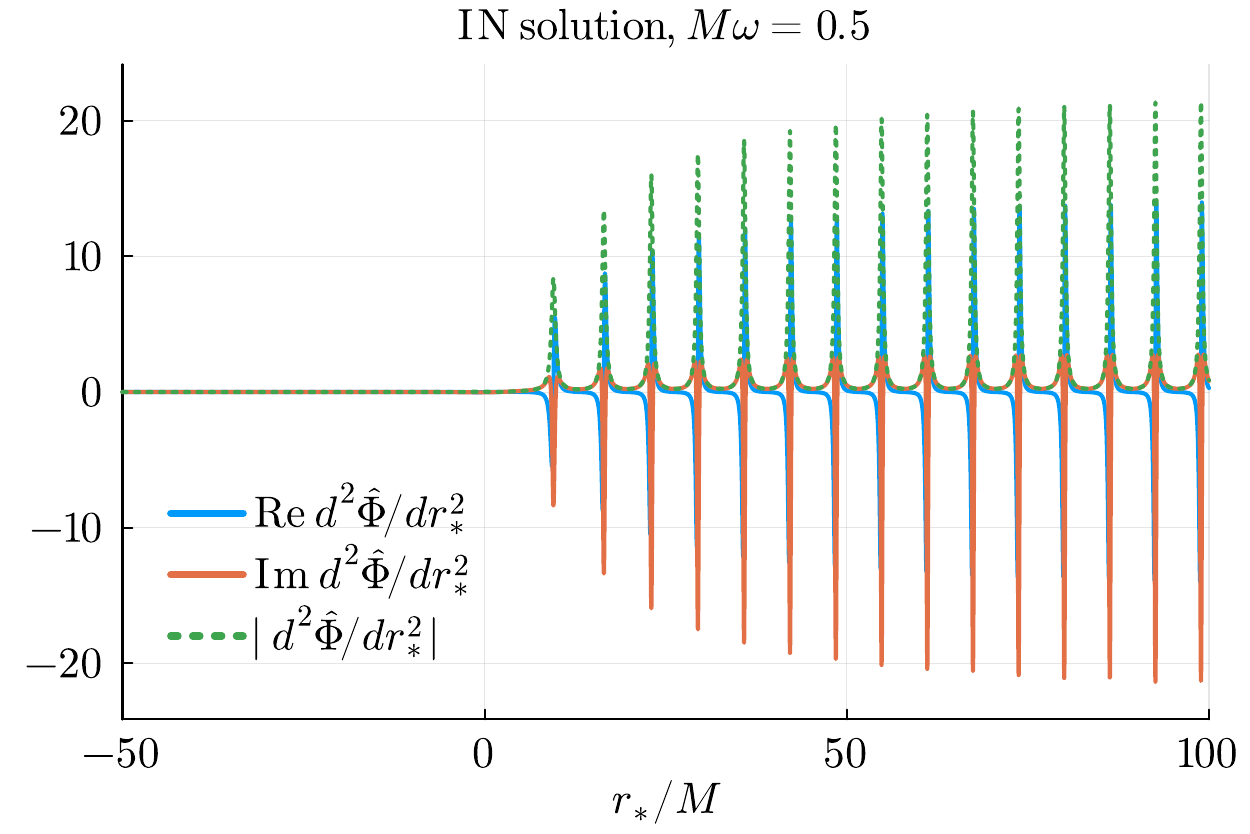}}\\
\subfloat[\label{fig:Phiin_Momega_1_secondD}]{\includegraphics[width=\columnwidth]{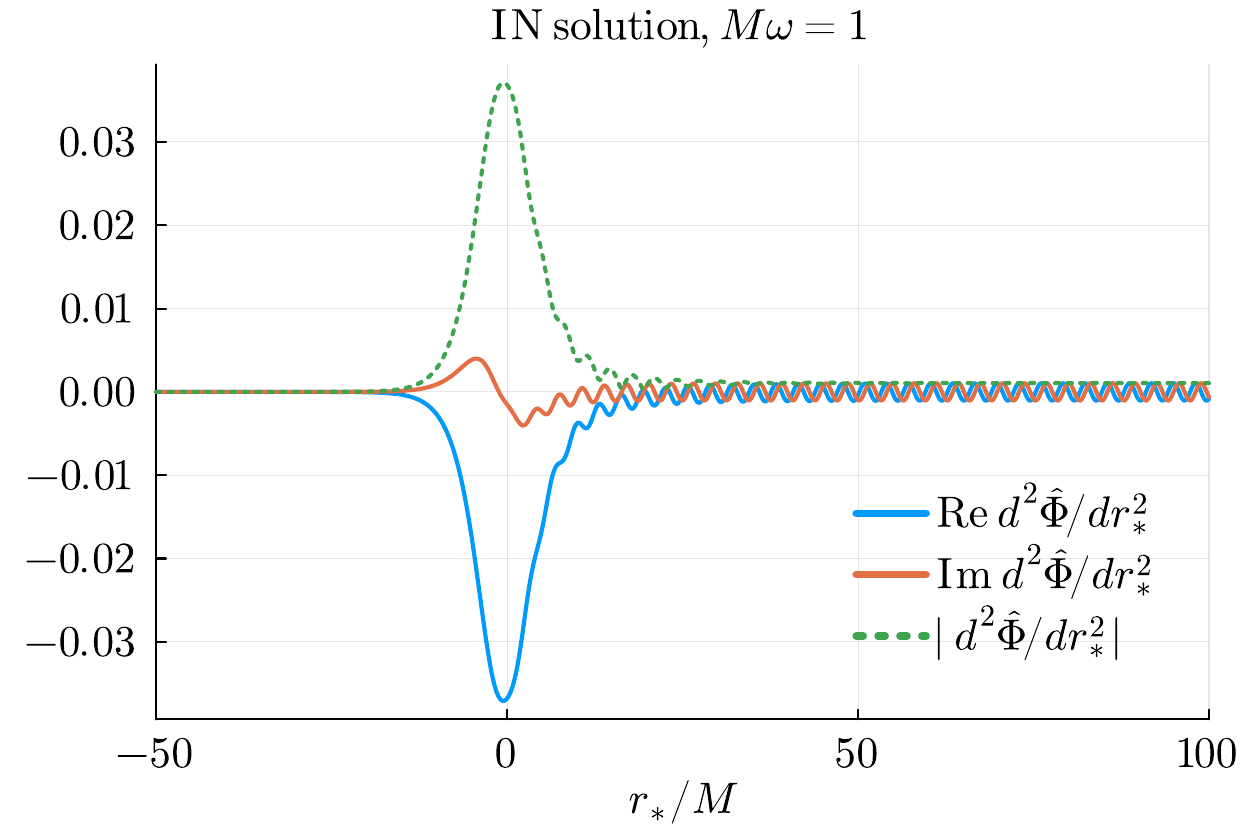}}
\caption{First derivative of the numerical solutions to the complex frequency function in Fig.~\ref{fig:Xin_vs_Phiin} [i.e., $d/dr_{*}(d\hat{\Phi}/dr_{*})$], computed using \gls{AD}, as indicators of how much the numerical solutions are changed locally as functions of $r_{*}$ [(a): $M\omega = 0.5$, (b): $M\omega = 1$]. \label{fig:Phiin_secondD}}
\end{figure}

This can be more easily seen in Fig.~\ref{fig:Phiin_secondD} where it shows the first derivative of the numerical IN solutions $d\hat{\Phi}/dr_{*}$, i.e., $d^2\hat{\Phi}/dr_{*}^2$, as indicators of how much they change locally as functions of $r_{*}$, for both the $M\omega=0.5$ and the $M\omega=1$ case. We compute the numerical derivatives using \gls{AD} on the interpolant of the numerical solutions of $d\hat{\Phi}/dr_{*}$ to avoid issues with using a finite difference method. We see from Fig.~\ref{fig:Phiin_Momega_0p5_secondD} that for $M\omega = 0.5$ the oscillation in $d\hat{\Phi}/dr_{*}$ is significant, while for $M\omega = 1$ we can see from Fig.~\ref{fig:Phiin_Momega_1_secondD} that the oscillation is much more minute. Note that the two panels have very different scales for their $y$ axes.

\begin{figure}[h]
\centering
\includegraphics[width=\columnwidth]{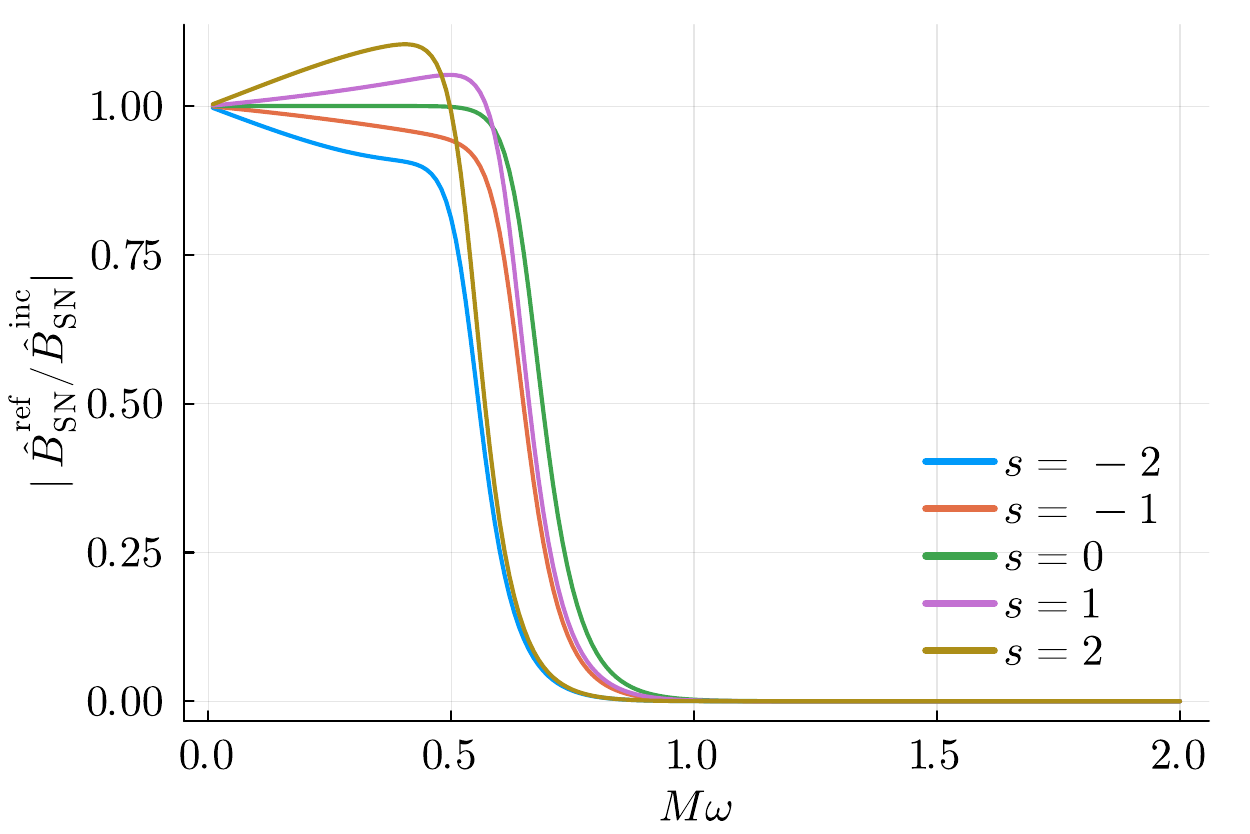}
\caption{Reflectivity $B^{\mathrm{ref}}_{\mathrm{SN}}/B^{\mathrm{inc}}_{\mathrm{SN}}$ of a Kerr \gls{BH} potential barrier in the \gls{GSN} formalism. We see that, for all the spin weights $s$ considered in this paper, the corresponding potential barriers are very permeable to high-frequency ($M\omega \gtrsim 1$) waves, meaning that the potentials will not reflect off the incidence waves and instead allow them to pass right through. In this figure, the \gls{BH} angular momentum was set to $a/M = 0.7$ but the same is true for other values of $a/M$ as well. \label{fig:reflectivity}}
\end{figure}

Physically this boils down to the fact that the potential barriers of a Kerr \gls{BH} for different types of radiation are all very permeable to waves at high frequencies. Figure~\ref{fig:reflectivity} shows the reflectivity of the potential barriers (for $s=0,\pm 1, \pm 2$ with $a/M=0.7$) as defined by $B^{\mathrm{ref}}_{\mathrm{SN}}/B^{\mathrm{inc}}_{\mathrm{SN}}$. This ratio compares the wave amplitude $B^{\mathrm{ref}}_{\mathrm{SN}}$ that is reflected off the potential barrier when a wave with an asymptotic amplitude $B^{\mathrm{inc}}_{\mathrm{SN}}$ is approaching the barrier from infinity. We see from Fig.~\ref{fig:reflectivity} that the reflectivities become zero when the wave frequency gets large (while we only show for the $a/M = 0.7$ case, the same is true for other values of $a/M$ as well). A low reflectivity means that the ratio of the left- and the rightgoing mode is going to be extreme.
Explicitly, for the case in Fig.~\ref{fig:reflectivity}, the rightgoing mode has an amplitude $|B^{\mathrm{ref}}_{\mathrm{SN}}|$ that is much smaller than the leftgoing mode $|B^{\mathrm{inc}}_{\mathrm{SN}}|$ when $M\omega \gtrsim 1$. The lack of beating in Fig.~\ref{fig:Xin_vs_Phiin_Momega_1} is a manifestation of this fact.

\begin{figure}[ht]
\centering
\subfloat[\label{fig:Xup_vs_Phiup_Momega_0p5}]{\includegraphics[width=\columnwidth]{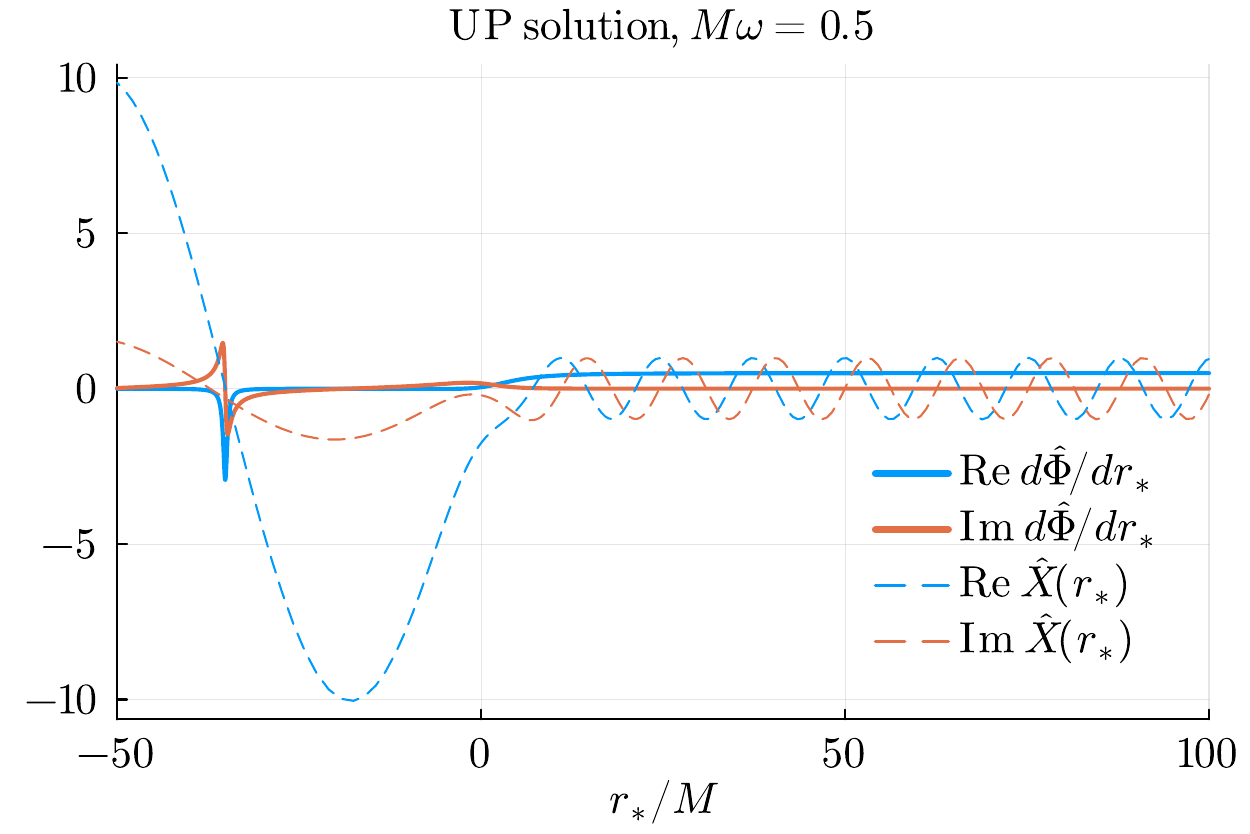}}\\
\subfloat[\label{fig:Xup_vs_Phiup_Momega_1}]{\includegraphics[width=\columnwidth]{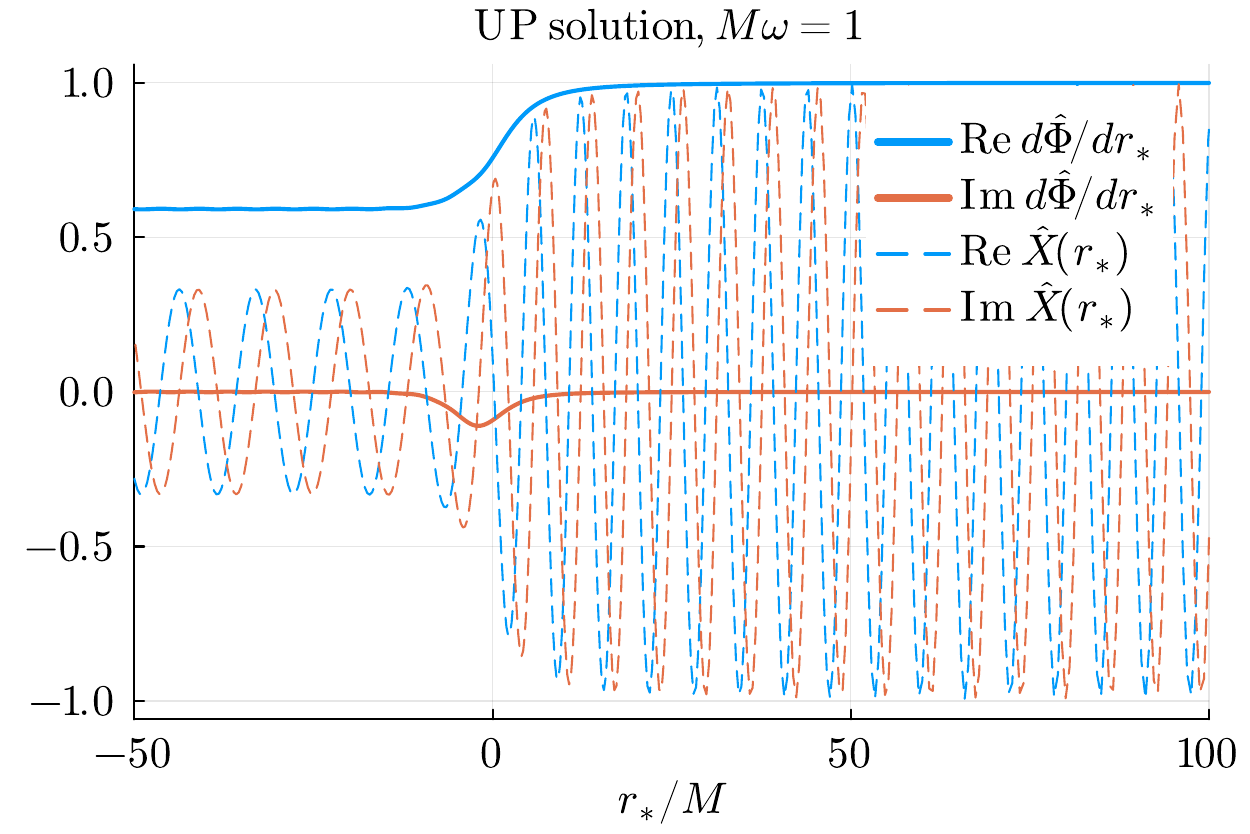}} 
\caption{\gls{GSN} UP solution of the $s=-2, \ell=2, m=2$ mode of a \gls{BH} with $a/M = 0.7$ and two different values of $\omega$ [(a): $M\omega = 0.5$, (b): $M\omega = 1$], in terms of the \gls{GSN} function $X$ and the complex frequency function $d\Phi/dr_{*}$. \label{fig:Xup_vs_Phiup}}
\end{figure}

\begin{figure}[h]
\centering
\includegraphics[width=\columnwidth]{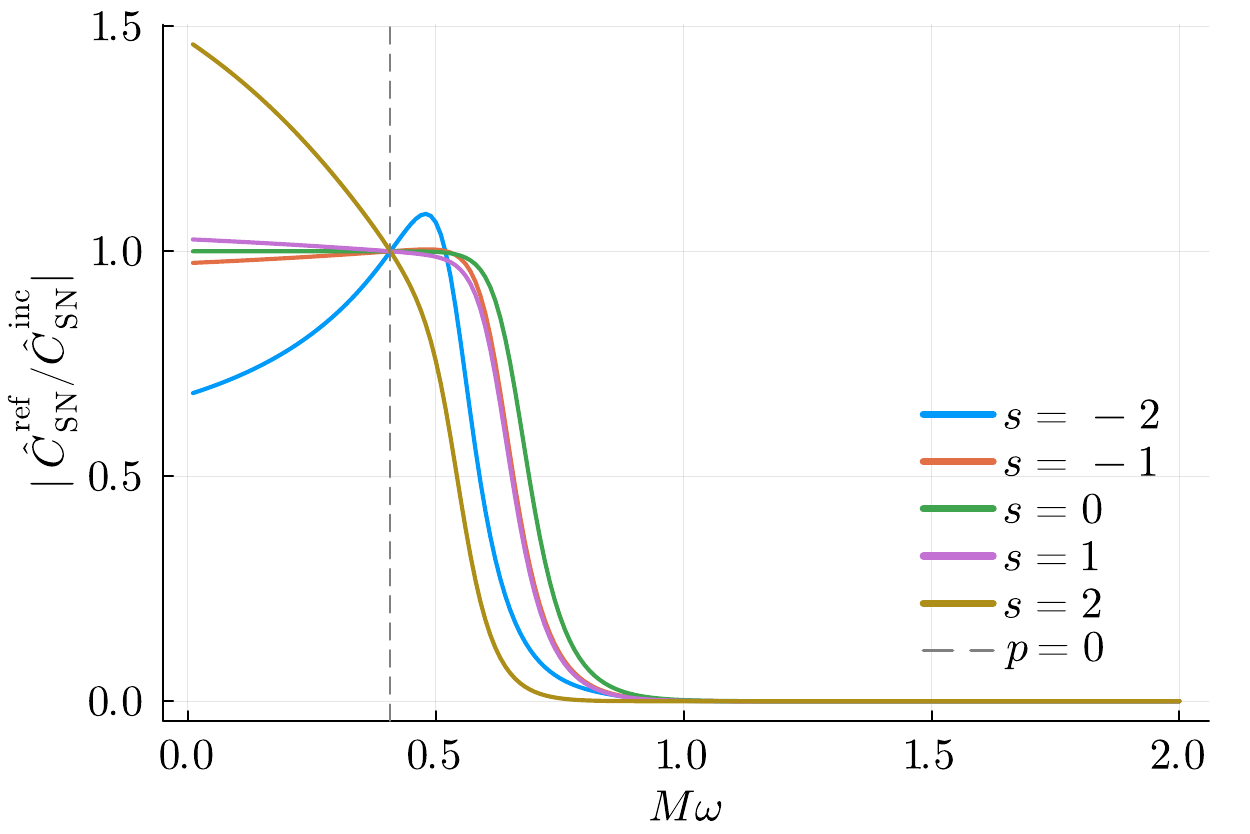}
\caption{The ratio $C^{\mathrm{ref}}_{\mathrm{SN}}/C^{\mathrm{inc}}_{\mathrm{SN}}$ of a Kerr \gls{BH} potential barrier in the \gls{GSN} formalism. This ratio can also be interpreted as the ``reflectivity'' for waves sent from the horizon toward infinity through the barrier. Again, we see that for all the spin weights $s$ considered in this paper, the potential barriers are very permeable to high-frequency ($M\omega \gtrsim 1$) waves, meaning that the potentials will not reflect off the incidence waves from the horizon and instead allow them to pass right through toward infinity. The \gls{BH} angular momentum was set to $a/M = 0.7$, but our conclusion here holds for other values of $a/M$ as well. \label{fig:reflectivity_from_hor}}
\end{figure}

Figure~\ref{fig:Xup_vs_Phiup} is similar to Fig.~\ref{fig:Xin_vs_Phiin}, but showing the UP solution instead. Recall that, for an UP solution, it is purely outgoing at infinity. Again, we see from the figure that, for both $M\omega = 0.5$ and $M\omega = 1$ (upper and lower panel, respectively), $d\hat{\Phi}/dr_{*}$ is flat and approaches the imposed asymptotic value $\omega^2$ as $r_{*} \to \infty$, while $\hat{X}$ is oscillating with the frequency $\omega$. Similar to the IN solutions shown in Fig.~\ref{fig:Xin_vs_Phiin}, since an UP solution is an admixture of the left- and the rightgoing modes near the horizon, depending on their relative amplitude $C^{\mathrm{ref}}_{\mathrm{SN}}/C^{\mathrm{inc}}_{\mathrm{SN}}$, both $\hat{X}$ and $d\hat{\Phi}/dr_{*}$ can be oscillatory near the horizon as shown in Fig.~\ref{fig:Xup_vs_Phiup_Momega_0p5}. When the frequency $\omega$ is sufficiently high, the beating in $d\hat{\Phi}/dr_{*}$ is suppressed while $\hat{X}$ remains oscillatory as shown in Fig.~\ref{fig:Xup_vs_Phiup_Momega_1}.
Again, this can be attributed to the property of a Kerr \gls{BH} that it allows high-frequency waves to pass through mostly unimpeded, as shown in Fig.~\ref{fig:reflectivity_from_hor} (like Fig.~\ref{fig:reflectivity}, showing only for the $a/M = 0.7$ case). At a special frequency where $\omega = m\Omega_{\rm H}$ (or, equivalently, $p = 0$, which is also indicated by the vertical dashed line in Fig.~\ref{fig:reflectivity_from_hor}), the ratio $|C^{\mathrm{ref}}_{\mathrm{SN}}/C^{\mathrm{inc}}_{\mathrm{SN}}|$ converges to unity even though individually $C^{\mathrm{ref, inc}}_{\mathrm{SN}}$ are not well defined at that particular frequency.

\subsubsection{Numerical accuracy \label{subsubsec:numerical_accuracy}}
As numerical solutions are only approximations to the true solutions, it is necessary to verify their accuracies. First, we need to show that the initial conditions $\hat{X}$ and $d\hat{X}/dr_{*}$ that we use are sufficiently accurate such that when solving for $\hat{X}^{\mathrm{in, up}}$ the corresponding asymptotic boundary forms are satisfied. Next, we need to show that the numerical solutions actually satisfy the \gls{GSN} equation inside the integration domain. In both cases, we can evaluate the residual $\varepsilon$, which is defined as
\begin{equation}
\label{eq:numerical_residual}
	\varepsilon = 	\left| \dfrac{d^2 \hat{X}}{d r_{*}^2} - \mathcal{F}(r)\dfrac{d\hat{X}}{dr_{*}} - \mathcal{U}(r)\hat{X} \right|,
\end{equation}
where a smaller value (ideally zero) means a better agreement of a numerical solution $\hat{X}$ with the \gls{GSN} equation.

\begin{figure}[ht]
\centering
\subfloat[for ansatz near infinity, $f^{\infty}_{\pm}$ \label{fig:epsilon_ansatz_f}]{\includegraphics[width=\columnwidth]{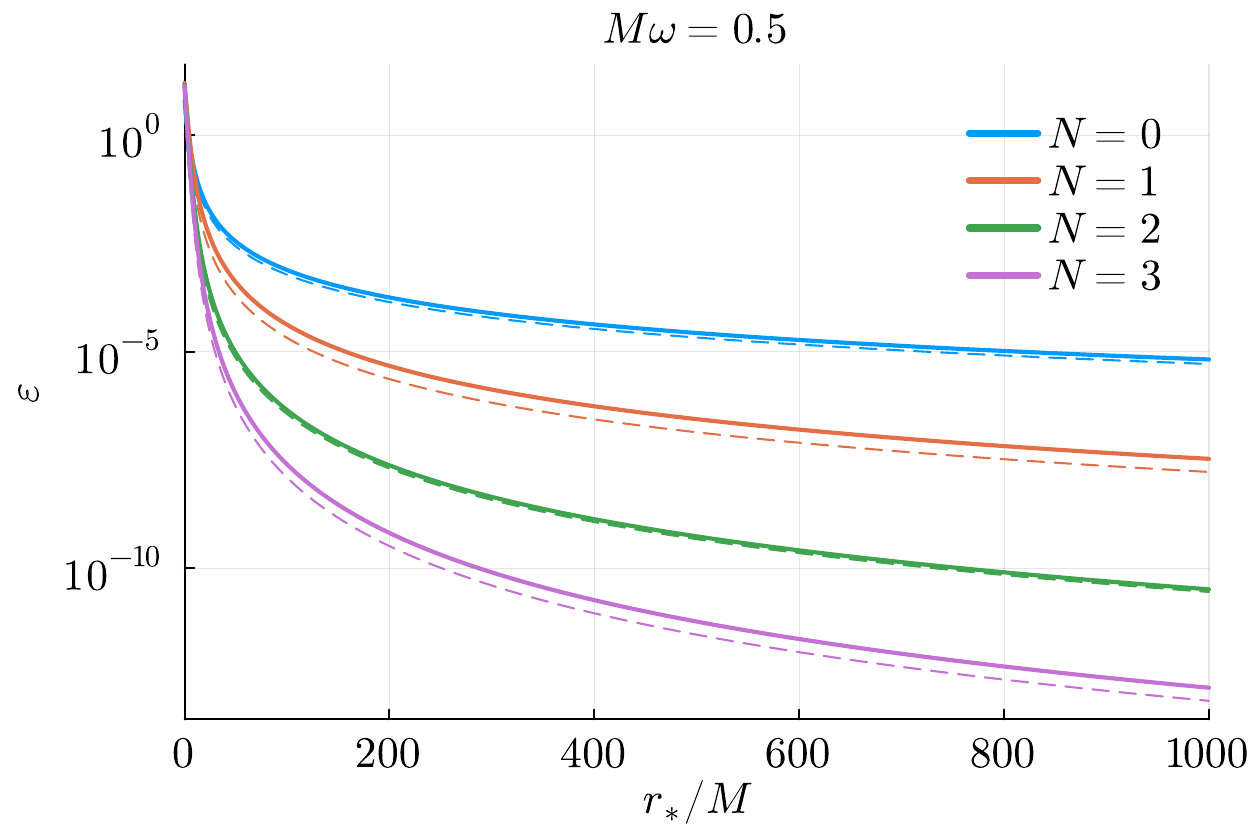}}\\
\subfloat[for ansatz near the horizon, $g^{\mathrm{H}}_{\pm}$ \label{fig:epsilon_ansatz_g}]{\includegraphics[width=\columnwidth]{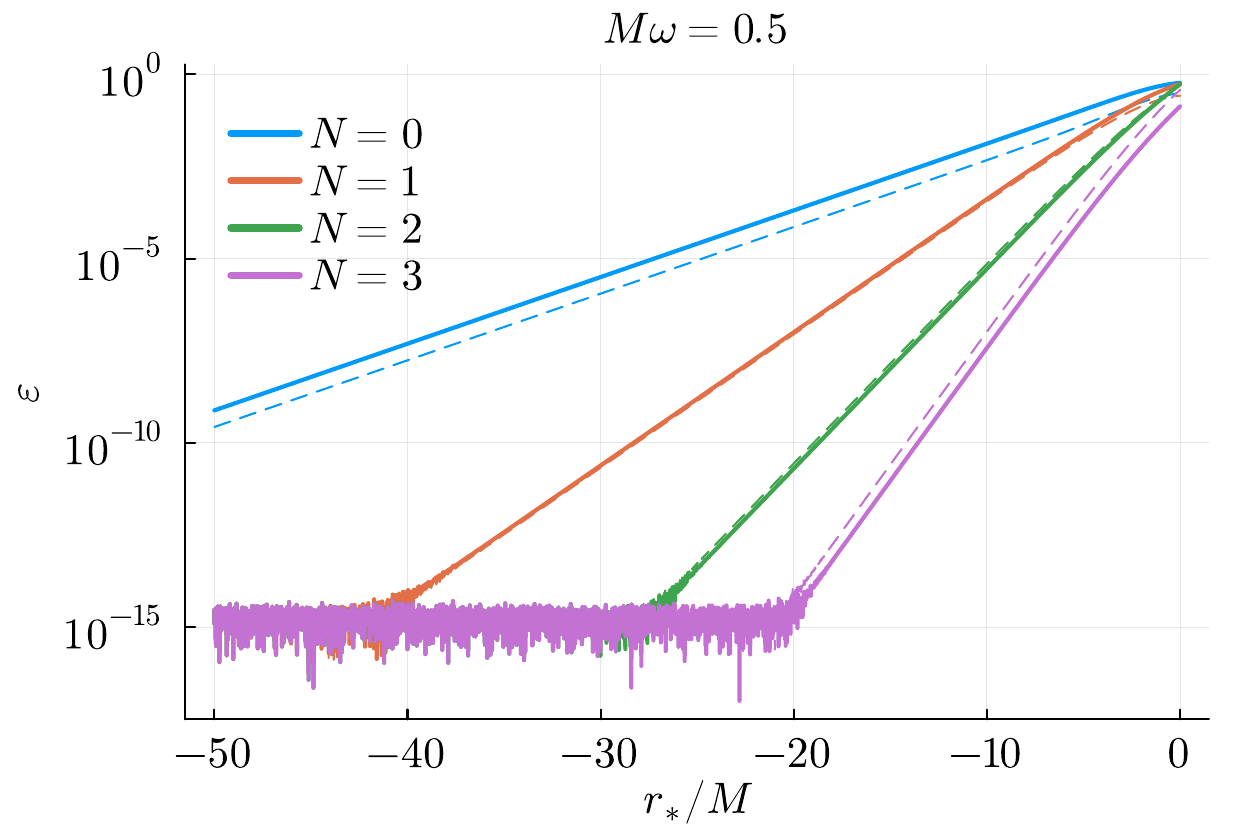}}
\caption{Residual $\varepsilon$ of the ans\"{a}tze [cf. Eq.~\eqref{eq:Xansatz}] $f^{\infty}_{\pm}$ near infinity (a) and $g^{\mathrm{H}}_{\pm}$ near the horizon (b) that we use in evaluating the initial conditions when solving for $X^{\mathrm{in, up}}$ and extracting the incidence and reflection amplitudes from the numerical solutions. In particular, we set $s=-2,\ell=2,m=2,a/M=0.7$ for the purpose of demonstration. For both plots, solid lines correspond to the outgoing ans\"{a}tze and dashed lines correspond to the ingoing ans\"{a}tze truncated to different orders $N=0,1,2,3$, respectively. \label{fig:epsilon_ansatz}}
\end{figure}

Figure~\ref{fig:epsilon_ansatz} shows the residual $\varepsilon$ of the ans\"{a}tze, $f^{\infty}_{\pm}$ near infinity (upper panel) and $g^{\mathrm{H}}_{\pm}$ near the horizon (lower panel) as functions of $r_{*}$. For both panels, solid lines correspond to the outgoing ans\"{a}tze and dashed lines correspond to the ingoing ans\"{a}tze truncated to different orders $N=0,1,2,3$, i.e., keeping the first $N+1$ terms in Eqs.~\eqref{eq:finfpm} and \eqref{eq:fhorpm}, respectively. Recall that for all the numerical results we have shown previously, we set the numerical outer boundary $r_{*}^{\mathrm{out}} = 1000M$ and truncate $f^{\infty}_{\pm}$ at $N = 3$ (i.e., including the first four terms). From Fig.~\ref{fig:epsilon_ansatz_f} we see that this corresponds to $\varepsilon \approx 10^{-13}$. As expected, for a fixed $r_{*} \gg 1$, the residual $\varepsilon$ decreases as one keeps more terms (i.e., higher $N$) in the summation in Eq.~\eqref{eq:finfpm}. Alternatively, for a fixed $N$, the residual $\varepsilon$ goes down as one has an numerical outer boundary $r_{*}^{\mathrm{out}}$ further away from the \gls{BH}.

As for the numerical inner boundary $r_{*}^{\mathrm{in}}$, recall that we set $r_{*}^{\mathrm{in}} = -50M$ and truncate $g^{\mathrm{H}}_{\pm}$ such that only the leading term is kept (i.e., $N = 0$). From Fig.~\ref{fig:epsilon_ansatz_g} we see that this corresponds to $\varepsilon \approx 10^{-10}$. Similar to $f^{\infty}_{\pm}$, the residual decreases with a higher $N$ in the summation of Eq.~\eqref{eq:fhorpm} for a fixed $r_{*}$ until the precision of a double-precision floating-point number (around $10^{-15}$) is reached and $\varepsilon$ plateaus. Again, for a fixed $N$, as one sets the inner boundary closer to the horizon, the residual drops until around $10^{-15}$.

\begin{figure}[h]
\centering
\includegraphics[width=\columnwidth]{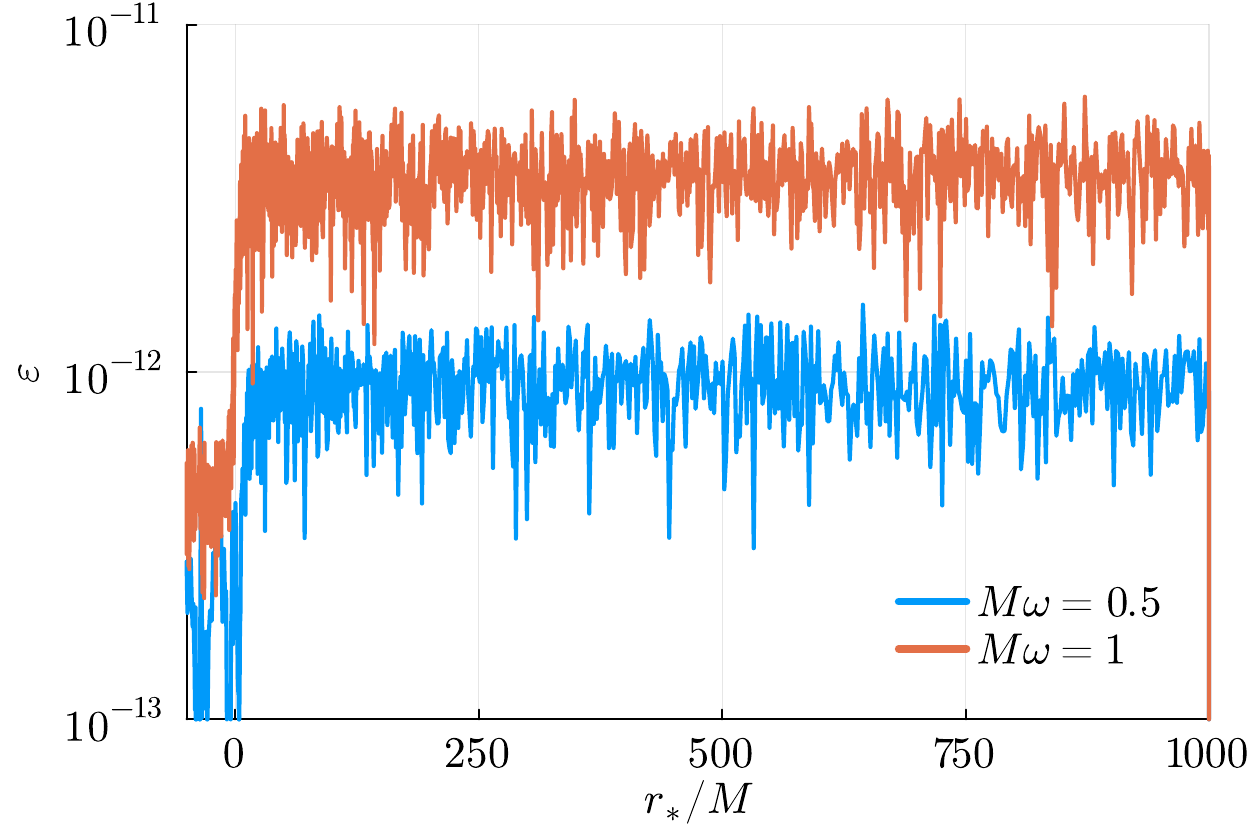}
\caption{Residual $\varepsilon$ [cf. Eq.~\eqref{eq:numerical_residual}] of the numerical \gls{GSN} UP solutions shown in Fig.~\ref{fig:Xup_vs_Phiup} for $M\omega = 0.5$ and $M\omega = 1$, respectively. Recall that both the absolute tolerance and the relative tolerance passed to the numerical \gls{ODE} solver are set to $10^{-12}$. \label{fig:epsilon_GSN_UP}}
\end{figure}

Figure~\ref{fig:epsilon_GSN_UP} shows the residual $\varepsilon$, defined in Eq.~\eqref{eq:numerical_residual}, for the numerical \gls{GSN} UP solutions in Fig.~\ref{fig:Xup_vs_Phiup} (with $s=-2,\ell=2,m=2$, and $a/M=0.7$), for both $M\omega=0.5$ and $M\omega = 1$.\footnote{Note that the second derivatives are computed using \gls{AD} on the numerical solution interpolants, instead of using a finite difference method that would introduce extra numerical artifacts.} We see that the residuals are indeed very small and stay roughly at $\varepsilon \approx 10^{-12}$, which is the absolute and relative tolerance given to the \gls{ODE} solver. As for the numerical \gls{GSN} IN solutions, the residuals are similar to that for the UP solutions. 

The scaled Wronskian $\mathcal{W}_{X}$ [cf. Eq.~\eqref{eq:WX_def}] can be used as a sanity check. Using again the numerical solutions in Fig.~\ref{fig:Xup_vs_Phiup} for the UP solution and Fig.~\ref{fig:Xin_vs_Phiin} for the IN solution with $M\omega = 0.5$ and $M\omega = 1$, we evaluate the magnitude of the complex scaled Wronskian $|\mathcal{W}_{X}|$, which should be constant, at four different values of $r_{*}/M = -50,0,50,1000$, respectively. The scaled Wronskian can also be computed using the asymptotic amplitudes at infinity [cf. Eq.~\eqref{eq:WX_at_inf}] and at the horizon [cf. Eq.~\eqref{eq:WX_at_hor}], respectively. The values are tabulated in Table~\ref{tab:scaledWX}. We see that the scaled Wronskians computed from the numerical solutions for the two values of $M\omega$ are indeed constant, at least up to the 11th digit, across the integration domain $r_{*} \in [-50M, 1000M]$. This means that our method for solving \gls{GSN} functions are numerically stable. The agreement of the scaled Wronskian evaluated at different locations in the integration domain and that evaluated using the asymptotic amplitudes at both boundaries also implies that our procedure of extracting incidence and reflection amplitudes from numerical solutions works. 

\begin{table}
\centering
\caption{\label{tab:scaledWX}Magnitude of the (complex) scaled Wronskian $|\mathcal{W}_{X}|$ of two frequencies, $M\omega = 0.5$ and $M\omega = 1$, evaluated at four different positions, $r_{*}/M = -50, 0, 50, 1000$, respectively, and evaluated using the asymptotic amplitudes, with the \texttt{GeneralizedSasakiNakamura.jl} code. Digits beyond the 11th digit are shown in brackets.}
\begin{ruledtabular}
\begin{tabular}{rcc}
  $r_{*}/M$ & $M\omega = 0.5$ & $M\omega = 1$ \\
  $-\infty$ & 0.06686918718(132409) & 0.09801150092(211632) \\
  $-50$ & 0.06686918718(132406) & 0.09801150092(220787) \\
  $0$ & 0.06686918718(135844) & 0.09801150092(220655) \\
  $50$ & 0.06686918718(137257) & 0.09801150092(220637) \\
  $1000$ & 0.06686918718(173902) & 0.09801150092(220587) \\
  $\infty$ & 0.06686918718(244163) & 0.09801150092(220785)
\end{tabular}
\end{ruledtabular}
\end{table}

\begin{figure}[h!]
\centering
\includegraphics[width=\columnwidth]{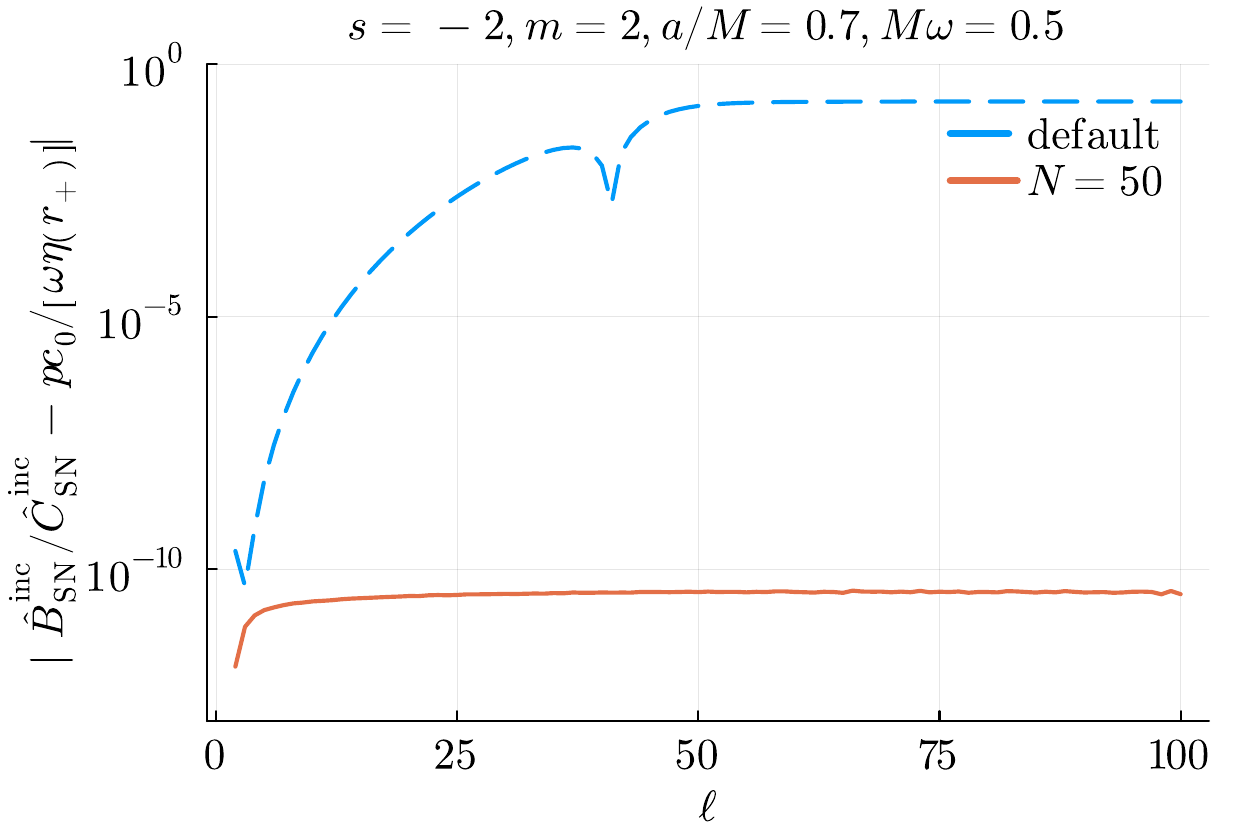}
\caption{\label{fig:high_l}Absolute error of the numerical estimate of $B^{\mathrm{inc}}_{\mathrm{SN}}/C^{\mathrm{inc}}_{\mathrm{SN}}$ as a function of $\ell$, as a sanity check of the numerical solutions. With the default settings used in this paper (i.e., taking only the leading term in the ansatz near the horizon and taking the third-order accurate ansatz near infinity), shown by the dashed curve, the error reaches $\bigO(1)$ after roughly $\ell \gtrsim 50$. However, when using 50-order-accurate (i.e., $N = 50$) ans\"{a}tze near both the horizon and infinity, the error plateaus at around $10^{-10}$ even when $\ell \sim 100$, as shown by the solid curve.}
\end{figure}

Our numerical implementation of the \gls{GSN} formalism is also robust when $\ell$ is large. This is demonstrated in Fig.~\ref{fig:high_l}, where we compare the value of $B^{\mathrm{inc}}_{\mathrm{SN}}/C^{\mathrm{inc}}_{\mathrm{SN}}$ extracted from the numerical $\hat{X}^{\mathrm{in, up}}$ solutions ($s=-2, m=2, a/M = 0.7, M\omega=0.5$) with the theoretical value given in Eq.~\eqref{eq:WX_identity} for $2 \leq \ell \leq 100$. In order for the numerical solutions to satisfy the identity in Eq.~\eqref{eq:WX_identity}, both the IN and the UP solutions need to be solved accurately. With the default settings used in this paper (cf. the first paragraph of Sec.~\ref{subsec:numerical_results}), the error reaches $\bigO(1)$ after roughly $\ell \gtrsim 50$, as shown by the dashed curve. However, when using ans\"{a}tze that are accurate to the 50th order (i.e., $N = 50$) near both the horizon and infinity, the error plateaus at around $10^{-10}$ even when $\ell \sim 100$, as shown by the solid curve. Thanks to the use of \gls{AD} in \texttt{GeneralizedSasakiNakamura.jl}, the extra computation needed for such a high-order ansatz is minimal.

\subsubsection{Comparisons with the Mano-Suzuki-Takasugi method \label{subsubsec:comparison_with_MST}}
As mentioned in Sec.~\ref{sec:intro}, there are other ways of computing homogeneous solutions to the radial Teukolsky equation and one of which is the \gls{MST} method. Using the \gls{MST} method, asymptotic amplitudes of Teukolsky functions (i.e., incidence and reflection amplitudes normalized by transmission amplitudes) can be determined accurately, together with the homogenous solutions themselves. Here we compare our numerical solutions and asymptotic amplitudes using the \gls{GSN} formalism with that using the \gls{MST} method. In particular, we use the implementation in the \texttt{Teukolsky} \cite{barry_wardell_2022_7037857} \textsc{Mathematica} package from the Black Hole Perturbation Toolkit \cite{BHPToolkit}.

We compute the scaled Wronskian $\mathcal{W}_{R}$ of the numerical solutions for $s=-2,\ell=2,m=2,a/M=0.7$ mode for both $M\omega=0.5$ and $M\omega=1$ (the same setup as in Table~\ref{tab:scaledWX}), using the \gls{MST} method.
Similar to the case for \gls{GSN} functions, we can compute $\mathcal{W}_{R}$ either from the numerical solutions $R^{\mathrm{in, up}}$ using Eq.~\eqref{eq:WR_def} or from the asymptotic amplitudes using Eqs.~\eqref{eq:WR_at_inf} or~\eqref{eq:WR_at_hor}, and they should agree. In addition, the values for $\mathcal{W}_{R}$ should be the same as $\mathcal{W}_{X}$.\footnote{Note that the \texttt{Teukolsky} package uses a normalization convention that $B^{\mathrm{trans}}_{\mathrm{T}} = C^{\mathrm{trans}}_{\mathrm{T}} = 1$, which is different from our \texttt{GeneralizedSasakiNakamura.jl} implementation. To account for the difference in the normalization convention, a factor of $\left(C^{\mathrm{trans}}_{\mathrm{T}}/C^{\mathrm{trans}}_{\mathrm{SN}}\right)\left(B^{\mathrm{trans}}_{\mathrm{T}}/B^{\mathrm{trans}}_{\mathrm{SN}}\right)$ is multiplied to $\mathcal{W}_{R}$ computed from the \texttt{Teukolsky} code.} The results are tabulated in Table~\ref{tab:scaledWR}. We see that the numbers shown in Table~\ref{tab:scaledWX}, which were computed using the \gls{GSN} formalism, agree with the numbers in Table~\ref{tab:scaledWR} at least up to the 11th digit, testifying the numerical accuracy and correctness of the solutions and the asymptotic amplitudes computed using \texttt{GeneralizedSasakiNakamura.jl}. It should also be remarked that the implementation of the \gls{MST} method in the \texttt{Teukolsky} package seems to be struggling either very close (e.g., $r_{*} = -50M$) or very far away (e.g., $r_{*} = 1000M$) from the \gls{BH}, and in general, the \gls{MST} method struggles more as $M\omega$ becomes larger\footnote{We performed the same set of calculations in Sec.~\ref{subsubsec:comparison_with_MST} using another \gls{MST}-based \texttt{Fortran} code described in Refs.~\cite{Han:2010tp, Han:2011qz} that uses machine-precision numbers. The same conclusion is reached.} while the \gls{GSN} formalism becomes more efficient instead.\footnote{More concretely, the authors of Ref.~\cite{10.1143/PTP.113.1165} gave explicit examples ($s=-2, \ell =2, a/M = 0, M\omega > 5$) where they found their \gls{MST} code was struggling to compute, while the \gls{GSN} formalism, for example, using our code, can handle these cases with ease.}

\begin{table}
\centering
\caption{\label{tab:scaledWR}Magnitude of the (complex) scaled Wronskian $|\mathcal{W}_{R}|$ of two frequencies, $M\omega = 0.5$ and $M\omega = 1$, evaluated at four different positions, $r_{*}/M = -50, 0, 50, 1000$, respectively, and evaluated using the asymptotic amplitudes, with the \gls{MST} method implemented in the \texttt{Teukolsky} code. Note that in the computations, we use the arbitrary-precision arithmetic in \textsc{Mathematica} (specifically 64-digit accurate). Digits beyond the 11th digit are shown in brackets and truncated to the 17th digit to match Table~\ref{tab:scaledWX}. The computations at $r_{*} = -50M$ for both cases were aborted after running for an hour.}
\begin{ruledtabular}
\begin{tabular}{rcc}
  $r_{*}/M$ & $M\omega = 0.5$ & $M\omega = 1$\\
  $-\infty$ & 0.06686918718(210336) & 0.09801150092(219980) \\
  $-50$ & Aborted & Aborted \\
  $0$ & 0.06686918718(210336) & 0.09801150092(219978) \\
  $50$ & 0.06686918718(210336) & Error \\
  $1000$ & Error & Error \\
  $\infty$ & 0.06686918718(210336) & 0.09801150092(219980)
\end{tabular}
\end{ruledtabular}
\end{table}

\section{Conclusion and future work \label{sec:conclusion_and_future_work}}
In this paper, we have revamped the \gls{GSN} formalism for computing homogeneous solutions to both the \gls{GSN} equation and the radial Teukolsky equation for scalar, electromagnetic, and gravitational perturbations. Specifically, we have provided explicit expressions for the transformations between the Teukolsky formalism and the \gls{GSN} formalism. We have also derived expressions for higher-order corrections to asymptotic solutions of the \gls{GSN} equation, as well as frequency-dependent conversion factors between asymptotic solutions in the Teukolsky and the \gls{GSN} formalism. Both are essential for using the \gls{GSN} formalism to perform numerical work. We have also described an open-source implementation of the now-complete \gls{GSN} formalism for solving homogeneous solutions, where the implementation reformulated the \gls{GSN} equation further into a Riccati equation so as to gain extra efficiency at high frequencies.

In the following we discuss two potential applications of the \gls{GSN} formalism in \gls{BH} perturbation theory, namely as an efficient procedure for computing gravitational radiation from \glspl{BH} and as an alternative method for \gls{QNM} determination.

\subsection{An efficient procedure for computing gravitational radiation from Kerr black holes \label{subsec:inhomogeneous_usingGSN}}
As we have demonstrated in Sec.~\ref{subsec:numerical_results}, the \gls{GSN} formalism is capable of producing accurate and stable numerical solutions to the homogenous \gls{GSN} equation, which can then be converted to numerical Teukolsky functions, across a wide range of $r_{*}/M$ when the \gls{MST} method tends to struggle when $r_{*}/M \ll 1$ and $r_{*}/M \gg 1$ as shown in Sec.~\ref{subsubsec:comparison_with_MST}. While we have only shown the numerical results for $M\omega = 0.5$ and $M\omega = 1$ explicitly, it is reasonable to expect the formalism to also work for other frequencies, if not even better at high frequencies when we gain extra efficiency by further transforming a \gls{GSN} function $X(r_{*})$ into a complex frequency function $d\Phi/dr_{*}$, while the \gls{MST} method requires a much higher working precision for computation. This can occur, for example, when computing a higher harmonic of an \acrlong{EMRI} waveform. For a generic orbit, the harmonic has a frequency $\omega$ given by \cite{Hughes:2021exa}
\begin{equation}
\label{eq:EMRI_wfm_freq}
	\omega = m \Omega_{\phi} + k \Omega_{\theta} + n \Omega_{r},
\end{equation}
where $\Omega_{\phi}, \Omega_{\theta}, \Omega_{r}$ are the fundamental orbital frequency for the $\phi$, $\theta$ and $r$ motion, respectively.

Indeed, we see from Sec.~\ref{subsubsec:numerical_solutions} that, in some regions of the parameter space, it is more efficient to solve for the complex frequency function $d\Phi/dr_{*}$ than to solve for the \gls{GSN} function $X$ itself. There are, however, cases where the reverse is true instead, especially at a lower wave frequency when the \gls{BH} potential barrier is less transmissive, since it is numerically more efficient (requiring fewer nodes) to track a less oscillatory function than a more oscillatory function (cf. Fig.~\ref{fig:Xin_vs_Phiin}). This means that a better numerical scheme solving for $X^{\mathrm{in, up}}$ (and by extension $R^{\mathrm{in, up}}$) can be formulated by first solving the first-order nonlinear \gls{ODE} for $d\Phi/dr_{*}$, and then ``intelligently'' switching to solving the second-order linear \gls{ODE} for $X$ when it is more efficient, for example, when $d/dr_{*}(d\hat{\Phi}/dr_{*})$ is above some predefined threshold. This hybrid approach is similar in spirit to some of the state-of-the-art solvers for oscillatory second-order linear \glspl{ODE} \cite{2022arXiv221206924A}.\footnote{As mentioned in both Refs.~\cite{Finn:2000sy} and~\cite{2022arXiv221206924A}, pseudospectral methods can be adopted instead of finite difference methods (like the \texttt{Vern9} algorithm that this paper uses) to achieve exponential convergence. We leave this as a future improvement to this work.}

While the \gls{GSN} formalism is a great alternative\footnote{Yet another great alternative would be the use of hyperboloidal coordinates \cite{Zenginoglu:2011jz, PanossoMacedo:2019npm}, which both give a short-ranged potential. Hence, both the \gls{GSN} and hyperboloidal formalisms have superior numerical behaviors compared to the Teukolsky formalism with the Boyer-Lindquist coordinates. Loosely speaking, the \gls{GSN} formalism can be thought of as an active transformation where wave functions themselves change, whereas the hyperboloidal formalism is a passive transformation where instead the coordinate system describing the wave functions changes. While the \gls{GSN} formalism as it stands is limited to only integer spin-weight fields, the hyperboloidal framework can deal with half-integer spin-weight fields such as neutrino waves.} to the \gls{MST} method for computing homogeneous solutions (i.e., $T = 0$) to the radial Teukolsky equation, the real strength of the \gls{GSN} formalism is the ability to also compute inhomogeneous solutions (i.e. $T \neq 0$).
Given an extended Teukolsky source term, such as a plunging test particle from infinity, the convolution integral with the Teukolsky functions can be divergent when using Green's function method to compute the inhomogeneous solution and regularization of the integral is needed \cite{Poisson:1996ya, Campanelli:1997sg}. In Ref.~\cite{10.1143/PTP.67.1788}, Sasaki and Nakamura had worked out a formalism, which was developed upon their \gls{SN} transformation, to compute the inhomogeneous solution for $s=-2$ where the new source term, constructed from the Teukolsky source term, is short ranged such that the convolution integral with the \gls{SN} functions is convergent when using Green's function method.

In a forthcoming paper, we show that their construction can also be extended to work for $s=2$, and the corresponding \gls{GSN} transformation, in a similar fashion, serves as the foundation of the method. This will be important for studying near-horizon physics \cite{OSullivan:2014ywd, Chen:2020htz, Xin:2021zir}, such as computing gravitational radiation from a point particle plunging toward a \gls{BH} as observed near the horizon, where the polarization contents are encoded in $\psi_0$ (with $s=2$) instead of $\psi_4$ (with $s=-2$). In particular, the Teukolsky-Starobinsky identities \cite{Starobinskil:1974nkd, Teukolsky:1974yv} are not valid in this case (since the source term does not vanish near the horizon) and we cannot use them to convert the asymptotic amplitude for $\psi_4$ to that for $\psi_0$.\footnote{Note that it is still possible to compute the asymptotic amplitude for $\psi_0$ using Green's function method constructed from the Teukolsky functions, but regularization is needed as the convolution integral is again divergent \cite{Srivastava:2021uku}.}

\subsection{An alternative method for quasinormal mode determination \label{subsec:QNMusingGSN}}
The reformulation of a Schrödinger-like equation into a Riccati equation introduced in Sec.~\ref{subsubsec:complexphasefunc} is not new and had actually been used previously, for instance, in the seminal work by Chandrasekhar and Detweiler on \glspl{QNM} of Schwarzschild \glspl{BH} \cite{doi:10.1098/rspa.1975.0112}. It was used [cf. Eq.~(5) of Ref.~\cite{doi:10.1098/rspa.1975.0112}\,] to alleviate the numerical instability associated with directly integrating the Zerilli equation and, equivalently, also the Regge-Wheeler equation to which the \gls{GSN} equation reduces in the nonspinning limit. Therefore, it is reasonable to expect that the reformulation to be useful for determining \gls{QNM} frequencies and their associated radial solutions.

Recall that a \gls{QNM} solution is both purely ingoing at the horizon and purely outgoing at infinity. In terms of the asymptotic amplitudes of the corresponding Teukolsky function [cf. Eqs.~\eqref{eq:Rin} and~\eqref{eq:Rup}] at a particular frequency $\omega_{\mathrm{QNM}}$, we have
\begin{equation}
\begin{aligned}
	& B^{\mathrm{inc}}_{\mathrm{T}}(\omega_{\mathrm{QNM}}) = C^{\mathrm{inc}}_{\mathrm{T}}(\omega_{\mathrm{QNM}}) = 0 \\
	& \Rightarrow \mathcal{W}_{R}(\omega_{\mathrm{QNM}}) = 0,
\end{aligned}
\end{equation}
where the second line uses Eq.~\eqref{eq:WR_at_inf}. This means that searching for \gls{QNM} frequencies is the same as searching for zeros of $\mathcal{W}_{R}$, the scaled Wronskian for Teukolsky functions. Also recall that, in Appendix~\ref{app:WR_WX_identity}, we proved that the scaled Wronskian for Teukolsky functions $\mathcal{W}_{R}$ and that for the corresponding \gls{GSN} functions $\mathcal{W}_{X}$ are the same, implying that the \gls{QNM} spectra for Teukolsky functions coincide with the \gls{QNM} spectra for \gls{GSN} functions.\footnote{The two equations, the radial Teukolsky equation and the \gls{GSN} equation, are therefore said to be isospectral.} Thus, we can use the \gls{GSN} equation, which has a short-ranged potential, instead of the Teukolsky equation for determining the \gls{QNM} frequencies and the corresponding excitation factors (after applying the conversion factors shown in Appendix~\ref{app:explicitGSN}).

Indeed, Glampedakis and Andersson proposed methods to calculate \gls{QNM} frequencies and excitation factors given a short-ranged potential \cite{Glampedakis:2003dn}, alternative to the Leaver's method \cite{Leaver:1985ax}. They demonstrated their methods by computing a few of the \gls{QNM} frequencies for scalar perturbations ($s=0$) and gravitational perturbations ($|s|=2$), as well as the \gls{QNM} excitation factors for scalar perturbations of Kerr \glspl{BH}. Together with the \gls{GSN} transformations and the asymptotic solutions from this paper, it is straightforward to compute the \gls{QNM} frequencies and their excitation factors for scalar, electromagnetic, and gravitational perturbations using the \gls{GSN} formalism.\footnote{The excitation factors for gravitational perturbations of Kerr \glspl{BH} have been calculated using a different method \cite{Berti:2006wq, Zhang:2013ksa, Oshita:2021iyn}, by explicitly computing the gravitational waveform from an infalling test particle and then extracting the amplitudes for each of the excited \glspl{QNM}.} We leave this for future work.

\begin{acknowledgments}
The author would like to thank Yanbei Chen, Manu Srivastava, Shuo Xin, Emanuele Berti, Scott Hughes, Aaron Johnson, Jonathan Thompson and Alan Weinstein for the valuable discussions and insights when preparing this work. The author would like to especially thank Manu Srivastava for the read of an early draft of this manuscript and Shuo Xin for performing the scaled Wronskian calculations using the \texttt{Fortran} code in Refs.~\cite{Han:2010tp, Han:2011qz}.
R.~K.~L.~L. acknowledges support from the National Science Foundation Awards No.~PHY-1912594 and No.~PHY-2207758.
R.~K.~L.~L. also acknowledges support from the research grant no.~VIL37766 and no.~VIL53101 by the Villum Fonden, the DNRF Chair program grant no.~DNRF162 by the Danish National Research Foundation, the European Union's Horizon 2020 research and the innovation programme under the Marie Sklodowska-Curie grant agreement No.~101131233.
\end{acknowledgments}

\section*{Data Availability}
\textsc{Mathematica} notebooks deriving and storing all the expressions shown here are publicly available from the Zenodo
repository \cite{10.5281/zenodo.8080242}.

\appendix

\section{Angular Teukolsky equation \label{app:swsh}}
After performing the separation of variables to the Teukolsky equation in Eq.~\eqref{eq:fullTeukolskyeqn} using an ansatz of the form
$\psi(t,r,\theta, \phi) = R(r) S(\theta, \phi)e^{-i\omega t}$,
the equation is separated into two parts: the angular part and the radial part. In this appendix, we focus only on solving the angular part (aptly named the angular Teukolsky equation) numerically, and the radial part is treated in the main text.

Let us define ${}_{s}S_{\ell m \omega}(\theta, \phi) \equiv {}_{s}S_{\ell m}(x \equiv \cos \theta; c \equiv a\omega) e^{im\phi}$, where the integer $m$ labels the (trivial) eigenfunctions that satisfy the azimuthal symmetry. The angular Teukolsky equation then reads
\begin{multline}
\label{eq:SWSH_eqn}
	\dfrac{d}{dx} \left[ (1 - x^2) \dfrac{d}{dx} {}_s S_{\ell m}(x;c)  \right] + \\
	\left[ (cx)^2 - 2csx + s + {}_{s} \mathcal{A}_{\ell m}(c) - \frac{(m+sx)^2}{1-x^2} \right] {}_s S_{\ell m}(x;c) = 0,
\end{multline}
where ${}_{s}\mathcal{A}_{\ell m}$ is the angular separation constant and it is related to $\lambda$ [cf. Eq.~\eqref{eq:VT}] by
\begin{equation}
\label{eq:lambda_const}
	\lambda = {}_{s}\mathcal{A}_{\ell m} + c^2 - 2mc.
\end{equation}
The angular Teukolsky equation is solved under the boundary conditions that the solutions at $x = \pm 1$ (or, equivalently, at $\theta = 0, \pi$) are finite, and the solutions are also known as the spin-weighted spheroidal harmonics, denoted by ${}_{s}S_{\ell m \omega}(\theta, \phi)$.

There are multiple methods for solving the angular Teukolsky equation numerically, such as Leaver's continued fraction method \cite{Leaver:1985ax}. A spectral decomposition method for solving the angular Teukolsky equation can be formulated \cite{Hughes:1999bq, Cook:2014cta} by writing a spin-weighted spheroidal harmonic ${}_{s} S_{\ell m \omega}(\theta, \phi)$ as a sum of spin-weighted spherical harmonics ${}_{s} Y_{\ell m}(\theta, \phi)$. The details for such a formulation can be found in, for example, Refs.~\cite{Hughes:1999bq} and~\cite{Cook:2014cta}. We briefly summarize the method here, mostly following and using the notations in Ref.~\cite{Cook:2014cta}, for the sake of completeness.

\subsection{Spectral decomposition method}
A spin-weighted spheroidal harmonic ${}_{s}S_{\ell m} (x; c)$ is expanded using spin-weighted spherical harmonics ${}_{s}Y_{\ell m}(\theta)$, or, equivalently, ${}_{s} S_{\ell m} (x; 0)$ as \cite{Cook:2014cta}
\begin{equation}
\label{eq:SWSH_spectral}
\begin{aligned}
	{}_{s} S_{\ell m} (x; c) & = \sum_{\ell' = \ell_{\mathrm{min}}}^{\infty} {}_{s}C_{\ell' \ell m}(c) \; {}_{s} S_{\ell' m} (x; 0) \\
	& = \left( \vec{C}_{\ell} \right)^{T} \vec{S}_{\ell},
\end{aligned}
\end{equation}
where $\ell_{\mathrm{min}} = \max(|m|, |s|)$ and ${}_{s} C_{\ell' \ell m}(c)$ is the expansion coefficient of the $\ell$th spheroidal harmonic with the $\ell'$th spherical harmonic (of the same value of $s$ and $m$ and we drop them in the subscripts hereafter), as a function of $c \equiv a\omega$.
Equivalently, we can define two column vectors $\vec{C}_{\ell}$ and $\vec{S}_{\ell}$, where the rows are labeled by the index $\ell'$. For example, the first row of the vectors (of index $\ell' = \ell_{\mathrm{min}}$) are ${}_{s}C_{\ell_{\mathrm{min}} \ell m}$ and ${}_{s} S_{\ell_{\mathrm{min}} m}(x;0)$, respectively. The index for the rows goes up to $\ell' = \ell_{\mathrm{max}} \to \infty$, and the vectors have a size of $\ell_{\mathrm{max}} - \ell_{\mathrm{min}} + 1$. Then the spin-weighted spheroidal harmonic $S_{\ell} (x; c)$ is the dot product of the two vectors.

Substituting Eq.~\eqref{eq:SWSH_spectral} into Eq.~\eqref{eq:SWSH_eqn}, we get an eigenvalue equation \cite{Cook:2014cta}
\begin{equation}
\label{eq:SWSH_matrix_equation}
	\mathbb{M} \vec{C}_{\ell} = \mathcal{A}_{\ell} \vec{C}_{\ell},
\end{equation}
where $\mathbb{M}$ is a $\left( \ell_{\mathrm{max}} - \ell_{\mathrm{min}} + 1 \right) \times \left( \ell_{\mathrm{max}} - \ell_{\mathrm{min}} + 1 \right)$ matrix, and recall that $\mathcal{A}_{\ell} \equiv {}_{s} \mathcal{A}_{\ell m}(c \equiv a\omega) $ is the angular separation constant (after writing back all the subscripts).
The matrix elements $\mathbb{M}_{\ell \ell'}$ are given by \cite{Cook:2014cta}
\begin{equation}
	\mathbb{M}_{\ell \ell'} = \begin{cases}
 		-c^2 \mathbb{A}_{\ell' m} & \mathrm{if}\;\; \ell' = \ell - 2, \\
 		-c^2 \mathbb{D}_{\ell' m} + 2cs\mathbb{F}_{\ell' m} & \mathrm{if}\;\; \ell' = \ell - 1, \\
 		\mathcal{A}_{\ell'}(0) - c^2 \mathbb{B}_{\ell' m} + 2cs\mathbb{H}_{\ell' m} & \mathrm{if}\;\; \ell' = \ell, \\
		-c^2 \mathbb{E}_{\ell' m} + 2cs\mathbb{G}_{\ell' m} & \mathrm{if}\;\; \ell' = \ell + 1, \\
		-c^2 \mathbb{C}_{\ell' m} & \mathrm{if}\;\; \ell' = \ell + 2, \\
		0 & \mathrm{otherwise,} \\
 \end{cases}
\end{equation}
where
\begin{subequations}
\begin{eqnarray}
	\mathbb{A}_{\ell m} & = & \mathbb{F}_{\ell m} \mathbb{F}_{(\ell + 1)m}, \\
	\mathbb{B}_{\ell m} & = & \mathbb{F}_{\ell m} \mathbb{G}_{(\ell + 1)m} + \mathbb{G}_{\ell m}\mathbb{F}_{(\ell - 1)m} + \mathbb{H}_{\ell m}^2, \\
	\mathbb{C}_{\ell m} & = & \mathbb{G}_{\ell m} \mathbb{G}_{(\ell - 1)m}, \\
	\mathbb{D}_{\ell m} & = & \mathbb{F}_{\ell m} \mathbb{H}_{(\ell + 1)m} + \mathbb{F}_{\ell m} \mathbb{H}_{\ell m}, \\
	\mathbb{E}_{\ell m} & = & \mathbb{G}_{\ell m} \mathbb{H}_{(\ell - 1)m} + \mathbb{G}_{\ell m} \mathbb{H}_{\ell m}, \\
	\mathbb{F}_{\ell m} & = & \sqrt{\dfrac{(\ell + 1)^2 - m^2}{(2\ell + 3)(2\ell + 1)}\dfrac{(\ell + 1)^2 - s^2}{(\ell + 1)^2}}, \\
	\mathbb{G}_{\ell m} & = & \begin{cases}
	\sqrt{\dfrac{\ell^2 - m^2}{4\ell^2 - 1} \dfrac{\ell^2 - s^2}{\ell^2}} & \mathrm{if}\;\; \ell \neq 0 \\
	0 & \mathrm{if}\;\; \ell = 0
	\end{cases},\\
	\mathbb{H}_{\ell m} & = & \begin{cases}
	-\dfrac{ms}{\ell(\ell+1)} & \mathrm{if}\;\; \ell \neq 0 \;\; \mathrm{and} \;\; s \neq 0 \\
	0 & \mathrm{if}\;\; \ell = 0 \;\; \mathrm{or} \;\; s=0
	\end{cases}, \\
	\mathcal{A}_{\ell}(0) & = & \ell(\ell + 1) - s(s+1) \label{eq:angular_sep_const_nonspinning}.
\end{eqnarray}
\end{subequations}

Solving the angular Teukolsky equation now amounts to solving the eigenvalue problem in Eq.~\eqref{eq:SWSH_matrix_equation} for the eigenvalue $\mathcal{A}_{\ell}$ and the eigenvector $\vec{C}_{\ell}$. The spin-weighted spheroidal harmonic can then be constructed using the eigenvector $\vec{C}_{\ell}$ and the corresponding spin-weight spherical harmonics with Eq.~\eqref{eq:SWSH_spectral}. In practice, we cannot solve a matrix eigenvalue problem of infinite size and we truncate the column vector $\vec{C}_{\ell}$ to have a finite value of $\ell_{\mathrm{max}}$. The accuracy of the numerical eigenvalue and eigenvector solution depends on the size of the truncated matrix. 

\texttt{SpinWeightedSpheroidalHarmonics.jl}\footnote{\url{https://github.com/ricokaloklo/SpinWeightedSpheroidalHarmonics.jl}} is our open-source implementation of the above-mentioned spectral decomposition method for solving spin-weighted spheroidal harmonics in \texttt{julia}. The code solves the truncated version of Eq.~\eqref{eq:SWSH_matrix_equation} to obtain the angular separation constant ${}_{s} \mathcal{A}_{\ell m}$ and the eigenvector ${}_{s} \vec{C}_{\ell m}$. Apart from the angular separation constant, the code can also compute the separation constant $\lambda$ [cf. Eq.~\eqref{eq:VT}], and evaluate numerical values of spin-weight spheroidal harmonics and their derivatives.\footnote{It should be noted that our code is also capable of handling complex $\omega$, which is necessary for carrying out quasinormal-mode related computations.} In particular, the code adopts the normalization convention for ${}_{s}S_{\ell m \omega}(\theta, \phi)$ that
\begin{equation}
\label{eq:SWSH_normalization}
	\int_0^{\pi} \left[ _{s} S_{\ell m}(\theta; c) \right]^{2} \sin(\theta) \, d\theta = \frac{1}{2\pi}.
\end{equation}
To evaluate numerical values of the spin-weighted spheroidal harmonics ${}_{s} S_{\ell m \omega}(\theta, \phi)$ and their derivatives, it is necessary to also be able to numerically (and possibly efficiently) evaluate the spin-weighted spheroidal harmonics ${}_{s} Y_{\ell m}(\theta, \phi)$.

\subsection{Evaluation of ${}_{s}S_{\ell m \omega}(\theta, \phi)$}
Recall from Eq.~\eqref{eq:SWSH_spectral} that the spin-weighted spheroidal harmonic ${}_{s}S_{\ell m \omega}(\theta, \phi)$ is expanded in terms of the spin-weighted spherical harmonics, i.e.,
\[
	{}_{s} S_{\ell m} (\theta, \phi; a\omega) = \sum_{\ell' = \ell_{\mathrm{min}}}^{\infty} {}_{s}C_{\ell' \ell m}(a\omega) \; {}_{s} Y_{\ell m}(\theta, \phi),
\]
and the spectral decomposition method solves for the expansion coefficients ${}_{s}C_{\ell' \ell m}(a\omega)$, which is only part of the ingredients. It is possible to evaluate ${}_{s} Y_{\ell m}(\theta, \phi)$ exactly, and the expression  is given by \cite{Goldberg:1966uu}
\begin{equation}
\label{eq:Y_exact}
\begin{aligned}
	{}_{s} Y_{\ell m} (\theta, \phi) & = (-1)^m e^{im\phi} \sqrt{\frac{(\ell+m)!(\ell-m)!(2\ell+1)}{4\pi (\ell+s)!(\ell-s)!}} \\
	& \times \sum_{r=r_{\text{min}}}^{r_{\text{max}}} \left[ \binom{\ell-s}{r} \binom{\ell+s}{r+s-m} (-1)^{\ell-r-s} \right. \\
	&  \times \left. \cos^{2r+s-m}\left(\frac{\theta}{2}\right) \sin^{2\ell-2r-s+m}\left(\frac{\theta}{2}\right) \vphantom{} \right]
\end{aligned},
\end{equation}
where $r_{\text{min}} = \max(0, m-s)$ and $r_{\text{max}} = \min(\ell-s, \ell+m)$.

In principle, obtaining the value of a spin-weighted spherical harmonic ${}_{s} Y_{\ell m}(\theta, \phi)$ is as simple as evaluating the sum as shown in Eq.~\eqref{eq:Y_exact}. However, when the index $\ell$ is big, evaluating the prefactor \[ \sqrt{\frac{(\ell+m)!(\ell-m)!}{(\ell+s)!(\ell-s)!}} \] in Eq.~\eqref{eq:Y_exact} on a machine can cause an overflow error because of the large factorials involved in the computation. Luckily, the expression for the prefactor can be simplified. In fact,
\begin{widetext}
\begin{equation}
\sqrt{\dfrac{(\ell+m)!(\ell-m)!}{(\ell+s)!(\ell-s)!}} = \begin{cases}
\sqrt{\dfrac{\left( \ell - m \right) \left( \ell - m - 1 \right) \dots \left( \ell - m - (s-m) + 1 \right)}{\left( \ell + m + (s-m) \right) \left( \ell + m + (s-m) - 1  \right) \dots \left( \ell + m + 1 \right) }} & \mathrm{if}\;\; s > m, \\
\sqrt{\dfrac{\left( \ell + s + (m-s) \right) \left( \ell + s + (m-s) - 1 \right) \dots \left( \ell + s + 1 \right)}{\left( \ell - s  \right) \left( \ell - s - 1  \right) \dots \left( \ell - s - (m - s) + 1 \right) }} & \mathrm{if}\;\; s < m, \\
 1 & \mathrm{if}\;\; |s| = |m|.
\end{cases}.
\end{equation}
\end{widetext}

\begin{figure}[h!]
\centering
\includegraphics[width=\columnwidth]{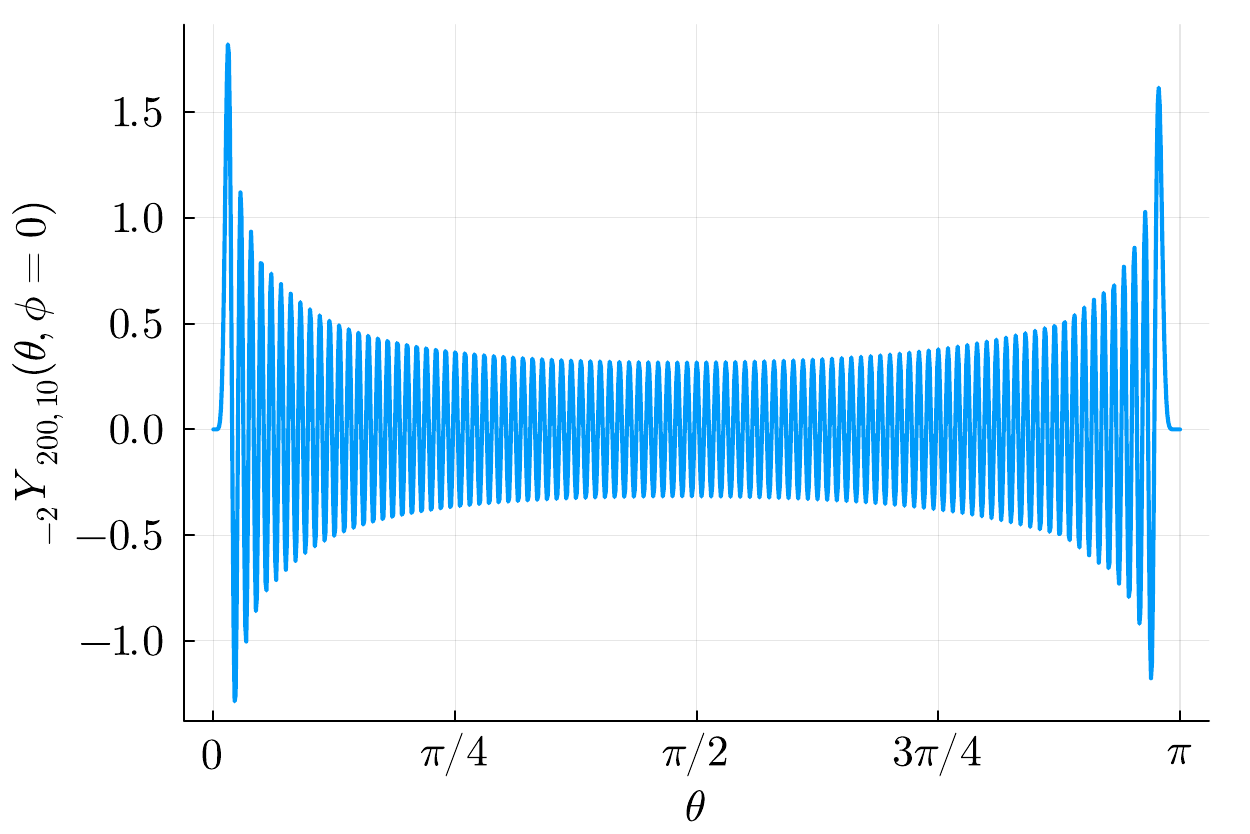}
\caption{\label{fig:Y_high_ell}Spin-weighted spherical harmonic with $s=-2, \ell = 200, m = 10$ as a function of $\theta$, obtained using the Chebyshev pseudospectral method. We see that the numerical solution does not suffer from any numerical instability.}
\end{figure}

Unfortunately, when $\ell \gtrsim 30$, the coefficient for each term in Eq.~\eqref{eq:Y_exact} has a wide dynamic range depending on the value of $r$, while the sum itself is always of the order of unity.\footnote{Recall that spin-weighted spherical harmonics are also normalized like Eq.~\eqref{eq:SWSH_normalization}.} Fortunately, many techniques have been developed to evaluate the sum in a numerically stable manner, such as Refs.~\cite{Gumerov2015, Feng:2015mqa}, in the context of Wigner's $d$ matrices. Here, we instead use the Chebyshev pseudospectral method \cite{boyd2001chebyshev} to solve for the boundary value problem as shown in Eq.~\eqref{eq:SWSH_eqn} with $c = 0$ using Chebyshev polynomials. Figure~\ref{fig:Y_high_ell} shows the spin-weighted spherical harmonic with a high value of $\ell$ solved using the Chebyshev pseudospectral method. We see that the numerical solution does not suffer from any numerical instability.

\subsection{Evaluation of $\partial_{\theta, \phi}^{n} \; {}_{s}S_{\ell m \omega}(\theta, \phi)$}
In order to evaluate partial derivatives of spin-weighted spheroidal harmonics, $\partial_{\theta, \phi}^{n} \; {}_{s}S_{\ell m \omega}(\theta, \phi)$, which are needed for evaluating source terms $T$ of the Teukolsky equation [cf. Eq.~\eqref{eq:fullTeukolskyeqn}], we can use the fact that the expansion coefficients ${}_{s} C_{\ell' \ell m}(c \equiv a\omega)$ in Eq.~\eqref{eq:SWSH_spectral} are independent of $\theta$ and $\phi$. This means that the partial derivatives $\partial_{\theta, \phi}^{n} \; {}_{s}S_{\ell m \omega}(\theta, \phi)$ are given by the sum of the partial derivatives of ${}_{s} Y_{\ell m}(\theta, \phi)$ with the same set of the expansion coefficients, i.e.,
\begin{multline}
	\left( \dfrac{\partial}{\partial \left\{ \theta, \phi \right\}}\right)^{n} \; {}_{s}S_{\ell m \omega}(\theta, \phi) = \\
	\sum_{\ell' = \ell_{\mathrm{min}}}^{\infty} \left[ {}_{s}C_{\ell' \ell m}(a\omega) \; \left( \dfrac{\partial}{\partial \left\{ \theta, \phi \right\}}\right)^{n} {}_{s} Y_{\ell m}(\theta, \phi) \right].
\end{multline}

In principle, we can evaluate the partial derivatives using \gls{AD}. However, the evaluation can be more efficient by noticing that the exact evaluation of the partial derivative with respect to $\phi$ is trivial because of the $e^{im\phi}$ dependence. Each partial differentiation with respect to $\phi$ gives a factor of $im$. As for the partial derivative of a spin-weighted spherical harmonic with respect to $\theta$, the computation scheme is less trivial. Note that each term in Eq.~\eqref{eq:Y_exact} is of the form $c_{r} \cos^{\alpha_r}(\theta/2) \sin^{\beta_r}(\theta/2)$,
where $r$ is the summation index and $c_{r}$ is the prefactor with $\alpha_r$ and $\beta_r$ being the exponent for the $\cos(\theta/2)$ and $\sin(\theta/2)$ factor, respectively. Each partial differentiation with respect to $\theta$ splits the term into two terms, one with $(c_{r}/2) \beta_r \cos^{\alpha_r + 1}(\theta/2) \sin^{\beta_r - 1}(\theta/2)$ and one with $(-c_{r}/2) \alpha_r \cos^{\alpha_r - 1}(\theta/2) \sin^{\beta_r + 1}(\theta/2)$.

\begin{figure*}[ht]
\includegraphics[width=\textwidth]{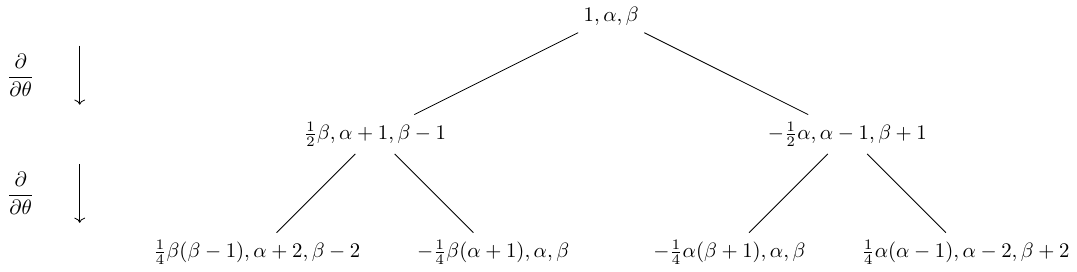}
\caption{Binary tree representation of a term and its partial derivatives with respect to $\theta$ in the summation of Eq.~\eqref{eq:Y_exact}. In each node, the three numbers correspond to the prefactor, the exponent for the $\cos(\theta/2)$ and the $\sin(\theta/2)$ factor respectively. A partial differentiation with respect to $\theta$ creates two leaf nodes with the prefactor and the exponents computed according to rules of partial differentiation. The $n$th partial derivative with respect to $\theta$ of the term in the root node can be evaluated by simply summing over all the nodes of depth $n$. \label{fig:binary_tree}}
\end{figure*}

We can keep track of the coefficients and the exponents for the cosine and the sine factor with the help of a binary tree. We represent each term in the summation with index $r$ in Eq.~\eqref{eq:Y_exact} as the root node of a tree (for an illustration, see Fig.~\ref{fig:binary_tree}) with an entry of three numbers $\left(c_{r}, \alpha_r, \beta_r \right)$. Each partial differentiation with respect to $\theta$ corresponds to adding two child nodes with the entry $\left(c_{r}\beta_r/2, \alpha_r+1, \beta_r-1 \right)$ and $\left(-c_{r}\alpha_r/2, \alpha_r-1, \beta_r+1 \right)$, respectively. Therefore, the $n$th order partial derivative of $\theta$ can be evaluated exactly by traversing all the nodes of depth $n$ and then summing over their contributions.

\section{Fast inversion from the tortoise coordinate $r_*$ to the Boyer-Lindquist coordinate $r$ \label{app:rstar_inversion}}
The tortoise coordinate $r_{*}$ (for Kerr \glspl{BH}) is defined by
\begin{equation}
\tag{\ref{eq:drstardr}}
	\dfrac{dr_{*}}{dr} = \dfrac{r^2+a^2}{\Delta} = \dfrac{r^2 + a^2}{\left( r - r_{+} \right) \left( r - r_{-} \right)}.
\end{equation}
Using Eq.~\eqref{eq:drstardr} one can generate different ``tortoise coordinates'' which differ from each other only by an integration constant. Here, and in most of the literature, we choose the integration constant such that
\begin{equation}
\tag{\ref{eq:rstar_from_r}}
    r_{*}(r) = r + \frac{2r_{+}}{r_{+}-r_{-}} \ln \left( \frac{r - r_+}{2} \right) - \frac{2r_{-}}{r_{+}-r_{-}} \ln \left( \frac{r - r_-}{2} \right).
\end{equation}
However, there is no simple analytical expression that gives $r = r(r_{*})$, and one will have to instead numerically invert Eq.~\eqref{eq:rstar_from_r}. Such an inversion scheme that is both fast and accurate is needed for our numerical implementation of the \gls{GSN} formalism because we numerically solve the \gls{GSN} equation in the $r_*$ coordinate instead of the Boyer-Lindquist $r$ coordinate, and yet the \gls{GSN} potentials, which will be evaluated at many different values of $r_{*}$ during the numerical integration, are written in terms of $r$.

This coordinate inversion is equivalent to a root-finding problem. Given a value of the tortoise coordinate $r_{*}^{0}$, we solve for $h^0 \equiv \left( r^0 - r_+ \right) > 0$ that satisfies
\begin{equation}
\label{eq:inversion_as_rootfinding}
    r_{*}^{0} - r_{*}(r_{+} + h^0) = 0,
\end{equation}
in order to find the corresponding Boyer-Lindquist coordinate $r^0 \equiv r_{+} + h^0$ that is outside the horizon \footnote{A similar construction (i.e., enforcing $h^0 < 0$) can be used to find the Boyer-Lindquist coordinate $r \in \left(r_{-}, r_{+}\right)$ that gives the same $r_{*}^0$.}.

\begin{figure}[h]
	\centering
	\includegraphics[width=\columnwidth]{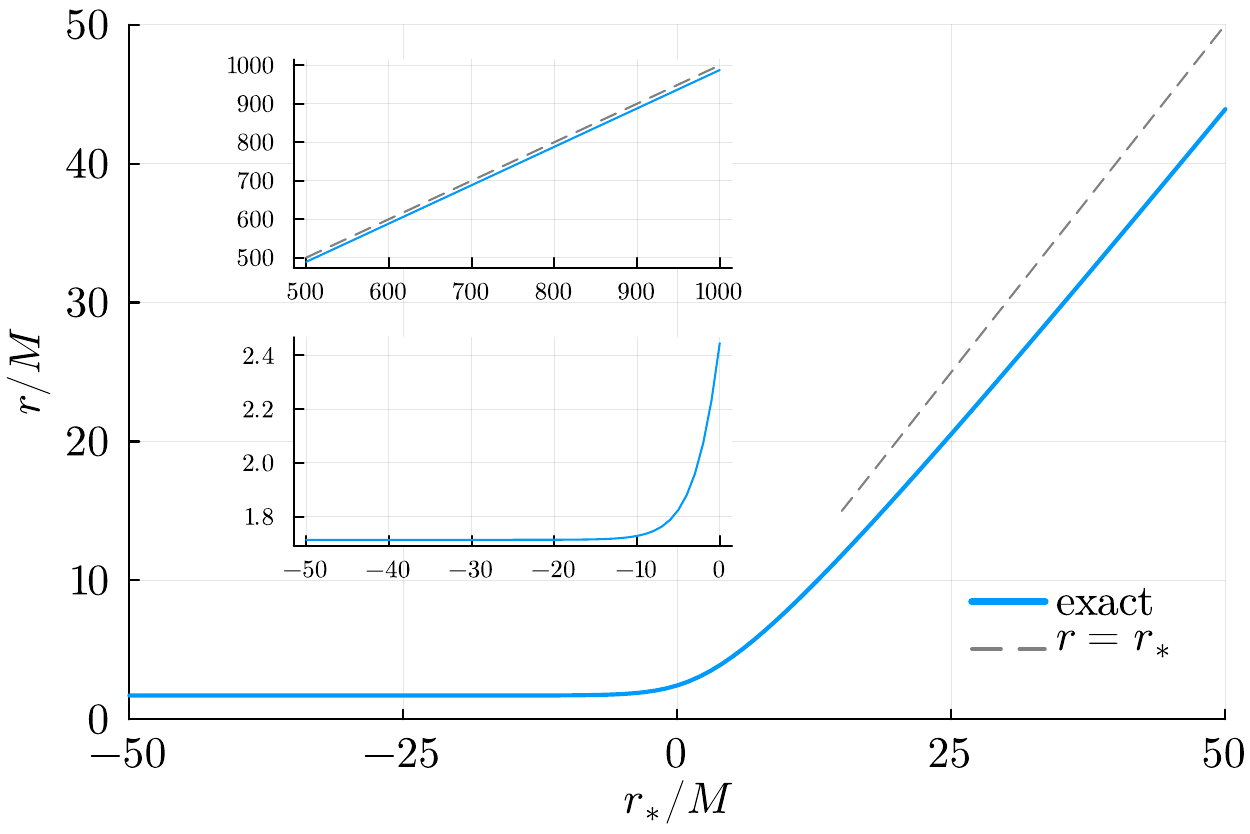}
	\caption{The Boyer-Lindquist $r$ coordinate as a function of the tortoise $r_{*}$ coordinate for $a/M = 0.7$. As the value of $r_{*}$ becomes larger (upper inset), the approximation $r(r_{*}) \approx r_{*}$ (dashed) gets increasingly better as $dr/dr_{*} \to 1$. Meanwhile, as the value of $r_{*}$ becomes more negative (lower inset), $r(r_{*})$ approaches $r = r_{+}$ as constructed and $dr/dr_{*} \to 0$. \label{fig:rrstar}}
\end{figure}

Figure~\ref{fig:rrstar} shows a plot of $r$ as a function of $r_{*}$ for $a/M = 0.7$. As the value of $r_{*}$ becomes larger, the simple approximation $r(r_{*}) \approx r_{*}$ works better. In fact, the slope $dr/dr_{*} \to 1$ as $r_{*} \gg 0$. Therefore, derivative-based methods such as the Newton-Raphson method and secant methods \cite{10.5555/1403886} are efficient in performing the coordinate inversion (since we can evaluate the derivatives exactly and cheaply). However, these methods are going to be inefficient for negative values of $r_*$ near the horizon since the slope tends to zero.

\begin{figure}[h]
	\centering
	\includegraphics[width=\columnwidth]{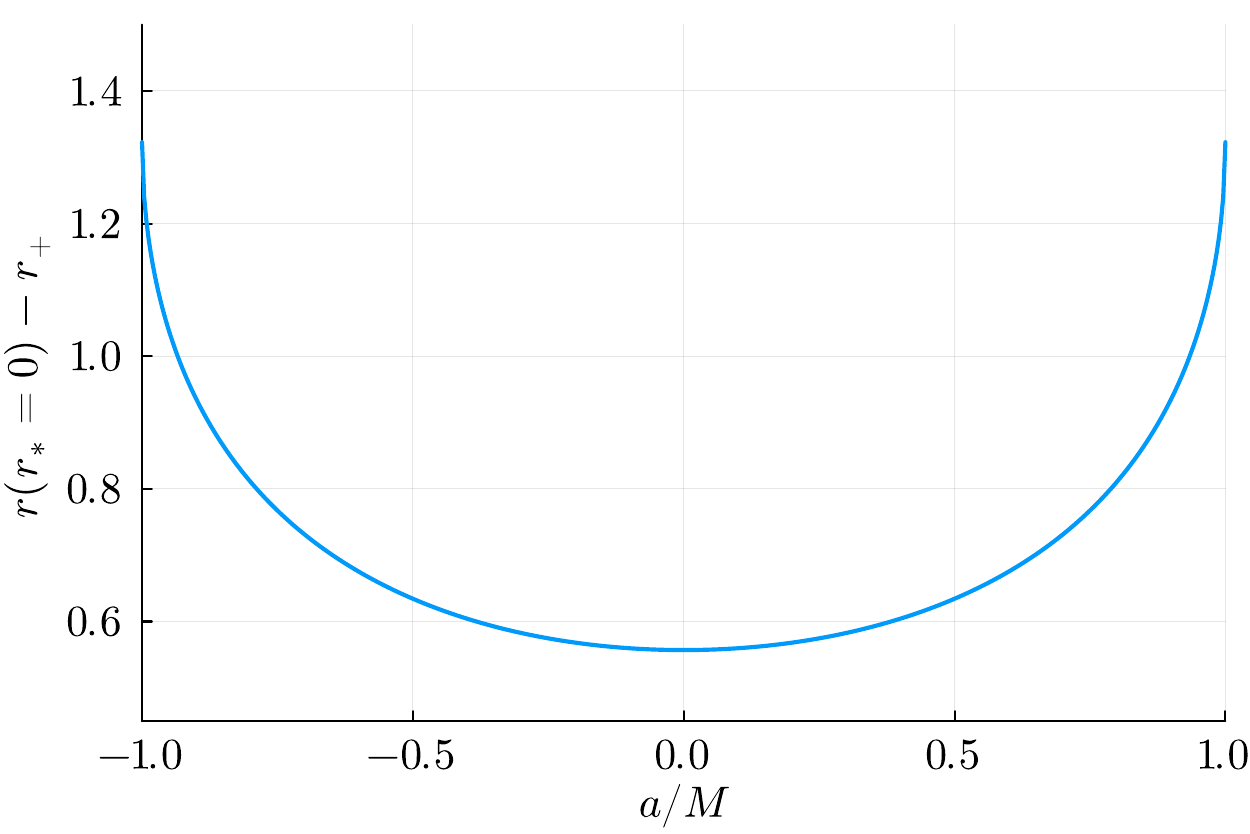}
	\caption{The difference between $r_{*} = 0$ and the horizon in the Boyer-Lindquist $r$ coordinate, $r(r_{*} = 0) - r_{+}$, as a function of the spin $a$ of the \gls{BH}. We see that the difference is monotonically increasing with $|a|$, and it is the smallest when $a = 0$ and the largest ($\approx 1.3$) when $|a| \to 1$. We can use this to construct an interval of $r$ that must contain $r = r(r_{*})$ for $r_{*} \leq 0$ when using the bisection method. \label{fig:h_upperbound}}
\end{figure}

In our numerical implementation, we use a hybrid of root-finding algorithms. For $r_{*}^{0} > 0$, we use the Newton-Raphson method \cite{10.5555/1403886} with an initial guess of $h = r_{*}^{0}$ and switch to using the bisection method \cite{10.5555/1403886} for $r_{*}^{0} \leq 0$. To use the bisection method, an interval of $h$ that contains the root of Eq.~\eqref{eq:inversion_as_rootfinding} is given to the algorithm as an initial guess. Since $r=r_+$ maps to $r_* \to -\infty$, a natural choice for the lower bound of the bracketing interval would be $h=0$. For the upper bracketing bound, from Fig.~\ref{fig:h_upperbound} we see that the value of $h$ that corresponds to $r_{*} = 0$ is a monotonically increasing function of the spin magnitude $|a|$. Therefore, we can simply choose the upper bound value to be (equal to or greater than) the limiting value of $h$ that corresponds to $r_* = 0$ when $|a| \to 1$. Explicitly, the numerical implementation in \texttt{GeneralizedSasakiNakamura.jl} uses the bracketing interval $0 < h < 1.4$.

\section{Exact solutions of static modes for the Teukolsky equation \label{app:static_modes_for_Teukolsky}}
Here we rederive exact solutions of static ($\omega = 0$) modes for the Teukolsky equation, conforming to the notation and normalization convention used throughout the text.
Note that these solutions are known in the literature (see, for example, Refs.~\cite{Barack:1999st, vandeMeent:2015lxa} for similar derivations with a different notation and normalization convention).

\subsection{Angular Teukolsky equation}
Since the angular Teukolsky equation depends on $\omega$ only through the combination $c \equiv a \omega$, therefore in the static case where $\omega = 0$, the angular separation constant ${}_{s}\mathcal{A}_{\ell m}$ and the solution to the equation ${}_{s}S_{\ell m \omega}(\theta, \phi)$ reduce to that for the corresponding nonspinning ($a = 0$) case. In particular, they are known exactly and can be found in Eq.~\eqref{eq:angular_sep_const_nonspinning} for the angular separation constant and in Eq.~\eqref{eq:Y_exact} for the solution, respectively.

\subsection{Radial Teukolsky equation}
As for the radial part, in the static case where $\omega = 0$, the potential associated with the radial Teukolsky equation $V_{\mathrm{T}}$ becomes
\begin{equation}
	V_{\mathrm{T}}(\omega = 0) = \ell(\ell + 1) - s(s+1) - \dfrac{a^2m^2 + 2i ams(r-1)}{(r-r_+)(r-r_-)}.
\end{equation}
Compared to a generic case where $\omega \neq 0$, the radial Teukolsky equation in the static case has three regular singular points, located at $r = r_{\pm}, \infty$, respectively, instead of having two regular singular points at $r = r_{\pm}$ and an irregular singular point at $r = \infty$ (for a more detailed discussion on the singularity structure of the \gls{ODE} in the $\omega \neq 0$ case, see Appendix~\ref{app:recurrence_relations}).

It is well known that a second-order linear \gls{ODE} with three regular singular points can be transformed into a hypergeometric differential equation \cite{ARFKEN2013329}, which is given by
\begin{equation}
\label{eq:hypergeometric}
	x(1-x) u''(x) + \left[ c - \left(a+b+1\right) x\right]u'(x) - ab u(x)=0,
\end{equation}
where $a,b,c$ are the three parameters characterizing the \gls{ODE} (and should not be confused with the angular momentum per unit mass $a$ and the combination $c \equiv a\omega$ introduced above).

Guided by the fact that the standard hypergeometric differential equation in Eq.~\eqref{eq:hypergeometric} has the three regular singular points at $x=0,1,\infty$, respectively, we first map the singular points for Eq.~\eqref{eq:radialTeukolskyeqn} in the static case by introducing the transformation
\begin{equation}
	x \equiv \dfrac{r_{+} - r}{r_{+} - r_{-}} \equiv \dfrac{r_{+} - r}{2\gamma},	
\end{equation}
where we define $\gamma \equiv \sqrt{1 - a^2}$ and allow $x$ to be a complex number. We see that this transformation maps $r = r_{-}$ to $x = 1$ and $r = r_{+}$ to $x = 0$, respectively, while leaving the singular point at $r = \infty$ remains at $x = \infty$. The static radial Teukolsky equation now reads
\begin{widetext}
\begin{equation}
\label{eq:static_radial_Teukolsky}
	x\left(1-x\right)R''(x) - \left(1+s\right)\left(2x-1\right)R'(x) + \dfrac{m^2 - 2iams\left(2x-1\right)\gamma - \left[m^2 + 4 \left(1-x\right) x \left(s - \ell\right) \left(1 + s + \ell\right)\right] \gamma^2 }{4x\left(1-x\right)\gamma^2}R(x) = 0,
\end{equation}
\end{widetext}
which is close but still not the same as Eq.~\eqref{eq:hypergeometric}. In fact, we need to solve Eq.~\eqref{eq:static_radial_Teukolsky} differently depending on whether or not $a$ or $m$ vanishes.

Once we have successfully mapped the static radial Teukolsky equation into a hypergeometric differential equation, the corresponding static solutions can then be expressed using hypergeometric functions. However, we still need to construct two linearly independent solutions.
In a nonstatic ($\omega \neq 0$) case, we can easily identify two linearly independent solutions at $r \to \infty$ (cf. Sec.~\ref{subsec:asymptotic_behaviors}), which are said to be left- and rightgoing, respectively. We can then construct the set of physically motivated solutions, $\left\{ R^{\mathrm{in}}, R^{\mathrm{up}}\right\}$, from these two solutions.
In a static ($\omega = 0$) case, we can use the fact that the effective wave frequency $p$ [cf. Eq.~\eqref{eq:wavefreqp}] at the horizon is nonvanishing even when $\omega$ vanishes, to uniquely identify $R^{\mathrm{in}}$ as the solution that goes like $\Delta^{-s} e^{-ipr_*}$ when $r \to r_+$. The other solution $R^{\mathrm{up}}$ is then constructed as the solution that is linearly independent of $R^{\mathrm{in}}$. These two solutions should also be rescaled properly to respect the unit ``transmission'' amplitude convention.

\subsubsection{Nonspinning ($a = 0$) or axisymmetric ($m = 0$) case}
When either $a$ or $m$ vanishes, the static radial Teukolsky equation in Eq.~\eqref{eq:static_radial_Teukolsky} reduces trivially to the form of a hypergeometric differential equation in Eq.~\eqref{eq:hypergeometric}. In fact, by comparing the two \glspl{ODE}, we see that choosing the parameters $a,b,c$ for the hypergeometric differential equation as
\begin{equation}
\begin{cases}
	a  = s - \ell \\
	b  = 1 + s + \ell \\
	c  = 1 + s
\end{cases}
\end{equation}
maps the static radial Teukolsky equation into a hypergeometric differential equation.

As $x \to 0$, the two linearly independent solutions to the hypergeometric differential equation are given by
\begin{align}
u_{1}(x) & = {}_{2}F_{1} \left( s - \ell, 1 + s + \ell; 1 + s; x  \right), \\
u_{2}(x) & = x^{-s} {}_{2}F_{1} \left( -\ell, 1 + \ell; 1-s; x  \right),
\end{align}
where ${}_{2}F_{1}\left(a,b;c;x\right)$ is a Gauss hypergeometric function.
On the other hand, $\Delta^{-s} e^{-ipr_*} \approx \left( -4 \gamma^2 x \right)^{-s}$ as $x \to 0$ when $m$ or $a$ vanishes. Therefore, we can construct $R^{\mathrm{in}}(r)$ using $u_2(x)$ as
\begin{equation}
\label{eq:Rin_static_am_0}
	R^{\mathrm{in}}(r(x)) = (-4\gamma^2)^{-s} x^{-s} {}_{2}F_{1} \left( -\ell, 1 + \ell; 1-s; x  \right).
\end{equation}

As $x \to \infty$, the two linearly independent solutions to the hypergeometric differential equation are
\begin{align}
u_{3}(x) & = x^{-s+\ell} {}_{2}F_{1} \left( s - \ell, -\ell; -2\ell; 1/x \right), \\
u_{4}(x) & = x^{-s-\ell-1} {}_{2}F_{1} \left( 1 + \ell, 1 + s + \ell; 2 + 2\ell; 1/x \right).
\end{align}
One can verify by an explicit calculation of the Wronskian (cf. Appendix~\ref{app:WR_WX_identity}) that $u_{4}$ is the solution that is linearly independent of $u_{2}$. Note that
\begin{equation}
\dfrac{1}{r} = -\dfrac{1}{2\gamma} \left( \dfrac{1}{x} \right) + \dots,
\end{equation}
when expanding $1/r$ in powers of $1/x$.
Therefore, we can construct the properly normalized $R^{\mathrm{up}}(r)$ using $u_4(x)$ as
\begin{multline}
\label{eq:Rup_static_am_0}
	R^{\mathrm{up}}(r(x)) = \\ (-2\gamma)^{-s-\ell-1} x^{-s-\ell-1} {}_{2}F_{1} \left( 1 + \ell, 1 + s + \ell; 2 + 2\ell; 1/x \right).
\end{multline}

\subsubsection{Generic ($ma \neq 0$) case}
When $ma$ is nonzero, a further substitution on $R(x)$ is needed. In particular, we use the following substitution
\begin{equation}
\label{eq:static_mode_sub}
	R(x) = x^{\rho} (1-x)^{\sigma} u(x),
\end{equation}
where we see that the singularity structure of the \gls{ODE} that $u(x)$ satisfies remains unaltered by this substitution.

The goal here is to find the appropriate values of $\rho, \sigma$ such that $u(x)$ satisfies the hypergeometric differential equation.
Substituting Eq.~\eqref{eq:static_mode_sub} into Eq.~\eqref{eq:static_radial_Teukolsky}, one will find that choosing
\begin{equation}
\begin{cases}
	\rho = -s + \kappa/2 \\
	\sigma = -s - \kappa/2
\end{cases},
\end{equation}
where $\kappa \equiv iam/\gamma$, gives the desired result. In fact, $u(x)$ now satisfies the \gls{ODE}
\begin{multline}
	x\left(1-x\right)u'' + \left[ 1 - s + \kappa - \left(2-2s\right)x  \right]u' \\ - \left(-1 + s -\ell \right) \left( s + \ell \right) u = 0,
\end{multline}
which is just the hypergeometric differential equation with the parameters $a, b, c$ chosen as
\begin{equation}
\begin{cases}
	a = - s - \ell \\
	b = 1 - s + \ell \\
	c = 1 - s + \kappa
\end{cases}.
\end{equation}

Just as before, the two linearly independent solutions to the hypergeometric differential equation when $x \to 0$ are given by
\begin{align}
u_{1}(x) & = {}_{2}F_{1} \left( - s - \ell, 1 - s + \ell; 1 - s + \kappa; x  \right), \\
u_{2}(x) & = x^{s - \kappa} \\
& \; {}_{2}F_{1} \left( -\ell - \kappa, 1 + \ell - \kappa; 1 + s -\kappa; x \right).
\end{align}
Meanwhile, when $x \to 0$,
\begin{equation}
	\Delta^{-s} e^{-ipr_*} \approx \left( -4 \gamma^2 x \right)^{-s} e^{\frac{iam}{2}} \gamma^{\frac{\kappa}{2}\frac{\gamma - 1}{\gamma + 1}}
 \left( -\gamma x\right)^{\kappa/2}.
 \end{equation}
Therefore, we see that here we should choose $u_{1}$ to construct $R^{\mathrm{in}}(r)$ as
\begin{multline}
\label{eq:Rin_static_generic}
	R^{\mathrm{in}}(r(x)) = \\ \left(-1\right)^{-\kappa/2} e^{\frac{iam}{2}} \gamma^{\frac{iam}{r_{+}}} \left( - 4 \gamma^2 \right)^{-s} x^{-s + \kappa/2} \left( 1 - x\right)^{\kappa/2} \\ {}_{2}F_{1} \left( -\ell + \kappa, 1 + \ell + \kappa; 1 - s + \kappa; x  \right),
\end{multline}
where we have used the identity for Gauss hypergeometric functions that
\begin{equation}
\label{eq:Gauss_2F1_Euler_identity}
	{}_{2}F_{1}\left(a,b;c;x\right) \equiv \left(1-x\right)^{c-a-b}{}_{2}F_{1}\left(c-a,c-b;c;x\right)
\end{equation}
to make the fact that Eq.~\eqref{eq:Rin_static_generic} reduces to Eq.~\eqref{eq:Rin_static_am_0} after putting $\kappa \equiv iam/\gamma = 0$ obvious.

Again, we construct $R^{\mathrm{up}}$ as the linearly independent solution of $R^{\mathrm{in}}$.
When $x \to \infty$, the two desired solutions to the hypergeometric differential equation are
\begin{align}
u_{3}(x) & = x^{s+\ell} {}_{2}F_{1} \left( - s - \ell, -\ell - \kappa; -2\ell; 1/x  \right), \\
u_{4}(x) & = x^{-1+s-\ell} {}_{2}F_{1} \left( 1-s+\ell,1+\ell-\kappa;2+2\ell;1/x\right).
\end{align}
By an explicit calculation, we see that this time $u_{4}$ is the linearly independent solution of $u_{1}$. Hence, we construct $R^{\mathrm{up}}(r)$ as
\begin{multline}
\label{eq:Rup_static_generic}
R^{\mathrm{up}}(r(x)) = \\ \left(-2\gamma\right)^{-s-\ell-1} x^{-s-\ell-1-\kappa/2} \left(x-1\right)^{\kappa/2} \\{}_{2}F_{1} \left( 1 + \ell + \kappa, 1 + s + \ell; 2 + 2\ell; 1/x \right),
\end{multline}
where we have again used the identity in Eq.~\eqref{eq:Gauss_2F1_Euler_identity} to make the fact that Eq.~\eqref{eq:Rup_static_generic} reduces to Eq.~\eqref{eq:Rup_static_am_0} after putting $\kappa \equiv iam/\gamma = 0$ obvious.

\section{Alternative pairs of linearly independent solutions of the radial Teukolsky and the generalized Sasaki-Nakamura equation\label{app:out-down}}
In the main text, the pairs of linearly independent solutions $\left\{R^{\mathrm{in}}, R^{\mathrm{up}}\right\}$ and $\left\{X^{\mathrm{in}}, X^{\mathrm{up}}\right\}$ are used for the radial Teukolsky and the \gls{GSN} equation, respectively.
These solutions behave like waves traveling ``to the left'' near the horizon and ``to the right'' near infinity, respectively, when $\omega \neq 0$.
Here we discuss an alternative pair of solutions, namely, $\left\{R^{\mathrm{out}}, R^{\mathrm{down}}\right\}$ and $\left\{X^{\mathrm{out}}, X^{\mathrm{down}}\right\}$ that behave instead like waves traveling to the right near the horizon and to the left near infinity, respectively.

Starting with the Teukolsky functions $R^{\mathrm{out, down}}$, we can construct these functions from linear combinations of $R^{\mathrm{in, up}}$ as
\begin{align}
	R^{\mathrm{out}} &= \dfrac{1}{C^{\mathrm{inc}}_{\mathrm{T}}} \left( R^{\mathrm{up}} - \dfrac{C^{\mathrm{ref}}_{\mathrm{T}}}{B^{\mathrm{trans}}_{\mathrm{T}}} R^{\mathrm{in}} \right), \label{eq:Rout}\\
	R^{\mathrm{down}} &= \dfrac{1}{B^{\mathrm{inc}}_{\mathrm{T}}} \left( R^{\mathrm{in}} - \dfrac{B^{\mathrm{ref}}_{\mathrm{T}}}{C^{\mathrm{trans}}_{\mathrm{T}}} R^{\mathrm{up}} \right).\label{eq:Rdown}
\end{align}
We can easily verify that they indeed satisfy the desired asymptotic behaviors. The same construction can also be applied to the \gls{GSN} functions as
\begin{align}
	X^{\mathrm{out}} &= \dfrac{1}{C^{\mathrm{inc}}_{\mathrm{SN}}} \left( X^{\mathrm{up}} - \dfrac{C^{\mathrm{ref}}_{\mathrm{SN}}}{B^{\mathrm{trans}}_{\mathrm{SN}}} X^{\mathrm{in}} \right),\label{eq:Xout} \\
	X^{\mathrm{down}} &= \dfrac{1}{B^{\mathrm{inc}}_{\mathrm{SN}}} \left( X^{\mathrm{in}} - \dfrac{B^{\mathrm{ref}}_{\mathrm{SN}}}{C^{\mathrm{trans}}_{\mathrm{SN}}} X^{\mathrm{up}} \right).\label{eq:Xdown}
\end{align}
These solutions comply with their respective ``unity transmission amplitude'' normalization convention. For instance, $X^{\mathrm{out}}(r_* \to -\infty) \sim e^{ipr_*}$ is simply a unit outgoing wave near the horizon, and $X^{\mathrm{down}}(r_* \to \infty) \sim e^{-i\omega r_*}$ is a unit ingoing wave near infinity.

Similarly, we can define some ``incidence'' and ``reflection'' amplitudes following Ref.~\cite{Pound2020}. With $X^{\mathrm{out}}$, for example, by substituting Eqs.~\eqref{eq:Xin} and \eqref{eq:Xup} when $r_* \to \infty$ into Eq.~\eqref{eq:Xout}, we see that
\begin{multline}
	X^{\mathrm{out}}(r_* \to \infty) = \underbrace{\left( \dfrac{C^{\mathrm{trans}}_{\mathrm{SN}}}{C^{\mathrm{inc}}_{\mathrm{SN}}} - \dfrac{B^{\mathrm{ref}}_{\mathrm{SN}}C^{\mathrm{ref}}_{\mathrm{SN}}}{B^{\mathrm{trans}}_{\mathrm{SN}}C^{\mathrm{inc}}_{\mathrm{SN}}} \right)}_{\mathrm{incidence\,amplitude}} e^{i\omega r_*} \\
	\underbrace{- \dfrac{B^{\mathrm{inc}}_{\mathrm{SN}}C^{\mathrm{ref}}_{\mathrm{SN}}}{B^{\mathrm{trans}}_{\mathrm{SN}}C^{\mathrm{inc}}_{\mathrm{SN}}}}_{\mathrm{reflection\,amplitude}} e^{-i\omega r_*}.
\end{multline}
As for $X^{\mathrm{down}}$, by substituting Eqs.~\eqref{eq:Xin} and \eqref{eq:Xup} when $r_* \to -\infty$ into Eq.~\eqref{eq:Xdown}, we have
\begin{multline}
	X^{\mathrm{down}}(r_* \to -\infty) = \underbrace{\left( \dfrac{B^{\mathrm{trans}}_{\mathrm{SN}}}{B^{\mathrm{inc}}_{\mathrm{SN}}} - \dfrac{B^{\mathrm{ref}}_{\mathrm{SN}}C^{\mathrm{ref}}_{\mathrm{SN}}}{B^{\mathrm{inc}}_{\mathrm{SN}}C^{\mathrm{trans}}_{\mathrm{SN}}} \right)}_{\mathrm{incidence\,amplitude}} e^{-i p r_*} \\
	\underbrace{- \dfrac{B^{\mathrm{ref}}_{\mathrm{SN}}C^{\mathrm{inc}}_{\mathrm{SN}}}{B^{\mathrm{inc}}_{\mathrm{SN}}C^{\mathrm{trans}}_{\mathrm{SN}}}}_{\mathrm{reflection\,amplitude}} e^{ip r_*}.
\end{multline}
The same definition applies to the Teukolsky functions $R^{\mathrm{out,down}}$ as well.

\section{Deriving the identity between the scaled Wronskians for Teukolsky functions and generalized Sasaki-Nakamura functions \label{app:WR_WX_identity}}
Recall that the scaled Wronskian $\mathcal{W}_{R}$ for the Teukolsky functions $R^{\mathrm{in, up}}$ is defined by
\begin{equation}
\tag{\ref{eq:WR_def}}
	\mathcal{W}_{R} = \Delta^{s+1} \left( R^{\mathrm{in}} {R^{\mathrm{up }}}^{\prime} - R^{\mathrm{up }}{R^{\mathrm{in}}}^{\prime}  \right),
\end{equation}
whereas the scaled Wronskian $\mathcal{W}_{X}$ for the \gls{GSN} functions $X^{\mathrm{in, up}}$ is defined by
\begin{equation}
\tag{\ref{eq:WX_def}}
	\mathcal{W}_{X} = \dfrac{1}{\eta} \left[ X^{\mathrm{in}} (dX^{\mathrm{up}}/dr_{*}) - (dX^{\mathrm{in}}/dr_{*}) X^{\mathrm{up}} \right].
\end{equation}
They are called scaled Wronskians because they are not the same as ``ordinary'' Wronskians. For a generic second-order linear \gls{ODE}
\begin{equation}
\label{eq:generic_2nd_ODE}
	\dfrac{d^2 y(x)}{dx^2} + p(x) \dfrac{dy(x)}{dx} + q(x) y(x) = 0,
\end{equation}
suppose it admits two linearly independent solutions $y_{1}(x)$ and $y_{2}(x)$, then the Wronskian $W(x)$ is defined by
\begin{equation}
\label{eq:W_def}
	W(x) = y_{1}\dfrac{dy_2}{dx} - y_{2}\dfrac{dy_1}{dx},
\end{equation}
which is a function of $x$ in general. It can be shown that $W(x)$ satisfies the \gls{ODE} \cite{ARFKEN2013329}
\begin{equation}
\label{eq:Wronskian_ODE}
	\dfrac{dW}{dx} + p(x)W = 0.	
\end{equation}
Let us define the scaled Wronskian $\mathcal{W}$ such that
\begin{equation}
	\mathcal{W} \equiv \exp\left(\int^{x} p(x') \; dx' \right) W(x),
\end{equation}
we see that $d\mathcal{W}/dx = 0$, i.e., $\mathcal{W}$ is a constant.

It is not immediately obvious that $\mathcal{W}_{X}$, evaluated using Eq.~\eqref{eq:WX_def}, is the same as $\mathcal{W}_{R}$, evaluated using Eq.~\eqref{eq:WR_def}. From Eq.~\eqref{eq:WX_def} and using Eq.~\eqref{eq:drstardr}, we have
\begin{equation}
	\mathcal{W}_{X} = \dfrac{\Delta}{(r^2 + a^2)\eta} \left( X^{\mathrm{in}}{X^{\mathrm{up}}}' - X^{\mathrm{up}}{X^{\mathrm{in}}}' \right).
\end{equation}
Recall that the \gls{GSN} function $X$ is transformed from a Teukolsky function $R$ using the ${}_{s} \Lambda$ operator that
\begin{equation}
\tag{\ref{eq:X_Lambda_R}}
\begin{aligned}
	X(r) & = {}_{s}\Lambda \left[ R(r) \right] \\
	& = \sqrt{\left( r^2 + a^2 \right) \Delta^{s}} \left[ \left( \alpha + \beta \Delta^{s+1} \dfrac{d}{dr} \right) R(r) \right].
\end{aligned}
\end{equation}
One can show that
\begin{equation}
\begin{aligned}
	& X^{\mathrm{in}}{X^{\mathrm{up}}}' - X^{\mathrm{up}}{X^{\mathrm{in}}}' \\
= & \left( r^2 + a^2 \right) \Delta^{s} \left\{ \eta - (s+1) \alpha \beta \Delta^{s} \left[ 2\left( r - 1\right) - \Delta' \right] \right\} \\
& \times \left( R^{\mathrm{in}}{R^{\mathrm{up}}}' - R^{\mathrm{up}}{R^{\mathrm{in}}}' \right) \\
= & \dfrac{\left( r^2 + a^2 \right) \eta}{\Delta} \Delta^{s+1} \left( R^{\mathrm{in}}{R^{\mathrm{up}}}' - R^{\mathrm{up}}{R^{\mathrm{in}}}' \right) \\
= & \dfrac{\left( r^2 + a^2 \right) \eta}{\Delta} \mathcal{W}_{R},
\end{aligned}
\end{equation}
using Eq.~\eqref{eq:Rpp_intermsof_RRp} and the fact that $\Delta' = 2(r-1)$. From here, we see that indeed
\begin{equation}
\tag{\ref{eq:WR_WX_identity}}
	\mathcal{W}_{X} = \mathcal{W}_{R}.
\end{equation}

\section{Recurrence relations for the higher order corrections to the asymptotic boundary conditions of the generalized Sasaki-Nakamura equation \label{app:recurrence_relations}}
In addition to the asymptotic boundary conditions to the leading order as shown in Eqs.~\eqref{eq:Xin} and \eqref{eq:Xup}, it is useful to also compute these boundary conditions to higher orders. To start off, we assume the following ansatz for the \gls{GSN} function
\begin{equation}
\tag{\ref{eq:Xansatz}}
	X(r_*) \sim \begin{cases}
		f^{\infty}_{\pm}(r) e^{\pm i\omega r_{*}}, & r_* \to \infty \\
		g^{\mathrm{H}}_{\pm}(r) e^{\pm i p r_{*}}, & r_* \to -\infty
		\end{cases}.
\end{equation}
By substituting the ansatz in Eq.~\eqref{eq:Xansatz} into the \gls{GSN} equation in Eq.~\eqref{eq:GSNeqn}, it can be shown that, as $r \to \infty$, the functions $f^{\infty}_{\pm}$ satisfy the following second-order \gls{ODE}:
\begin{equation}
\label{eq:ODE_for_fansatz}
		{f^{\infty}_{\pm}}'' + P^{\infty}_{\pm}(r) {f^{\infty}_{\pm}}' + Q_{\pm}^{\infty}(r) {f^{\infty}_{\pm}} = 0,
\end{equation}
where we define the functions
\begin{align}
\label{eq:PpmInf}
	P_{\pm}^{\infty}(r) & = \left( \dfrac{r^2 + a^2}{\Delta} \right)\left[ \left( \dfrac{\Delta}{r^2 + a^2} \right)' \pm 2i\omega - \mathcal{F} \right], \\
\label{eq:QpmInf}
	Q_{\pm}^{\infty}(r) & = \left( \dfrac{r^2 + a^2}{\Delta} \right)^2 \left( -\omega^2 \mp i \omega \mathcal{F} - \mathcal{U} \right).
\end{align}
As $r \to r_{+}$, the functions $g^{\mathrm{H}}_{\pm}$ satisfy the following second-order \gls{ODE}:
\begin{equation}
\label{eq:ODE_for_gansatz}
		{g^{\mathrm{H}}_{\pm}}'' + P^{\mathrm{H}}_{\pm}(r) {g^{\mathrm{H}}_{\pm}}' + Q_{\pm}^{\mathrm{H}}(r) {g^{\mathrm{H}}_{\pm}} = 0,
\end{equation}
where we define the functions
\begin{align}
\label{eq:PpmHor}
	P_{\pm}^{{\mathrm{H}}}(r) & = \left( \dfrac{r^2 + a^2}{\Delta} \right)\left[ \left( \dfrac{\Delta}{r^2 + a^2} \right)' \pm 2ip - \mathcal{F} \right], \\
\label{eq:QpmHor}
	Q_{\pm}^{{\mathrm{H}}}(r) & = \left( \dfrac{r^2 + a^2}{\Delta} \right)^2 \left( -p^2 \mp i p \mathcal{F} - \mathcal{U} \right).
\end{align}
We look for formal series expansions of the solutions $f^{\infty}_{\pm}$ at infinity and $g^{\mathrm{H}}_{\pm}$ at the horizon, respectively. We then truncate these expansions at an arbitrary order and use them to set the boundary conditions when solving the \gls{GSN} equation on a numerically finite interval.

\subsection{Formal series expansion about infinity}
Inspecting Eq.~\eqref{eq:ODE_for_fansatz} with $P_{\pm}^{\infty}(r)$ and $Q_{\pm}^{\infty}(r)$ defined in Eqs.~\eqref{eq:PpmInf} and \eqref{eq:QpmInf}, respectively and performing the standard change of variable $z \equiv 1/r$, we see that infinity (i.e., $z = 0$) is an irregular singular point of rank $1$.
We expand $P_{\pm}^{\infty}(r)$ and $Q_{\pm}^{\infty}(r)$ as $r \to \infty$ with
\begin{align}
	P_{\pm}^{\infty}(r) & = \sum_{j=0}^{\infty} \dfrac{P_{\pm, j}^{\infty}}{r^{j}},\\
	Q_{\pm}^{\infty}(r) & = \sum_{j=0}^{\infty} \dfrac{Q_{\pm, j}^{\infty}}{r^{j}}.
\end{align}
In particular, we find that $Q_{\pm, 0}^{\infty}$ and $Q_{\pm, 1}^{\infty}$ are zero.
Using these facts, the functions $f^{\infty}_{\pm}$ have the following formal series expansions near infinity as \cite{NIST:DLMF}
\begin{equation}
\label{eq:finfpm_generalsoln}
	f^{\infty}_{\pm}(r) = e^{\nu_{\pm} r} r^{\kappa_{\pm}} \sum_{j = 0}^{\infty} \dfrac{a_{\pm, j}}{r^{j}}
\end{equation}
(note that we suppress the $\infty$ superscript on the rhs since the context is clear), where $\kappa_{\pm}$ is given by
\begin{equation}
	\kappa_{\pm} = -\dfrac{P_{\pm, 1}\nu_{\pm} + Q_{\pm, 1}}{P_{\pm, 0} + 2\nu_{\pm}},
\end{equation}
and $\nu_{\pm}$ is a solution to the characteristic equation
\begin{equation}
 \nu_{\pm}^2 - P_{\pm, 0} \nu_{\pm} = 0.
\end{equation}
There are two solutions to the characteristic equation: $\nu_{\pm} = 0$ or $\nu_{\pm} = P_{\pm, 0}$. We pick $\nu_{+} = \nu_{-} = 0$ as this gives the desired form for the series expansions and as a result we have both $\kappa_{+} = \kappa_{-} = 0$ (recall that $Q_{\pm, 1} = 0$).
The expansion coefficients $a_{\pm, j}$ can be evaluated using the recurrence relation \cite{NIST:DLMF}
\begin{equation}
	P_{0} j a_j = j(j-1)a_{j-1} + \sum_{k=1}^{j} \left[ Q_{k+1} - \left( j - k \right)	P_k \right] a_{j-k},
\end{equation}
where we further suppress the $\pm$ subscript (both the out- and the ingoing mode have the same form above for the recurrence relations), and we set $a_{0} = 1$. As an example, the coefficient $a_{1}$ is given by $ a_{1} = Q_{2}/P_{0}$.
Comparing Eq.~\eqref{eq:finfpm} with Eq.~\eqref{eq:finfpm_generalsoln}, we have
\begin{equation}
	\mathcal{C}^{\infty}_{\pm, j} = \omega^j a^{\infty}_{\pm, j}.	
\end{equation}

 \subsection{Formal series expansion about the horizon}
 Inspecting Eq.~\eqref{eq:ODE_for_gansatz} with $P_{\pm}^{\mathrm{H}}(r)$ and $Q_{\pm}^{\mathrm{H}}(r)$ defined in Eqs.~\eqref{eq:PpmHor} and~\eqref{eq:QpmHor}, respectively, we see that $r = r_+$ is a regular singular point.
 In particular, $P_{\pm}^{\mathrm{H}}(r)\left(r - r_{+}\right)$ and $Q_{\pm}^{\mathrm{H}}(r)\left(r - r_{+}\right)^2$ are analytic at $r = r_{+}$ since
\[
\begin{aligned}
	P_{\pm}^{\mathrm{H}}(r)\left(r - r_{+}\right) & = \left( \dfrac{r^2 + a^2}{r - r_{-}} \right)\left[ \left( \dfrac{\Delta}{r^2 + a^2} \right)' \pm 2ip - \mathcal{F} \right], \\
	Q_{\pm}^{\mathrm{H}}(r)\left(r - r_{+}\right)^2 & = \left( \dfrac{r^2 + a^2}{r - r_{-}} \right)^2 \left( -p^2 \mp i p \mathcal{F} - \mathcal{U} \right).
\end{aligned}
\]
 A formal series expansion near the horizon can be obtained using the Frobenius method. We expand $P_{\pm}^{\mathrm{H}}(r)$ and $Q_{\pm}^{\mathrm{H}}(r)$ near $r = r_{+}$ as
 \begin{align}
	P_{\pm}^{\mathrm{H}}(r) & = \sum_{j=0}^{\infty} P_{\pm, j}^{\mathrm{H}} (r - r_{+})^{j-1},\\
	Q_{\pm}^{\mathrm{H}}(r) & = \sum_{j=0}^{\infty} Q_{\pm, j}^{\mathrm{H}} (r - r_{+})^{j-2}.
\end{align}
The functions $g^{\mathrm{H}}_{\pm}(r)$ again have the formal series expansions near the horizon as \cite{NIST:DLMF}
 \begin{equation}
 \label{eq:fhorpm_generalsoln}
 	g^{\mathrm{H}}_{\pm}(r) = (r - r_{+})^{\nu_{\pm}} \sum_{j=0}^{\infty} a_{\pm, j} (r - r_{+})^j
 \end{equation}
(note that we again suppress the ${\mathrm{H}}$ superscript on the rhs since the context is clear), where $\nu_{\pm}$ is a root to the indicial polynomial $I(\nu_{\pm})$, which is given by \cite{NIST:DLMF}
 \begin{equation}
 	I(\nu_{\pm}) = \nu_{\pm}(\nu_{\pm} - 1) + P_{\pm, 0} \nu_{\pm} + Q_{\pm, 0}.
 \end{equation}
Note that we have $Q_{\pm, 0} = 0$, therefore the indicial equation $I(\nu_{\pm}) = 0$ has two solutions: $\nu_{\pm} = 0$ or $\nu_{\pm} = \left( 1 - P_{\pm, 0} \right)$. Again we pick $\nu_{+} = \nu_{-} =0$ as this gives the desired expansions. The expansion coefficients $a_{\pm, j}$ can be evaluated again using a recurrence relation as \cite{NIST:DLMF}
 \begin{equation}
 	I(j) a_{j} = - \sum_{k=0}^{j-1} \left( k P_{j-k} + Q_{j-k} \right) a_{k},
 \end{equation}
where we again further suppress the $\pm$ subscript (both the out- and the ingoing mode have the same form above for the recurrence relations), and we set $a_{0} = 1$.
For example, explicitly $a_1 = -Q_{1}/P_{0}$. Comparing Eq.~\eqref{eq:fhorpm} with Eq.~\eqref{eq:fhorpm_generalsoln}, we have
\begin{equation}
	\mathcal{C}^{\mathrm{H}}_{\pm, j} = \omega^{-j} a^{\mathrm{H}}_{\pm, j}.	
\end{equation}

\section{Explicit generalized Sasaki-Nakamura transformations for physically relevant radiation fields \label{app:explicitGSN}}
In this appendix, we explicitly show our choices of $g_{i}(r)$ for radiation fields with spin weight $s=0,\pm 1,\pm 2$ that we use to construct the \gls{GSN} transformation. For each transformation, we give explicit expressions for the weighting functions $\alpha(r), \beta(r)$, the determinant of the transformation matrix $\eta(r)$, the asymptotic solutions to the \gls{GSN} equation at infinity and at the horizon for both the in- and the outgoing mode, and the conversion factors for transforming the asymptotic amplitudes between the Teukolsky function $R$ and the \gls{SN} function $X$. Together with Sec.~\ref{sec:formalism} and this appendix, one should have all the necessary ingredients to use the \gls{GSN} formalism to numerically solve the homogenous radial Teukolsky equation for physically relevant radiation fields ($s = 0$ for scalar radiation, $s=\pm 1$ for electromagnetic radiation, and $s=\pm 2$ for gravitational radiation).

Despite being long-winded, we opt to show the expressions explicitly for the sake of completeness. Accompanying this paper are \textsc{Mathematica} notebooks deriving and storing all the expressions shown here, and they can be found on Zenodo \cite{10.5281/zenodo.8080242}.
While the \gls{GSN} formalism was proposed to facilitate numerical computations, all the expressions in this appendix and Sec. \ref{sec:formalism} are exact. In particular, we do not assume that $\omega$ is real when deriving expressions shown here and they can be used in \gls{QNM} calculations with the \gls{GSN} formalism 
(such as Ref.~\cite{PhysRevD.102.044032} using the parametrized \gls{BH} quasinormal ringdown formalism \cite{PhysRevD.99.104077, PhysRevD.100.044061, PhysRevD.101.064031} to compute semianalytical corrections from \gls{QNM} frequencies for a nonrotating \gls{BH}, and Sec.~\ref{subsec:QNMusingGSN}). We also do not use the identities shown in Eqs.~\eqref{eq:CtransBinc_identity} and~\eqref{eq:BtransCinc_identity} to simplify the expressions for the conversion factors below.

\subsection{Scalar radiation $s=0$}
By choosing $g_0(r) = 1$, we have the weighting functions
\begin{subequations}
\begin{eqnarray}
	\alpha(r) & = & 1 , \\
	\beta(r) & = & 0 .
\end{eqnarray}
\end{subequations}
The determinant of the transformation matrix $\eta(r)$ can be written as
\[ \eta = c_0 + c_1/r + c_2/r^2 + c_3/r^3 + c_4/r^4 \]
with the coefficients
\begin{subequations}
\begin{eqnarray}
	c_0 & = & 1 , \\
	c_{1,2,3,4} & = & 0 .
\end{eqnarray}
\end{subequations}

The asymptotic outgoing mode of $X$ when $r_* \to \infty$ is given by
\[ X(r_* \to \infty) \propto f^{\infty}_{+}(r) e^{i\omega r_{*}} = e^{i\omega r_{*}} \left( 1 + \sum_{j=1}^{\infty} \dfrac{\mathcal{C}^{\infty}_{+, j}}{r^j} \right) \]
with the first three expansion coefficients
\begin{subequations}
\begin{eqnarray}
	\mathcal{C}^{\infty}_{+, 1} & = & \dfrac{1}{2} i \left(\lambda +2 a m \omega \right) , \\
	\mathcal{C}^{\infty}_{+, 2} & = & \dfrac{1}{8} \left\{ -\lambda ^2+\lambda  \left(2-4 a m \omega \right) \right. \\
	& & \left. +4 \omega  \left[ i-a^2m^2 \omega +a \left( m+2 i m \omega \right) \right] \right\}, \nonumber \\
	\mathcal{C}^{\infty}_{+, 3} & = & -\dfrac{1}{48} i \left\{ \lambda ^3+\lambda ^2 (-8+6 a m \omega ) \right. \\
	& & \left. +4 \lambda  \left[ 3-\left(9 i+8 a m\right)\omega +a \left(2 a-6 i m+3 a m^2\right) \omega ^2\right] \right. \nonumber \\
	& & \left. +8 \omega  \left[3 i+a^2 \left(-1+m^2 (-3-6 i \omega )\right) \omega \right. \right. \nonumber \\
	& & \left. \left. +a^3 m \left(2+m^2\right) \omega ^2+a m \left(3-3 i \omega -8 \omega ^2\right)\right] \right\} \nonumber.
\end{eqnarray}
\end{subequations}
The asymptotic ingoing mode of $X$ when $r_* \to \infty$ is given by
\[ X(r_* \to \infty) \propto f^{\infty}_{-}(r) e^{-i\omega r_{*}} = e^{-i\omega r_{*}} \left( 1 + \sum_{j=1}^{\infty} \dfrac{\mathcal{C}^{\infty}_{-, j}}{r^j} \right) \]
with the first three expansion coefficients
\begin{subequations}
\begin{eqnarray}
	\mathcal{C}^{\infty}_{-, 1} & = & -\dfrac{1}{2} i \left(\lambda +2 a m \omega \right) , \\
	\mathcal{C}^{\infty}_{-, 2} & = & \dfrac{1}{8} \left\{ -\lambda ^2+\lambda  \left( 2-4 a m \omega \right) \right. \\
	& & \left. -4 \omega  \left[ i+am \left(-1+2 i \omega \right)+a^2 m^2 \omega \right] \right\} , \nonumber \\
	\mathcal{C}^{\infty}_{-, 3} & = & \dfrac{1}{48} i \left\{ \lambda ^3+\lambda ^2 \left(-8+6 a m \omega \right) \right. \\
	& & \left. +4 \lambda  \left[ 3+ \left(9 i-8 am \right) \omega +a \left(2 a+6 i m+3 a m^2\right) \omega ^2 \right] \right. \nonumber \\
	& & \left. +8 \omega  \left[ -3 i+a^2 \left(-1+m^2 (-3+6 i \omega )\right) \omega \right. \right. \nonumber \\
	& & \left. \left. +a^3 m \left(2+m^2\right)\omega ^2+a m \left(3+3 i \omega -8 \omega ^2\right)\right] \right\} \nonumber.
\end{eqnarray}
\end{subequations}
These expressions (except for $\mathcal{C}^{\infty}_{+, j}$) match with those found in Ref.~\cite{Hughes:2000pf}. Note that $\mathcal{C}^{\infty}_{+, j} = \left( \mathcal{C}^{\infty}_{-, j} \right)^{*}$ as claimed in Ref.~\cite{Hughes:2000pf} is true only for real $\omega$ since the \gls{GSN} potentials $\mathcal{F},\mathcal{U}$ are real valued in this case.

The conversion factors between the \gls{GSN} and the Teukolsky formalism are found to be
\begin{subequations}
\begin{eqnarray}
	\dfrac{B^{\mathrm{ref}}_{\mathrm{T}}}{B^{\mathrm{ref}}_{\mathrm{SN}}} = \dfrac{C^{\mathrm{trans}}_{\mathrm{T}}}{C^{\mathrm{trans}}_{\mathrm{SN}}} & = & 1, \\
	\dfrac{B^{\mathrm{inc}}_{\mathrm{T}}}{B^{\mathrm{inc}}_{\mathrm{SN}}}  & = & 1, \\
	\dfrac{C^{\mathrm{inc}}_{\mathrm{T}}}{C^{\mathrm{inc}}_{\mathrm{SN}}} & = & \dfrac{1}{\sqrt{2r_+}}, \\
	\dfrac{B^{\mathrm{trans}}_{\mathrm{T}}}{B^{\mathrm{trans}}_{\mathrm{SN}}} = \dfrac{C^{\mathrm{ref}}_{\mathrm{T}}}{C^{\mathrm{ref}}_{\mathrm{SN}}} & = & \dfrac{1}{\sqrt{2r_+}}.
\end{eqnarray}
\end{subequations}
Note that these conversion factors are frequency independent.

\subsection{Electromagnetic radiation}
\subsubsection{$s=+1$}
By choosing $g_0(r) = \dfrac{r^2 + a^2}{r^2}$ and $g_1(r) = 1$, we have the weighting functions
\begin{subequations}
\begin{eqnarray}
	\alpha(r) & = & \dfrac{1}{r^2 \sqrt{\Delta}} \left[ -ia^3 m - iam r^2 + ia^4 \omega \right. \\
	& & \left. + r^3 \left( 1 + ir\omega \right) + a^2 \left( -2 + r + 2ir^2 \omega \right) \right], \nonumber \\
	\beta(r) & = &  \dfrac{\left( r^2 + a^2 \right)}{r^2 \Delta^{3/2}} .
\end{eqnarray}
\end{subequations}

The determinant of the transformation matrix $\eta(r)$ can be written as
\[ \eta = c_0 + c_1/r + c_2/r^2 + c_3/r^3 + c_4/r^4 \]
with the coefficients
\begin{subequations}
\begin{eqnarray}
	c_0 & = & -\left( 2+\lambda \right), \\
	c_1 & = & 2 i a m, \\
	c_2 & = & -a^2 \left( 3 + 2\lambda \right), \\
	c_3 & = & -2a^2 \left( 1 - iam \right), \\
	c_4 & = & -a^4\left(1+\lambda\right).
\end{eqnarray}
\end{subequations}

The asymptotic outgoing mode of $X$ when $r_* \to \infty$ is given by
\[ X(r_* \to \infty) \propto f^{\infty}_{+}(r) e^{i\omega r_{*}} = e^{i\omega r_{*}} \left( 1 + \sum_{j=1}^{\infty} \dfrac{\mathcal{C}^{\infty}_{+, j}}{r^j} \right) \]
with the first three expansion coefficients
\begin{subequations}
\begin{eqnarray}
	\mathcal{C}^{\infty}_{+, 1} & = & \dfrac{1}{2} i \left(2+\lambda +2 a m \omega \right) , \\
	\mathcal{C}^{\infty}_{+, 2} & = & \dfrac{1}{8} \left[ -\lambda ^2-2 \lambda  \left(1+2 a m \omega \right) \right. \\
	& & \left. -4 a \omega  \left(m-2a \omega -2 i m \omega +a m^2 \omega \right) \right], \nonumber \\
	\mathcal{C}^{\infty}_{+, 3} & = & -\dfrac{1}{48} i \left\{ \lambda ^3+\lambda ^2 \left(-2+6 a m \omega \right) \right. \\
	& & \left. +4 \lambda  \left[ -2-2 \left(3i+a m\right) \omega \right. \right. \nonumber \\
	& & \left. \left. +a \left(-4 a-6 i m+3 a m^2\right) \omega ^2\right] \right. \nonumber \\
	& & \left. +8 \omega  \left[ -6 i+a^3 m \left(-4+m^2\right) \omega ^2 \right. \right. \nonumber \\
	& & \left. \left. +3 a^2 \omega  \left(-1-2i m^2 \omega \right)-a m \left(3+6 i \omega +8 \omega ^2\right)\right] \right\} \nonumber.
\end{eqnarray}
\end{subequations}
The asymptotic ingoing mode of $X$ when $r_* \to \infty$ is given by
\[ X(r_* \to \infty) \propto f^{\infty}_{-}(r) e^{-i\omega r_{*}} = e^{-i\omega r_{*}} \left( 1 + \sum_{j=1}^{\infty} \dfrac{\mathcal{C}^{\infty}_{-, j}}{r^j} \right) \]
with the first three expansion coefficients
\begin{subequations}
\begin{eqnarray}
	\mathcal{C}^{\infty}_{-, 1} & = & \dfrac{1}{2 c_{0}} i \left[ 4+\lambda ^2+8 a m \omega +2 \lambda  \left(2+a m \omega \right) \right] , \\
	\mathcal{C}^{\infty}_{-, 2} & = & \dfrac{1}{8 c_{0}} \left\{ \lambda ^3+4
\lambda ^2 (1+a m \omega ) \right. \\
& & \left. +8 a \omega  \left[ m \left(2+2 i \omega \right)-a \omega +3 a m^2 \omega \right] \right. \nonumber \\
& & \left. +4 \lambda  \left[ 1+a m \left(5+2 i \omega \right) \omega +a^2 \left(-2+m^2\right) \omega ^2\right] \right\}, \nonumber \\
	\mathcal{C}^{\infty}_{-, 3} & = & -\dfrac{1}{48 c_{0}}i \left\{ \lambda ^4+6 a m \lambda ^3 \omega \right. \\
	& & \left. +4 \lambda ^2 \left[-3+\left(6i+4 a m\right) \omega \right. \right. \\
	& & \left. \left. +a \left(-4 a+6 i m+3 a m^2\right) \omega ^2\right] \right. \nonumber \\
	& & \left. +8 \lambda  \left[ -2+\left(12 i-5 a m\right) \omega \right. \right. \nonumber \\
	& & \left. \left. +a \left(12 i m+a (-4+9 m^2)\right) \omega ^2 \right. \right. \nonumber \\
& & \left. \left. +a m \left(-8+6 i a m+a^2 (-4+m^2)\right) \omega ^3\right] \right. \nonumber \\
& & \left. +16 \omega  \left[ 6 i+a^3 m \left(-1+4 m^2\right) \omega ^2 \right. \right. \nonumber \\
& & \left. \left. +3 i a^2\omega  \left(i-2 \omega +4 m^2 \omega \right) \right. \right. \nonumber \\
& & \left. \left. +a m \left(-3+12 i \omega -8 \omega ^2\right) \right] \right\} \nonumber.
\end{eqnarray}
\end{subequations}

The conversion factors between the \gls{GSN} and the Teukolsky formalism are found to be
\begin{subequations}
\begin{eqnarray}
	\dfrac{B^{\mathrm{ref}}_{\mathrm{T}}}{B^{\mathrm{ref}}_{\mathrm{SN}}} = \dfrac{C^{\mathrm{trans}}_{\mathrm{T}}}{C^{\mathrm{trans}}_{\mathrm{SN}}} & = & \dfrac{1}{2i\omega}, \\
	\dfrac{B^{\mathrm{inc}}_{\mathrm{T}}}{B^{\mathrm{inc}}_{\mathrm{SN}}} & = & \dfrac{2i\omega}{c_0}, \\
	\dfrac{C^{\mathrm{inc}}_{\mathrm{T}}}{C^{\mathrm{inc}}_{\mathrm{SN}}} & = & \dfrac{r_{+}^{3/2}}{4\sqrt{2}}, \\
	&  & \times \left[ r_+(1+4i\omega-iam) - a^2(1+2i\omega) \right]^{-1} \nonumber \\
	\dfrac{B^{\mathrm{trans}}_{\mathrm{T}}}{B^{\mathrm{trans}}_{\mathrm{SN}}} = \dfrac{C^{\mathrm{ref}}_{\mathrm{T}}}{C^{\mathrm{ref}}_{\mathrm{SN}}} & = & \sqrt{2r_+} \dfrac{2 r_+ \omega - a m }{2am + 2i (2+\lambda)}.
\end{eqnarray}
\end{subequations}

\subsubsection{$s=-1$}
By choosing $g_0(r) = \dfrac{r^2+a^2}{r^2}$ and $g_1(r) = 1$, we have the weighting functions
\begin{subequations}
\begin{eqnarray}
	\alpha(r) & = & -\dfrac{\sqrt{\Delta}}{r^2} \left[ r + i\dfrac{\left(r^2+a^2\right)K}{\Delta} \right], \\
	\beta(r) & = &  \dfrac{\sqrt{\Delta}\left(r^2 + a^2\right)}{r^2}.
\end{eqnarray}
\end{subequations}

The determinant of the transformation matrix $\eta(r)$ can be written as
\[ \eta = c_0 + c_1/r + c_2/r^2 + c_3/r^3 + c_4/r^4 \]
with the coefficients
\begin{subequations}
\begin{eqnarray}
	c_0 & = & -\lambda, \\
	c_1 & = & -2iam, \\
	c_2 & = &  a^2 \left( 1 - 2\lambda \right), \\
	c_3 & = &  -2a^2 \left( 1 + iam \right), \\
	c_4 & = & a^4 \left( 1 - \lambda \right).
\end{eqnarray}
\end{subequations}

The asymptotic outgoing mode of $X$ when $r_* \to \infty$ is given by
\[ X(r_* \to \infty) \propto f^{\infty}_{+}(r) e^{i\omega r_{*}} = e^{i\omega r_{*}} \left( 1 + \sum_{j=1}^{\infty} \dfrac{\mathcal{C}^{\infty}_{+, j}}{r^j} \right) \]
with the first three expansion coefficients
\begin{subequations}
\begin{eqnarray}
	\mathcal{C}^{\infty}_{+, 1} & = & -\frac{1}{2 c_{0}}i \left(\lambda^2+4 a m \omega +2 a m \lambda  \omega \right) , \\
	\mathcal{C}^{\infty}_{+, 2} & = & \frac{1}{8 c_{0}} \left[ \lambda^3 - \lambda^2
 \left(2-4 a m \omega \right) - \right. \\
 & & \left. 8 a \omega  \left(m-a \omega -2 a m^2 \omega \right) \right. \nonumber \\
 & & \left. + 4 a \omega \lambda  \left(m-2 a \omega -2 i m \omega +a m^2 \omega \right)\right], \nonumber \\
	\mathcal{C}^{\infty}_{+, 3} & = & \dfrac{1}{48 c_{0}}i \left\{ \lambda ^4+\lambda ^3 (-8+6 a m \omega ) \right. \\
 & & \left. +4 \lambda ^2 \left[3-\left(6 i+5 a m\right) \omega \right. \right. \nonumber \\
 & & \left. \left. +a \left(-4a-6 i m+3 a m^2 \right) \omega ^2\right] \right. \nonumber \\
 & & \left. +48 a \omega  \left[a (-1+2 i \omega ) \omega +a^2 m^3 \omega^2 \right. \right. \nonumber \\
 & & \left. \left. -2 i a m^2 \omega  (-i+\omega )+m \left(1-2i \omega +a^2 \omega ^2\right)\right] \right. \nonumber \\
 & & \left. +8 a \lambda  \omega  \left[4 a \omega +3 a m^2 (1-2 i \omega ) \omega +a^2 m^3 \omega^2 \right. \right. \nonumber \\
 & & \left. \left. -4 m \left(1+ (2+a^2)\omega^2\right)\right] \right\}. \nonumber
\end{eqnarray}
\end{subequations}
The asymptotic ingoing mode of $X$ when $r_* \to \infty$ is given by
\[ X(r_* \to \infty) \propto f^{\infty}_{-}(r) e^{-i\omega r_{*}} = e^{-i\omega r_{*}} \left( 1 + \sum_{j=1}^{\infty} \dfrac{\mathcal{C}^{\infty}_{-, j}}{r^j} \right) \]
with the first three expansion coefficients
\begin{subequations}
\begin{eqnarray}
	\mathcal{C}^{\infty}_{-, 1} & = & -\dfrac{1}{2} i \left(\lambda +2 a m \omega \right), \\
	\mathcal{C}^{\infty}_{-, 2} & = & \dfrac{1}{8} \left[-\lambda^2 + \lambda \left(2-4 a m \omega \right) \right. \\
	& & \left. +4 a \omega  \left(m+2a \omega -2 i m \omega -a m^2 \omega \right)\right], \nonumber \\
	\mathcal{C}^{\infty}_{-, 3} & = & \dfrac{1}{48} i \left\{ \lambda ^3+\lambda ^2 (-8+6 a m \omega ) \right. \\
	& & \left. +4 \lambda  \left[3+(6 i-8a m) \omega +a \left(-4 a+6 i m+3 a m^2\right) \omega ^2\right] \right. \nonumber \\
	& & \left. +8 a \omega  \left[ a \omega +3 a m^2 \left(-1+2 i \omega \right) \omega +a^2 m^3 \omega ^2 \right. \right. \nonumber \\
	& & \left. \left. +m \left(2-4 \left(2+a^2\right) \omega ^2\right)\right] \right\}. \nonumber
\end{eqnarray}
\end{subequations}
These expressions (except for $\mathcal{C}^{\infty}_{+, j}$) match with those found in Ref.~\cite{Hughes:2000pf}. Note that $\mathcal{C}^{\infty}_{+, j} = \left( \mathcal{C}^{\infty}_{-, j} \right)^{*}$ as claimed in Ref.~\cite{Hughes:2000pf} is not true even for real $\omega$ since the \gls{GSN} potentials $\mathcal{F},\mathcal{U}$ are, in general, complex valued.

The conversion factors between the \gls{GSN} and the Teukolsky formalism are found to be
\begin{subequations}
\begin{eqnarray}
	\dfrac{B^{\mathrm{ref}}_{\mathrm{T}}}{B^{\mathrm{ref}}_{\mathrm{SN}}} = \dfrac{C^{\mathrm{trans}}_{\mathrm{T}}}{C^{\mathrm{trans}}_{\mathrm{SN}}} & = & -\dfrac{2i\omega}{c_0}, \\
	\dfrac{B^{\mathrm{inc}}_{\mathrm{T}}}{B^{\mathrm{inc}}_{\mathrm{SN}}} & = & -\dfrac{1}{2i\omega}, \\
	\dfrac{C^{\mathrm{inc}}_{\mathrm{T}}}{C^{\mathrm{inc}}_{\mathrm{SN}}} & = & -\dfrac{\sqrt{r_{+}}\left[ \left( a m - 4\omega \right) r_{+} + 2 a^2 \omega \right]}{\sqrt{2} \left( a m - i \lambda \right)}, \\
	\dfrac{B^{\mathrm{trans}}_{\mathrm{T}}}{B^{\mathrm{trans}}_{\mathrm{SN}}} = \dfrac{C^{\mathrm{ref}}_{\mathrm{T}}}{C^{\mathrm{ref}}_{\mathrm{SN}}} & = & \dfrac{r_{+}^{3/2}}{4\sqrt{2}} \\
	& & \times \left[ \left( 1 + i a m - 4 i \omega \right)r_{+} - a^2 \left( 1 - 2 i \omega \right) \right]^{-1}. \nonumber
\end{eqnarray}
\end{subequations}

\begin{widetext}
\subsection{Gravitational radiation}
\subsubsection{$s=+2$}
By choosing $g_0(r) = \dfrac{r^2}{r^2 + a^2}$, $g_1(r) = 1$, and $g_2(r) = \dfrac{r^2 + a^2}{r^2}$ , we have the weighting functions
\begin{subequations}
\begin{eqnarray}
	\alpha(r) & = & \dfrac{1}{r^2 \Delta} \left\{ 4a^3 m r \left( i + r\omega\right) + 2amr^2 \left( i - 3ir + 2r^2\omega\right) - 2a^4 \left( -3 + 2ir\omega + r^2 \omega^2 \right) \right. \\
	& & \left. + r^3 \left[ -2\lambda + r\left( 2 + \lambda + 10 i\omega \right) - 2r^3 \omega^2 \right] - a^2 r \left( 8 + 2m^2 r - r\lambda + 2ir\omega + 4ir^2 \omega + 4r^3 \omega^2 \right) \right\}, \nonumber \\
	\beta(r) & = & \dfrac{1}{r\Delta^{3}} \left[ -2iamr + a^2 \left( -4 + 2ir\omega \right) + 2r \left( 3 - r + ir^2 \omega \right) \right].
\end{eqnarray}
\end{subequations}

The determinant of the transformation matrix $\eta(r)$ can be written as
\[ \eta = c_0 + c_1/r + c_2/r^2 + c_3/r^3 + c_4/r^4 \]
with the coefficients
\begin{subequations}
\begin{eqnarray}
	c_0 & = & 24 + 12 i\omega + \lambda(10 + \lambda) -12a\omega \left( a\omega - m \right), \\
	c_1 & = & -32iam - 8iam\lambda + 8ia^2 \omega( 1 + \lambda), \\
	c_2 & = & 12a^2 - 24 iam - 24a^2 m^2 + 24i a^2 \omega + 48 a^3 m \omega - 24a^4 \omega^2, \\
	c_3 & = & -24 i a^3 \left( a \omega - m \right) - 24a^2, \\
	c_4 & = & 12a^4.
\end{eqnarray}
\end{subequations}

The asymptotic outgoing mode of $X$ when $r_* \to \infty$
\[ X(r_* \to \infty) \propto f^{\infty}_{+}(r) e^{i\omega r_{*}} = e^{i\omega r_{*}} \left( 1 + \sum_{j=1}^{\infty} \dfrac{\mathcal{C}^{\infty}_{+, j}}{r^j} \right) \]
with the first three expansion coefficients
\begin{subequations}
\begin{eqnarray}
	\mathcal{C}^{\infty}_{+, 1} & = & \dfrac{1}{2} i \left( 6 + \lambda + 2 a m \omega \right), \\
	\mathcal{C}^{\infty}_{+, 2} & = & -\dfrac{1}{8} \left\{ \lambda^2 + 2\lambda \left( 5 + 2 a m \omega \right) + 4 \left[ 6 + \left( 3 i + 5 am \right)\omega + am \left( -2i + am \right)\omega^2 \right] \right\}, \\
	\mathcal{C}^{\infty}_{+, 3} & = & -\frac{1}{48} i \left\{ \lambda ^3+2 \lambda ^2 \left( 5+3 a m \omega \right) + 4 \lambda  \left[ 6+ \left(3i+10 a m \right) \omega +a \left(2 a-6 i m+3 a m^2\right) \omega^2 \right] \right. \\
	& & \left. + 8 a \omega  \left[ a \omega +6 a m^2 \left(1-i \omega \right) \omega +a^2 m^3 \omega ^2+m
\left(2-9 i \omega +2 (-4+a^2) \omega^2 \right) \right] \right\}.
\end{eqnarray}
\end{subequations}

The asymptotic ingoing mode of $X$ when $r_* \to \infty$
\[ X(r_* \to \infty) \propto f^{\infty}_{-}(r) e^{-i\omega r_{*}} = e^{-i\omega r_{*}} \left( 1 + \sum_{j=1}^{\infty} \dfrac{\mathcal{C}^{\infty}_{-, j}}{r^j} \right) \]
with the first three expansion coefficients
\begin{subequations}
\begin{eqnarray}
	\mathcal{C}^{\infty}_{-, 1} & = & \dfrac{1}{2c_{0}}i \left\{ -\lambda ^3-2 \lambda ^2 (8+a m \omega )+4 \lambda  \left[ -21-3 \left( i+4 a m \right) \omega +7 a^2 \omega ^2\right] \right. \\
	& & \left. + 8\left[ -18-\left(9 i+23 a m\right) \omega +a \left(11 a-3 i m-3 a m^2\right) \omega ^2+3 a^3 m \omega ^3\right] \right\}, \nonumber \\
	\mathcal{C}^{\infty}_{-, 2} & = & -\dfrac{1}{8c_{0}} \left\{ \lambda^4 + 4 \lambda ^3 \left( 5+a m \omega \right) +4 \lambda ^2 \left[ 37+2 a m \left(13+i \omega \right) \omega +a^2 \left( -11+m^2 \right) \omega^2 \right] \right. \\
& & \left. -8 \lambda  \left[ -60 + 8a m \left(-11-2 i \omega \right) \omega +a^2 \left(39-19 m^2\right) \omega ^2+14 a^3 m \omega ^3\right] \right. \nonumber \\
& & \left. -16 \left[ a^2 \left( 34+m^2 (-49-9 i \omega ) + 3 i \omega \right) \omega ^2+a^3 m \left(43-3 m^2+6 i \omega \right) \omega ^3+3 a^4 \left(-4+m^2\right) \omega^4 \right. \right. \nonumber \\
& & \left. \left. -9 \left(4+\omega ^2\right) +2 a m \omega \left(-44-15 i \omega +3 \omega ^2\right)\right] \right\}, \nonumber \\
	\mathcal{C}^{\infty}_{-, 3} & = & -\dfrac{1}{48c_{0}}i \left\{ -\lambda ^5-2 \lambda ^4 (10+3 a m \omega )-4 \lambda ^3 \left[ 37+2 a m \left(20+3 i \omega \right) \omega +a^2 \left(-13+3 m^2\right) \omega ^2\right] \right. \\
	& & \left. -8\lambda ^2 \left[ 60+2 a^2 \left( -29+3 m^2 (9+i \omega ) \right) \omega ^2+a^3 m \left(-31+m^2\right) \omega ^3+a m \omega  \left(157+48 i \omega -8
\omega ^2\right)\right] \right. \nonumber \\
& & \left. + 16\lambda  \left[ a^2 \left(91+m^2 (-210-81 i \omega )+9 i \omega \right) \omega ^2+2 a^3 m \left(73-13 m^2+21 i \omega \right)
\omega ^3+3 a^4 \left(-10+7 m^2\right) \omega ^4 \right. \right. \nonumber \\
& & \left. \left. -9 \left(4+\omega ^2\right)+2 a m \omega  \left(-116-63 i \omega +29 \omega ^2\right) \right] +96 a \omega  \left[-a^3 m^4 \omega ^3+a \omega  \left(18+9 i \omega -11 a^2 \omega ^2\right) \right. \right. \nonumber \\
& & \left. \left. +a^2 m^3 \omega ^2 \left(-28-7 i \omega +a^2 \omega ^2\right) + am^2 \omega  \left(-70-55 i \omega +2 (7+15 a^2) \omega ^2+6 i a^2 \omega ^3\right) \right. \right. \nonumber \\
& & \left. \left. +m \left( -36-36 i \omega + (25+47 a^2) \omega
^2+i (8+23 a^2) \omega ^3-2 a^2 (4+5 a^2) \omega ^4\right)\right] \right\} \nonumber.
\end{eqnarray}
\end{subequations}

The conversion factors
\begin{subequations}
\begin{eqnarray}
	\dfrac{B^{\mathrm{ref}}_{\mathrm{T}}}{B^{\mathrm{ref}}_{\mathrm{SN}}} = \dfrac{C^{\mathrm{trans}}_{\mathrm{T}}}{C^{\mathrm{trans}}_{\mathrm{SN}}} & = & - \dfrac{1}{4\omega^2}, \\
	\dfrac{B^{\mathrm{inc}}_{\mathrm{T}}}{B^{\mathrm{inc}}_{\mathrm{SN}}} & = & - \dfrac{4\omega^2}{c_0}, \\
	\dfrac{C^{\mathrm{inc}}_{\mathrm{T}}}{C^{\mathrm{inc}}_{\mathrm{SN}}} & = & - \dfrac{r_{+}^{3/2}}{4\sqrt{2}} \left\{ \left[ 2 \left( -1 - 6i\omega + 8\omega^2 \right) + a^2 \left( 2 + m^2 + 9i\omega - 8\omega^2 \right) + am \left( 3i - 8\omega \right) \right] r_{+}^2 \right. \\
	& & \left. + a^3 \left(-3i + 4\omega \right) \left(m r_{+} -a\omega \right) \right\}^{-1}, \nonumber \\
	\dfrac{B^{\mathrm{trans}}_{\mathrm{T}}}{B^{\mathrm{trans}}_{\mathrm{SN}}} = \dfrac{C^{\mathrm{ref}}_{\mathrm{T}}}{C^{\mathrm{ref}}_{\mathrm{SN}}} & = & 2\sqrt{2} r_{+}^{3/2} \\
	& & \times \left\{ \left[ 4\omega \left( i - 4 \omega \right) - a m \left(i - 8 \omega \right) - a^2 \left( m^2 + 2i\omega - 4 \omega^2\right) \right]r_{+}^2 + a^2 \left( i - 4\omega \right) \left( a m - 2\omega \right)r_{+} \right\} \nonumber \\
	& & \times \left\{ 2r_{+}^3 \left( 24 + 10\lambda + \lambda^2 + 12 i \omega \right) - r_{+}^2 \left[ 8iam \left( 11 + 2\lambda + 6 i \omega  \right) \right. \right. \nonumber \\
	& & \left. \left. + a^2 \left( 24 + 24m^2 + 10 \lambda + \lambda^2 - 28 i \omega - 16 i \lambda \omega + 48 \omega^2 \right) \right] \right. \nonumber \\
	& & \left. + 8ia^3 r_{+} \left[ m \left( 7 + \lambda - 6i\omega \right) - a\omega \left( 4 + \lambda \right) \right] + 12 a^5 \omega \left( a \omega - 3 m \right) \right\}^{-1}. \nonumber
\end{eqnarray}
\end{subequations}

\subsubsection{$s=-2$}
By choosing $g_0(r) = \dfrac{r^2}{r^2 + a^2}$, $g_1(r) = 1$, and $g_2(r) = \dfrac{r^2 + a^2}{r^2}$ \footnote{Note that $g_0, g_1, g_2$ here are not the same as the $f,g,h$ in Ref.~\cite{10.1143/PTP.67.1788}. In fact, we see that $g = g_{1} = 1$ and $h = g_{2} = \dfrac{r^2 + a^2}{r^2}$, but $f = g_{0}g_{1}g_{2}=1$.}, we have the weighting functions
\begin{subequations}
\begin{eqnarray}
	\alpha(r) & = & \dfrac{1}{r^2\Delta} \left\{ 4a^3 mr\left( -i + r\omega \right) + 2amr^2 \left( 3i - ir + 2r^2 \omega \right) + a^4 \left( 6 + 4ir\omega -2r^2 \omega^2 \right) \right. \\
	& & + a^2 r \left[ -24 + r\left( 12 - 2m^2 + \lambda -6i\omega \right) + 12ir^2\omega - 4r^3\omega^2 \right] \nonumber \\
	& & \left. + r^2 \left[ 24 - 2r \left( 12 + \lambda \right) + r^2 \left( 6 + \lambda -18 i\omega \right) + 8ir^3 \omega - 2r^4\omega^2 \right] \right\}, \nonumber \\
	\beta(r) & = & \dfrac{2\Delta}{r} \left[ iamr + a^2\left( -2 - ir\omega \right) + r \left( 3 - r - i r^2 \omega \right)\right].
\end{eqnarray}
\end{subequations}

The determinant of the transformation matrix $\eta(r)$ can be written as
\[ \eta = c_0 + c_1/r + c_2/r^2 + c_3/r^3 + c_4/r^4 \]
with the coefficients
\begin{subequations}
\begin{eqnarray}
	c_0 & = & -12i\omega + \lambda(2 + \lambda) -12 a\omega \left( a\omega - m \right), \\
	c_1 & = & 8iam\lambda + 8ia^2 \omega( 3 - \lambda), \\
	c_2 & = & -24 i a \left( a \omega - m\right) + 12 a^2 \left[ 1 - 2 \left( a \omega - m \right)^2 \right], \\
	c_3 & = & 24 i a^3 \left( a \omega - m \right) - 24a^2, \\
	c_4 & = & 12a^4.
\end{eqnarray}
\end{subequations}

The asymptotic outgoing mode of $X$ when $r_* \to \infty$
\[ X(r_* \to \infty) \propto f^{\infty}_{+}(r) e^{i\omega r_{*}} = e^{i\omega r_{*}} \left( 1 + \sum_{j=1}^{\infty} \dfrac{\mathcal{C}^{\infty}_{+, j}}{r^j} \right) \]
with the first three expansion coefficients
\begin{subequations}
\label{eq:CplusInf_minus2}
\begin{eqnarray}
	\mathcal{C}^{\infty}_{+, 1} & = & -\dfrac{1}{2 c_{0}}i \left\{-\lambda ^3-2 \lambda ^2 (2+a m \omega )+4 \lambda  \left[-1+\left(3 i-8 a m\right) \omega +7 a^2 \omega ^2\right] \right. \\
	& & \left. + 24 \omega  \left[ i-a^2 \left(1+m^2\right) \omega +a^3 m \omega ^2+i a m (i+\omega )\right]\right\}, \nonumber \\
	\mathcal{C}^{\infty}_{+, 2} & = & -\dfrac{1}{8 c_{0}} \left\{ \lambda ^4+4 \lambda ^3 (1+a m \omega )+4 \lambda ^2 \left[ 1+2 a m \left( 7-i\omega \right) \omega +a^2 \left(-11+m^2\right) \omega ^2\right] \right. \\
	& & \left. -8 a \lambda  \omega \left[ -5 a \omega -15 a m^2 \omega +2 m \left(-4+4 i \omega +7 a^2
\omega ^2\right)\right] \right. \nonumber \\
& & \left. -48 \omega ^2 \left[ -3-a^3 m \left(-5+m^2+2 i \omega \right) \omega +a^4 \left(-4+m^2\right) \omega ^2+2 a m \left(i+\omega \right)+ia^2 \left(-\omega +  5 im^2+3 m^2\omega \right)\right]\right\}, \nonumber \\
	\mathcal{C}^{\infty}_{+, 3} & = & \dfrac{1}{48c_{0}} i
\left\{ -\lambda ^5-6 a m \lambda ^4 \omega -4 \lambda ^3 \left[ -3+2 a m \left(8-3 i \omega \right) \omega +a^2 \left(-13+3 m^2\right) \omega ^2\right] \right. \\
& & \left. -8\lambda ^2 \left[ -2+2 a^2 \left(10+3 m^2 (6-i \omega )\right) \omega ^2+a^3 m \left(-31+m^2\right) \omega ^3-a m \omega  \left(11+12 i \omega +8
\omega ^2\right)\right] \right. \nonumber \\
& & +16 \lambda  \omega  \left[ -9 \omega +3 a^2 \left(5+m^2 (-10+19 i \omega )-3 i \omega \right) \omega -2 a^3 m \left(-11+11m^2+21 i \omega \right) \omega ^2+3 a^4 \left(-10+7 m^2\right) \omega ^3 \right. \nonumber \\
& & \left. \left. +2 a m \left(6+3 i \omega +13 \omega ^2\right)\right] +96 \omega ^2 \left[ 6+am (-3-8 i \omega ) \omega +a^4 \left(9-m^4+2 m^2 (8-3 i \omega )\right) \omega ^2+a^5 m \left(-10+m^2\right) \omega ^3 \right. \right. \nonumber \\
& & \left. \left. +a^3 m \omega  \left(-9+m^2(-12+7 i \omega )+5 i \omega -8 \omega ^2\right)+a^2 \left(-3 i \omega +m^2 \left(6+9 i \omega +14 \omega ^2\right)\right)\right] \right\} \nonumber
.
\end{eqnarray}
\end{subequations}

The asymptotic ingoing mode of $X$ when $r_* \to \infty$
\[ X(r_* \to \infty) \propto f^{\infty}_{-}(r) e^{-i\omega r_{*}} = e^{-i\omega r_{*}} \left( 1 + \sum_{j=1}^{\infty} \dfrac{\mathcal{C}^{\infty}_{-, j}}{r^j} \right) \]
with the first three expansion coefficients
\begin{subequations}
\label{eq:CminusInf_minus2}
\begin{eqnarray}
	\mathcal{C}^{\infty}_{-, 1} & = & -\dfrac{1}{2} i \left( 2 + \lambda +2 a m \omega \right), \\
	\mathcal{C}^{\infty}_{-, 2} & = & \dfrac{1}{8} \left\{ -\lambda ^2-2 \lambda  (1+2 a m \omega )-4 \omega  \left[-3
i+a^2 m^2 \omega +a \left( m+2 i m \omega \right)\right]\right\}, \\
	\mathcal{C}^{\infty}_{-, 3} & = & \dfrac{1}{48} i \left\{ \lambda^3 + \lambda^2 \left( -2+6 a m \omega \right) +4 \lambda  \left[-2-\left(3i + 2am \right) \omega +a \left(2 a+6 i m+3 a m^2\right) \omega ^2\right] \right. \\
	& & \left. + 8 \omega  \left[ 6 i+a^3 m \left(2+m^2\right) \omega ^2+3 a^2 \omega  \left(-1+2 i
m^2 \omega \right)-a m \left(6+3 i \omega +8 \omega ^2\right)\right]\right\}. \nonumber
\end{eqnarray}
\end{subequations}

The conversion factors
\begin{subequations}
\begin{eqnarray}
	\dfrac{B^{\mathrm{ref}}_{\mathrm{T}}}{B^{\mathrm{ref}}_{\mathrm{SN}}} = \dfrac{C^{\mathrm{trans}}_{\mathrm{T}}}{C^{\mathrm{trans}}_{\mathrm{SN}}} & = & -\dfrac{4\omega^2}{c_{0}}, \\
	\dfrac{B^{\mathrm{inc}}_{\mathrm{T}}}{B^{\mathrm{inc}}_{\mathrm{SN}}} & = & -\dfrac{1}{4\omega^2}, \\
	\dfrac{C^{\mathrm{inc}}_{\mathrm{T}}}{C^{\mathrm{inc}}_{\mathrm{SN}}} & = & -\dfrac{4p\sqrt{2r_{+}}}{\eta\left(r_{+}\right)} \left[ 2pr_{+} + i \left( r_{+} - 1 \right)\right],  \\
	\dfrac{B^{\mathrm{trans}}_{\mathrm{T}}}{B^{\mathrm{trans}}_{\mathrm{SN}}} = \dfrac{C^{\mathrm{ref}}_{\mathrm{T}}}{C^{\mathrm{ref}}_{\mathrm{SN}}} & = &  \dfrac{1}{\sqrt{2r_{+}}} \left[ \left( 8 - 24i\omega - 16 \omega^2 \right) r_{+}^2 \right. \\
	& & \left. + \left( 12iam - 16 + 16am\omega + 24i\omega \right) r_{+}  + \left( - 4a^2 m^2 - 12 iam + 8 \right) \right]^{-1}. \nonumber
\end{eqnarray}
\end{subequations}
\end{widetext}

These expressions match those found in literature, for example, Refs.~\cite{Hughes:1999bq, Dolan:2008kf, Piovano:2020zin}. Note again that $\mathcal{C}^{\infty}_{+, j} \neq \left( \mathcal{C}^{\infty}_{-, j} \right)^{*}$ even for real $\omega$ since the \gls{GSN} potentials $\mathcal{F},\mathcal{U}$ are, in general, complex valued.\footnote{This was corrected in the erratum \cite{PhysRevD.78.109902} for Ref.~\cite{Hughes:1999bq}. In both Refs.~\cite{Dolan:2008kf, PhysRevD.78.109902}, expressions for $\mathcal{C}^{\infty}_{+,j}$ written in a form much more concise than that in Eq.~\eqref{eq:CplusInf_minus2} were shown by relating them with the complex conjugate of $\mathcal{C}^{\infty}_{-,j}$. Those expressions are valid only for real $\omega$. We opt to not make such an assumption when deriving the expressions and hence not many simplifications can be made.}

\bibliography{BHPT}%

\end{document}